%% file: main.tex
\documentclass[10pt,journal,compsoc]{IEEEtran}

\usepackage{bm}
\usepackage{subcaption}
\usepackage{amsmath,amsfonts}
\usepackage{algorithm}
\usepackage{array}

\usepackage{textcomp}
\usepackage{stfloats}
\usepackage{url}
\usepackage{verbatim}
\usepackage{graphicx}
\usepackage{cite}
\usepackage{xcolor}         % colors
\usepackage{hyperref}
\hypersetup{colorlinks=true,
            urlcolor=[RGB]{5, 125, 238},
            linkcolor=[RGB]{5, 125, 238},
            citecolor=[RGB]{5, 125, 238}}

\usepackage{algorithmicx}
\usepackage[noend]{algpseudocode}
\usepackage{setspace}

\newcommand{\ie}{\textit{i}.\textit{e}., }

\usepackage{enumitem}
\setlist{nosep}

\begin{document}

% \title{Complete Molecule Generation \\ Based on Joint 2D \& 3D Diffusion}
\title{Learning Joint 2D \& 3D Diffusion Models for Complete Molecule Generation}

\author{Han~Huang, Leilei~Sun, Bowen~Du, Weifeng~Lv \IEEEcompsocitemizethanks{\IEEEcompsocthanksitem H. Huang, L. Sun, B. Du and W. Lv are with the State Key Laboratory of Software Development Environment, Beihang University, Beijing, 100191, China.\protect\\
E-mail: \{h-huang,leileisun,dubowen,lwf\}@buaa.edu.cn
}
        % <-this % stops a space
\thanks{(Corresponding author: Leilei Sun.)}}

% \markboth{Journal of \LaTeX\ Class Files,~Vol.~XX, No.~X, XX~XXXX}%
% {Huang \MakeLowercase{\textit{et al.}}: Learning Joint 2D \& 3D Diffusion Models for Complete Molecule Generation}


\IEEEtitleabstractindextext{%
\begin{abstract}
Designing new molecules is essential for drug discovery and material science. Recently, deep generative models that aim to model molecule distribution have made promising progress in narrowing down the chemical research space and generating high-fidelity molecules. However, current generative models only focus on modeling either 2D bonding graphs or 3D geometries, which are two complementary descriptors for molecules. The lack of ability to jointly model both limits the improvement of generation quality and further downstream applications.
In this paper, we propose a new joint 2D and 3D diffusion model (JODO) that generates complete molecules with atom types, formal charges, bond information, and 3D coordinates. To capture the correlation between molecular graphs and geometries in the diffusion process, we develop a Diffusion Graph Transformer to parameterize the data prediction model that recovers the original data from noisy data. The Diffusion Graph Transformer interacts node and edge representations based on our relational attention mechanism, while simultaneously propagating and updating scalar features and geometric vectors.
Our model can also be extended for inverse molecular design targeting single or multiple quantum properties. In our comprehensive evaluation pipeline for unconditional joint generation, the results of the experiment show that JODO remarkably outperforms the baselines on the QM9 and GEOM-Drugs datasets. Furthermore, our model excels in few-step fast sampling, as well as in inverse molecule design and molecular graph generation.
Our code is provided in \href{https://github.com/GRAPH-0/JODO}{https://github.com/GRAPH-0/JODO}. 
\end{abstract}

}

\maketitle

\input{introdution}

\input{related_work}

\input{methods}

\input{experiments}

\section{Conclusion}

We present a joint 2D and 3D diffusion model based on the Diffusion Graph Transformer to generate complete molecules. 
Our model leverages complementary molecule descriptors to capture accurate molecule distribution and achieves state-of-the-art performance in unconditional generation.
Moreover, our model displays remarkable improvement in conditional generation for quantum properties, demonstrating its ability to generate molecules with desired attributes.   
In future work, we aim to extend our model to challenging scenarios such as generating molecules that fit into protein pockets and RNA structures.
We also intend to enhance the sampling efficiency of equivariant diffusion models to increase the practical utility of our model. 

\section*{Acknowledgments}
This work was supported by the National Natural Science Foundation of China (62272023, 51991391, 51991395).

\bibliographystyle{IEEEtran}
\bibliography{reference}

\clearpage
\input{appendix}

\end{document}

%% file: introdution.tex
\section{Introduction}

Machine learning has been increasingly integrated with molecular science and has made a significant impact, as exemplified by AlphaFold \cite{jumper2021highly} and de novo drug design \cite{zhavoronkov2019deep}. Among the various applications of machine learning for the analysis, design, and simulation of molecules \cite{butler2018machine}, discovering novel molecules with desired properties, is a long-standing challenge that facilitates drug and material design. To avoid brute-force searching in an astronomical number of pharmacologically-sensible molecules, deep generative models provide a powerful approach to narrowing down the chemical search space \cite{polykovskiy20MOSES, du2022molgensurvey}.

Molecules can be represented by different descriptors, leading to different types of generative models. A typical representation is the molecular graph, which describes the 2D bonding topology of a molecule by using nodes for atoms and edges for covalent bonds. This representation is convenient for chemical synthesis, molecular dynamics simulation, etc. Therefore, many graph generative models \cite{Liu18CGVAE, LuoGraphdf21, huangCDGS23, JoLH22GDSS, vignacDigress22} aim to generate realistic and valid molecular structures. However, molecules exist in 3D physical space, and their geometries affect their quantum properties, which are hard to estimate accurately from 2D graphs. Molecular graph generative models may produce unstable molecules or require additional simulations to find low-energy conformers, which hinders inverse molecular design and optimization for quantum properties. Moreover, valuable application scenarios, such as structure-based drug design \cite{anderson2003SBDD}, also depend on the 3D geometry of molecules.

\begin{figure}[t]
    \centering
    \includegraphics[width=0.95\columnwidth]{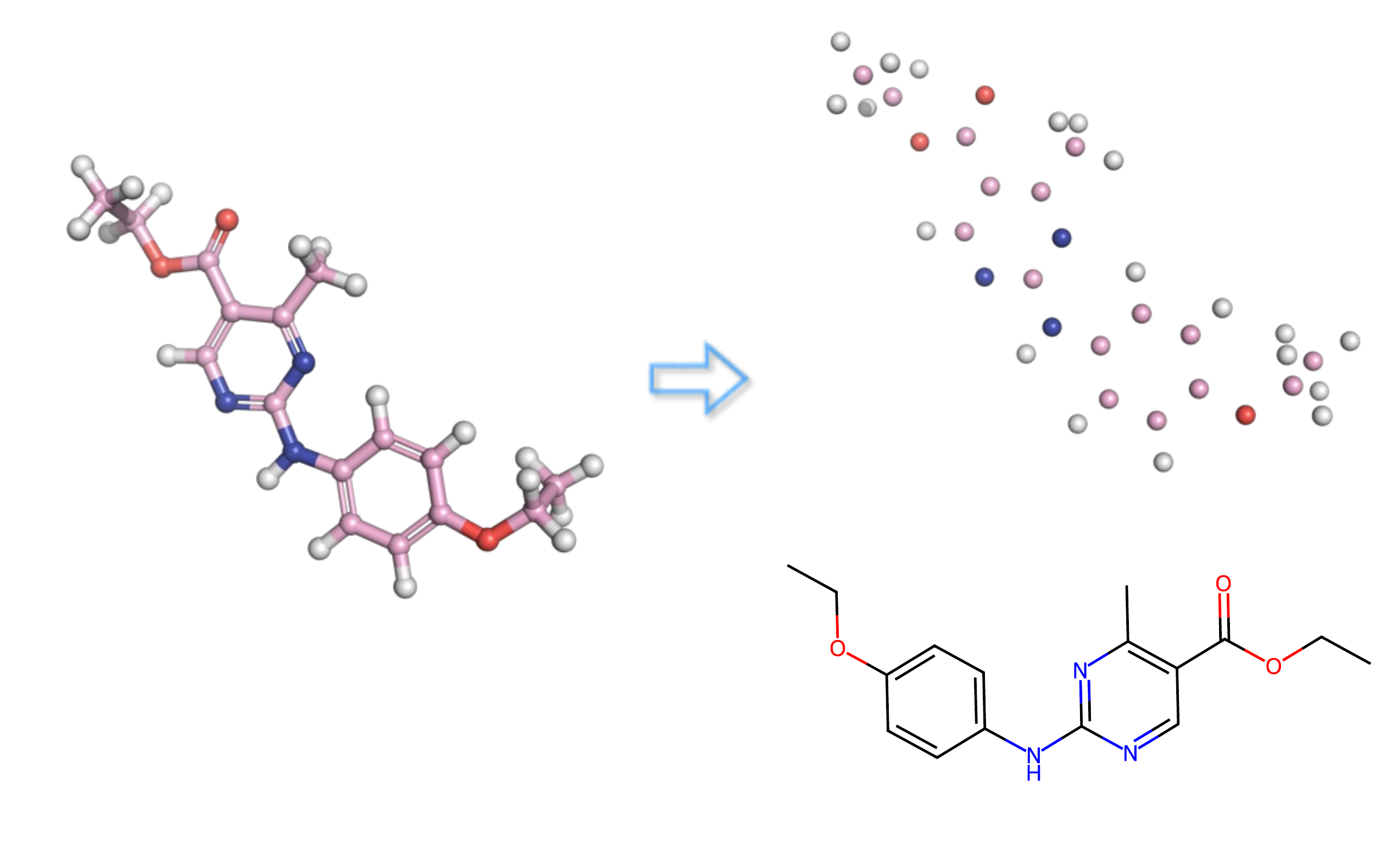}
    \caption{Molecules can be described by complementary 2D bonding graphs and 3D point clouds.}
    \label{fig:exp_geom}
\end{figure}

Generative models that generate atom types and coordinates have received increasing attention in the community \cite{gebauerG-sch19, simm2020symmetry, simm20RL, gebauer2022inverse, garciaE-NF21, luoG-sphere22, HoogeboomEDM22, wu22diffusionbridge, baoEEGSDE22}, in addition to 2D graph generation. By exploiting geometric symmetries such as translations and rotations in Euclidean space (referred to as the SE(3) group), these models manage to generate realistic and stable 3D geometries of small molecules. 
However, atomic 3D positions alone do not contain the bonding information of molecules, which poses some limitations. 
On the one hand, bonding information is essential for the quality evaluation of molecule generation and many downstream applications. The post-processing steps to construct molecular graphs from 3D geometries \cite{garciaE-NF21, HoogeboomEDM22} may introduce errors and are suboptimal. 
On the other hand, directly generating high-quality point clouds of larger drug-like molecules from the GEOM-Drugs dataset \cite{axelrod2022geom} is challenging, while modeling 2D molecular graphs could provide useful guidance for geometry generation intuitively. 
Therefore, we propose that developing generative models that co-design 2D graphs and 3D geometries is a promising direction, which could improve sample quality and facilitate further applications. 
With only a few exploratory works \cite{nesterov20203dmolnet, roney22GEN3D} having attempted joint generation, this area remains largely underexplored not only in model design but also in benchmark pipeline construction.

In this paper, we propose a \textbf{JO}int 2D and 3D \textbf{D}iffusion m\textbf{O}del (JODO) for generating complete molecules that include atom types, formal charges, bond information, and 3D coordinates.
Diffusion models \cite{sohl2015deep, ho2020denoising, song2021score} have shown flexibility in network architecture and have been successfully applied to model the permutation-invariant distribution of graphs \cite{huangGraphgdp22, JoLH22GDSS, huangCDGS23, vignacDigress22}, and the SE(3)-invariant distribution of 3D geometries \cite{HoogeboomEDM22}.
However, constructing diffusion models to learn the complex joint distribution of complete molecules is still nontrivial, as it requires appropriate diffusion mechanisms and powerful networks to handle multimodal noised data in the generative diffusion process.

First, we define our diffusion model as continuous in both time and data space, utilizing a unified noise schedule to gradually corrupt the distribution of the molecule components and their correlation. 
Although some improvements have been made to graph generation by defining diffusion models in discrete space \cite{vignacDigress22}, the preservation of continuity enables the potential advantages of fast sampler design, guidance methods, uncertainty modeling, etc., which motivates \cite{chen2022analog, dieleman2022continuous} to continuousize the discrete input in diffusion models.
Inspired by the self-conditioning technique introduced by \cite{chen2022analog}, we train a data prediction model instead of a noise prediction model to recover original data from corrupted data directly, and exploit the model predictions from the last sampling step as an additional condition to capture accurate graph discreteness and enhance the utilization of model capacity.

Second, we propose parameterizing the data prediction model with a novel Diffusion Graph Transformer that thoroughly interacts node and edge representations via our relational attention mechanism. 
The geometric coordinates also join the propagation through scalarization and update equivariantly along with the scalar features following \cite{satorras21EGNN}. 
We can conveniently plug the extra conditional information, such as noise level or target property label, into our network via adaptive normalization \cite{perez2018film}, similar to the previous diffusion model design \cite{dhariwal2021diffusion, peebles2022scalable}. 

Moreover, with the aim of modeling molecule distribution and generating chemically valid and geometrically stable molecules, we evaluate our models from the perspective of molecular graphs, 3D geometry, and their alignments in unconditional generation. 
Experimental results show that our model outperforms existing baselines in unconditional joint generation.
The proposed model also leads in performance on 2D molecular graph generation and inverse molecule design targeted at quantum properties.

Our main contributions are summarized as follows:

\begin{itemize}[leftmargin=*]
\item
We introduce a joint 2D and 3D diffusion-based end-to-end generative model for complete molecule generation that leverages two complementary descriptors of molecules and expedites further applications.

\item
We develop an effective Diffusion Graph Transformer based on the relational attention mechanism that explicitly interacts node and edge features and captures the correlation between bonding graphs and geometries.

\item
We demonstrate the superior performance of our model in various aspects on the QM9 and Geom-Drug datasets under our comprehensive evaluation pipeline. Our model also excels in few-step sampling, inverse molecule design, and molecular graph generation.
\end{itemize}

We have extended our preliminary work \cite{huangCDGS23} in several important ways:
\textbf{(\romannumeral1)} We examine the complementary effects of 2D bonding graphs and 3D geometry in molecule generation, model their joint distribution, and conduct experiments on larger drug-sized molecules from the GEOM-Drugs Dataset;
\textbf{(\romannumeral2)} We employ the graph transformer architecture taking into account the SE(3) equivariance as a data prediction network in diffusion models;
\textbf{(\romannumeral3)}We capture more accurate graph discreteness from previous predicted data rather than intermediate noised data; 
\textbf{(\romannumeral4)} We enable conditional generation of molecule quantum properties and achieve significant improvement.

%% file: related_work.tex
\section{Related Work}

\subsection{Molecule Generation}

Graph generative models are widely used to generate novel molecular structures based on graph representations. Depending on the sampling process, autoregressive generation constructs molecular graphs step by step with decision sequences \cite{Jin18JT-VAE, Liu18CGVAE, Youpolicy18, ShiGraphaf20, LuoGraphdf21}, while one-shot generation builds all graph components simultaneously \cite{zangMoflow20, lippeGraphcnf21, JoLH22GDSS, huangCDGS23, vignacDigress22}. These models learn graph distributions using different types of generative models, such as variational auto-encoders \cite{simonovsky2018graphvae, Liu18CGVAE, Jin18JT-VAE}, generative adversarial networks \cite{de2018molgan, DEFactor18}, and normalizing flows \cite{ShiGraphaf20, LuoGraphdf21, lippeGraphcnf21, zangMoflow20}. Recently, diffusion-based models, another family of generative models, have shown great potential to generate images \cite{sohl2015deep, ho2020denoising, song2021score}, texts \cite{HoogeboomArgmax21, AustinD3PM21, dieleman2022continuous}, time series \cite{tashiro2021csdi, liu2023pristi}, and diverse domains \cite{yang2022diffusion}.
\cite{JoLH22GDSS, huangCDGS23, vignacDigress22} apply diffusion models for one-shot molecular graph generation without relying on node orderings, achieving performance comparable to autoregressive models. However, these models do not consider the crucial 3D geometry of molecules in the generation process, which limits their further applications.

Generative models for 3D molecules that are represented as attributed point clouds have been explored in different ways. Some methods employ an autoregressive pipeline to sequentially place atoms in 3D space \cite{gebauerG-sch19, luoG-sphere22}, or use reinforcement learning with an autoregressive policy for 3D molecular design \cite{simm2020symmetry, simm20RL}. Another one-shot method \cite{garciaE-NF21} develops an equivariant continuous-time normalizing flow model to generate all 3D coordinates but at a high training cost. EDM \cite{HoogeboomEDM22} utilizes diffusion models to iteratively refine the geometries of molecules sampled from the prior Gaussian distribution in a linear subspace, achieving a significant improvement in generation quality. Based on EDM, some works supplement extra local geometry models \cite{huangMDM22}, or incorporate dataset priors into the generative process \cite{wu22diffusionbridge}. Another line of research focuses on conditional geometry generation for inverse molecule design \cite{gebauer2022inverse, baoEEGSDE22}. However, these models do not model bonding information, which leads to post-processing errors and limited quality on larger drug-like molecules, hindering various downstream tasks. Conformation generation models \cite{xu2022geodiff, torsional22} sample 3D geometries from given molecular graphs, but they cannot sample new molecules and are less challenging than generating from scratch. In contrast to these methods, we propose a joint 2D and 3D generative model for molecule generation. 

\subsection{Diffusion Models for Molecule Science}

Diffusion models have a wide range of applications in various fields of molecular science, not only for designing drug-like small molecules with up to 200 atoms. This section briefly reviews some of their uses in protein design, antibody design, and molecular docking. Protein design is a tough and impactful biological problem that aims to create proteins with desired functions. Diffusion models have been successfully integrated with protein design and have achieved remarkable results \cite{anand2022protein, wu2022protein, ingraham2022illuminating, watson2022broadly}. Proteins are a special type of molecule, consisting of a sequence of amino acids. Therefore, protein generation does not need to predict the 3D coordinates of all atoms and can scale up more easily with fewer degrees of freedom than edge generation in graphs. \cite{luo2022antigen} applies diffusion models for antibody design to generate the amino acid type, position, and orientation in the antigen-antibody complex. Unlike typical generation tasks, \cite{corso2022diffdock} formulates molecular docking as a generative problem to handle uncertainty. It models the ligand pose distribution using a diffusion model over a non-Euclidean manifold. Our model can handle both graphs and geometries, which complements the use of diffusion models in molecular science and inspires further applications.

\subsection{Geometric Graph Neural Networks}

Geometric graph neural networks (GNNs) are core tools for modeling physical objects that have both relational structures and geometries, serving as the foundation for various tasks such as prediction, generation, and simulation. These networks can be divided into two main types: invariant models and equivariant models. Invariant models propagate local invariant scalar geometric features, such as relative distances \cite{schutt2018schnet}, angles \cite{gasteiger2020directional}, and dihedral angles \cite{gasteiger2021gemnet, liu2022spherical}. 
Equivariant models preserve geometric quantities in the message passing along with scalar features, such as Cartesian vectors \cite{satorras21EGNN, schutt2021equivariant, jing2020learning}. 
Recently, \cite{shi2022benchmarking} has extended the 2D graph Transformer \cite{graphformer21} to 3D domains by using 3D distance encoding as attention bias, which can be seen as transmitting messages over fully connected geometric graphs. This model has stronger expressive power than invariant models on local message passing \cite{joshi2023expressive}, and achieves empirical success on large-scale molecular property prediction tasks. It can also use an SE(3)-equivariant prediction head for more general tasks.

Transformer-M \cite{luo2022one} is the most related architecture to ours, as it builds on \cite{graphformer21} and \cite{shi2022benchmarking} and develops separate channels to jointly learn from 2D and 3D molecular data. However, unlike Transformer-M which uses an SE(3)-equivariant prediction head, our Diffusion Graph Transformer propagates and updates scalar features and geometric vectors for each block, while using different relational attention and incorporating extra conditional input from the diffusion process.

%% file: methods.tex
\section{Preliminaries} 

\subsection{Diffusion Models}
% Variational Diffusion Models Definition.

The first step in constructing diffusion models \cite{sohl2015deep, ho2020denoising, song2021score, VDM2021} is to define a forward diffusion process that perturbs data with a sequence of noise until the marginal distribution matches a known prior distribution. 
Let $\mathrm{x}_0 \in \mathbb{R}^{d}$ be a continuous random variable and $\{\mathrm{x}_t\}_{t \in [0,T]}$ be a well-defined forward process.  We have a Gaussian transition kernel as 
\begin{equation}
q_{0t}(\mathrm{x}_t|\mathrm{x}_0) = \mathcal{N}(\mathrm{x}_t | \alpha_t \mathrm{x}_0, {\sigma}^2_t \bm{I}) \ ,
\label{eq:q0t}
\end{equation}
where $\alpha_t, \sigma_t \in \mathbb{R}^{+} $ are time-dependent differentiable functions.
$\alpha_t$ and $\sigma_t$ are usually chosen to ensure that
$q_T(\mathrm{x}_T) \approx \mathcal{N}(\bm{0}, \bm{I})$ with the strictly decreasing signal-to-noise ratio (SNR) $\alpha^2_t / \sigma^2_t$ \cite{VDM2021}. 

By learning to reverse this process, the diffusion model generates new samples from the prior distribution. 
The reverse-time generative diffusion process can be described by the stochastic differential equation (SDE) from time $T$ to $0$ \cite{song2021score, VDM2021} as
\begin{equation}
    \mathrm{d} \mathrm{x}_t = [f(t) \mathrm{x}_t - g^2(t) \nabla_{\mathrm{x}}\log q_t (\mathrm{x}_t)] \mathrm{d}_t 
    + g(t) \mathrm{d} \bar{\bm{w}}_t \ ,
\end{equation}
where $f(t)=\frac{\mathrm{d} \log \alpha_t} {\mathrm{d}t} $ is the drift coefficient,  $g^2(t)= \frac{\mathrm{d}\sigma^2_t}{\mathrm{d}t} - 2\frac{\mathrm{d} \log \alpha_t} {\mathrm{d}t}\sigma^2_t$ is the diffusion coefficient, $\nabla_{x}\log q_t (x_t)$ is the score function and $\bar{\bm{w}}_t$ is the reverse-time standard Wiener process. 
We usually apply a neural network to parameterize the variants of the score function, which can be defined in two alternative ways.
The noise prediction model $\bm{\epsilon}_{\bm{\theta}}(\mathrm{x}_t, t)$ aims to predict the adding noise from $x_t$, equivalent to parameterize $-\sigma_t\nabla_{\mathrm{x}}\log q_t (\mathrm{x}_t)$, while the data prediction model $\bm{d}_{\bm{\theta}}(\mathrm{x}_t, t)$ attempts to directly recover the original data $\mathrm{x}_0$ from $\mathrm{x}_t$.
With the relationship of $\bm{d}_{\bm{\theta}}=(\mathrm{x}_t - \sigma_t\bm{\epsilon}_{\bm{\theta}})/\alpha_t$, both definitions are widely used in diffusion models. 

\subsection{Geometric Graph Representation of Molecules}

A molecule with $N$ atoms can be formulated as a geometric graph $\bm{G} = (\bm{A}, \bm{x}, \bm{h})$, where $\bm{x} = (\bm{x}^1, \cdots, \bm{x}^N) \in \mathbb{R}^{N \times 3}$ denotes the atom coordinates that determine the molecular conformation, $\bm{h} = (\bm{h}^1, \cdots, \bm{h}^N) \in \mathbb{R}^{N \times d1}$ represents the atom features including one-hot encoding of atom types and formal charges of the integer value.
Bond information is encoded in $\bm{A} \in \mathbb{R}^{N \times N \times d2}$, where we typically use one channel for bond existence, one channel for aromatic bond existence, and another channel for bond orders. 
Since the graphs in this paper are undirected and have no self-loops,
the adjacency matrices are symmetric and have zero diagonal entries.
We focus only on the lower triangles of the adjacency matrices in the subspace that can be linearly transformed as $\mathbb{R}^{N(N-1)/2 \times d2}$. 
The subscript $t$ of $\bm{G}_t = (\bm{A}_t, \bm{x}_t, \bm{h}_t)$ indicates the time in the diffusion or generative process.

\subsection{Equivariant Diffusion Models}
Generative modeling of molecules is challenging as the likelihood function of geometric graphs should be invariant to rotations, translations, and permutation. 
We first explain the concepts of equivariance and invariance.
For a transformation $\bm{R}$, a distribution $p(y)$ is invariant to $\bm{R}$ if $p(y) = p(\bm{R}y)$ holds for all $y$, and a conditional distribution $p(y|x)$ is equivariant to $\bm{R}$ if $p(y|x) = p(\bm{R}y|\bm{R}x)$.
A function $f$ is equivariant to $\bm{R}$ when its output is transformed equivalently according to the transformation applied to its input, denoted $\bm{R}f(x) = f(\bm{R}x)$.

It is proved that an SE(3)-invariant prior distribution and an SE(3)-equivariant neural network to parameterize the transition kernels in the diffusion model ensure the marginal distribution is SE(3)-invariant, which is desired for 3D molecule generation.
Refer to \cite{xu2022geodiff, HoogeboomEDM22, baoEEGSDE22} for more detailed analyses.
However, there is no prior nonzero distribution with translation invariance in the full space $\mathbb{R}^{N \times 3}$ \cite{garciaE-NF21}. 
Current methods \cite{xu2022geodiff, HoogeboomEDM22, baoEEGSDE22} use a normal distribution over a linear subspace where the Euclidean variable $\bm{x} = (\bm{x}^1, \cdots, \bm{x}^N)$ satisfies $\sum_i \bm{x}^i = \bm{0}$, that is, the element with zero center of mass (CoM) \cite{kohler2020equivariant}.
The zero CoM subspace is denoted as $X = \{\bm{x} \in \mathbb{R}^{N \times 3}: \frac{1}{N} \sum_{i=1}^{N} \bm{x}^i = \bm{0}\}$.
As the normal distributions are isotropic with rotation invariance and the CoM-free system ensures translation invariance, we construct a desired prior distribution for 3D coordinates.

\section{Methodology}

\begin{figure*}[t]
    \centering
    \includegraphics[width=\textwidth]{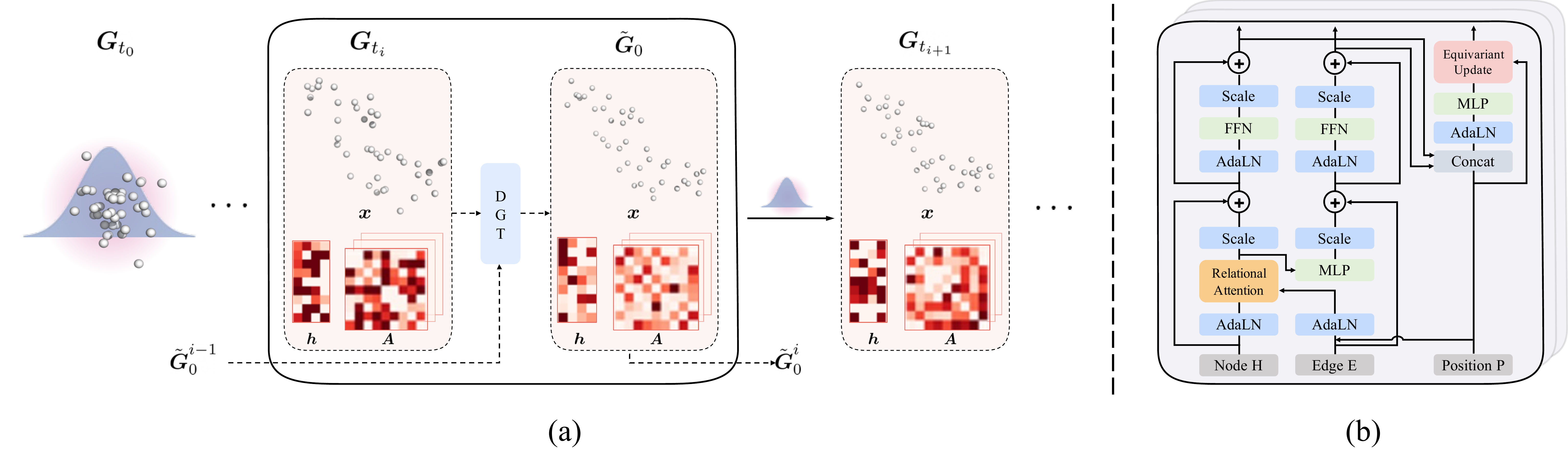}
    
    % \begin{subfigure}{0.39\textwidth}
    % \centering
    % \includegraphics[width=\textwidth]{figs/exp/DGT.pdf}
    % \end{subfigure}
    % \begin{subfigure}{0.6\textwidth}
    % \centering
    % \includegraphics[width=\textwidth]{figs/exp/sampling_exp.pdf}
    % \end{subfigure}
    
    \caption{
    (a) Illustration of the generative diffusion process.
    Molecules are represented by geometric graphs with node attributes $\bm{h}$, edge attributes $\bm{A}$, and 3D coordinates $\bm{x}$.
    Starting from the prior distribution, new molecules are generated iteratively by a data prediction model that takes the previous estimate $\tilde{\bm{G}}_0^{i-1}$ and the current state $\bm{G}_{t_i}$ as input.
    (b) The Diffusion Graph Transformer (DGT) architecture. Three distinct representations interact with each other and are updated within a block.
    Conditional information such as noise level and target quantum properties can be incorporated through the functions $\mathrm{AdaLN}$ and $\mathrm{Scale}$.
    }
    \label{fig:pipeline}
\end{figure*}

In this section, we present JODO, a joint diffusion model for 2D bonding graphs and 3D geometries of molecules. We first define the diffusion process on the geometric graphs and introduce our equivariant Diffusion Graph Transformer that facilitates the generative diffusion process.

\subsection{Joint 2D Graph and 3D Geometry Diffusion}

Representing molecules as $\bm{G} = (\bm{A}, \bm{x}, \bm{h})$,  we construct a continuous-time forward diffusion process in the product space $\mathbb{R}^{N(N-1)/2 \times d2} \times X \times \mathbb{R}^{N \times d1}$ to gradually perturb the distribution of molecular components and their correlations. 
Such a process can be described by a forward SDE  with $t \in [0, T]$ as 
\begin{equation}
\mathrm{d} \bm{G}_t = f(t) \bm{G}_t \mathrm{d}t + g(t) \mathrm{d}(\bm{w}_{\bm{A}}, \bm{w}_{\bm{x}}, \bm{w}_{\bm{h}}) \ ,
\label{eq:forward_SDE}
\end{equation}
where $\bm{w}_{\bm{A}}, \bm{w}_{\bm{x}}, \bm{w}_{\bm{h}}$ are independent standard Wiener processes in the three spaces, respectively. 
We ensure continuity in both time and data space, enabling the use of fast samplers, guidance methods, etc. 
This forward SDE has a linear Gaussian transition kernel in Eq. (\ref{eq:q0t}) to conveniently sample $\bm{G}_t = \alpha_t \bm{G}_0 + \sigma_t \bm{\epsilon_{G}}$ at any time $t$, where $\bm{\epsilon_{G}}$ is the Gaussian noise in the product space.
The corresponding reverse-time SDE from time T to 0 is given by
\begin{equation}
    \mathrm{d} \bm{G}_t = [f(t) - g^2(t) \nabla_{\bm{G}}\log q_t (\bm{G}_t)] \mathrm{d}t 
    + g(t) \mathrm{d}(\bar{\bm{w}}_{\bm{A}}, \bar{\bm{w}}_{\bm{x}}, \bar{\bm{w}}_{\bm{h}}) \ ,
\label{eq:reverse_SDE}
\end{equation}
where $q_t(\bm{G}_t)$ denotes the marginal distribution of data, and $\bar{\bm{w}}_{\bm{A}}$, $\bar{\bm{w}}_{\bm{x}}$, and $\bar{\bm{w}}_{\bm{h}}$ are independent reverse-time standard Wiener processes.
$\nabla_{\bm{G}}\log q_t (\bm{G}_t)$ represents the gradient field of the logarithmic marginal distribution, a.k.a. the score function, which consists of $\nabla_{\bm{A}}\log q_t (\bm{G}_t)$, $\nabla_{\bm{x}}\log q_t (\bm{G}_t) - \overline{\nabla_{\bm{x}}\log q_t (\bm{G}_t)}$, and $\nabla_{\bm{h}}\log q_t (\bm{G}_t)$.
$\overline{\nabla_{\bm{x}}\log q_t (\bm{G}_t)} = \frac{1}{N} \sum_{i=1}^{M} \nabla_{\bm{x}^i}\log q_t (\bm{G}_t)$ is the center of mass (CoM) of the score of geometric variables, which we subtract to keep $\bm{x}_t$ in the CoM-free system \cite{baoEEGSDE22}.

To parameterize the score function variant and facilitate molecule generation via the reversed-time SDE, we train a data prediction model that iteratively predicts the original data in order to transform the random noise into high-fidelity data. 
% to parameterize the score function variant and facilitate molecule generation via the reversed-time SDE.
% In the sampling process of diffusion models, the data prediction model iteratively predicts the original data in order to transform the random noise into high-fidelity data.
Inspired by the self-conditioning technique introduced by \cite{chen2022analog}, we feed the noised data $\bm{G}_t$, the previously estimated $\tilde{\bm{G}}_0$, and the noise level to the data prediction model $\bm{d_\theta}(\bm{G}_t, \tilde{\bm{G}}_0, \log(\alpha^2_t / \sigma^2_t))$. 
Our data prediction model aims to refine previous predictions rather than solely recover original data from the noised data. 
This approach enhances the model capacity utilization and allows us to conveniently use more accurate information like graph discreteness from the end-point data, instead of from the intermediate data as in \cite{huangCDGS23, vignacDigress22}. 

The data prediction model $\bm{d_\theta}(\bm{G}_t, \tilde{\bm{G}}_0, \log(\alpha^2_t / \sigma^2_t))$ produces three outputs in the product space $\mathbb{R}^{N(N-1)/2 \times d2} \times X \times \mathbb{R}^{N \times d1}$, denoted as $(\bm{d_{\theta}^A}, \bm{d_{\theta}^x}, \bm{d_{\theta}^h})$. 
We optimize the model by minimizing the following objective function:
\begin{equation}
\begin{aligned}
    \min_{\bm{\theta}} \mathbb{E}_t\{ \sqrt{\frac{\alpha_t}{\sigma_t}} \mathbb{E}_{\bm{G}_0} \mathbb{E}_{\bm{G}_t|\bm{G}_0} 
    [ \lambda_1 ||\bm{d_{\theta}^A} - \bm{A}_0||^2_2 + \\
    \lambda_2 ||\bm{d_{\theta}^x} - \hat{\bm{x}}_0||^2_2 +
    \lambda_3 ||\bm{d_{\theta}^h} - \bm{h}_0||^2_2]  \} \ ,
\end{aligned}
\end{equation}
where $\lambda_1, \lambda_2, \lambda_3$ are the loss weights for the three outputs.  
The weighting term $\sqrt{\frac{\alpha_t}{\sigma_t}}$ makes the training objective equivalent to the simple noise prediction loss used in \cite{ho2020denoising, song2021score}.
As $\bm{d}_\theta$ is supposed to be SE(3)-equivariant, we follow \cite{xu2022geodiff} to make the geometric supervision signal $\bm{x}_0$ equivariant with $\bm{x}_t$.
We first transform both $\bm{x}_0$ and $\bm{x}_t$ into the CoM-free system, apply the Kabsch alignment algorithm \cite{kabsch1976solution} to find the optimal rotation matrix, and finally obtain $\hat{\bm{x}}_0$ after alignment.
Furthermore, our diffusion model supports both the linear schedule \cite{ho2020denoising} and the cosine schedule \cite{nichol2021improved} to control the SNR. 
We show a more detailed training procedure in Algorithm \ref{alg:training}.

% training
\begin{algorithm}[t]
\setstretch{1.3}
\caption{JODO Training.}
\label{alg:training}
% \textbf{Require}: X

\begin{algorithmic}[1]
    \State $t \sim \mathcal{U}(0,1]$
    \State $\bm{G}_0 = (\bm{A}_0, \bm{x}_0, \bm{h}_0) \sim \mathrm{Training \ Set}$
    \State $\bm{x}_0 \gets \bm{x}_0 - \overline{\bm{x}_0}$
    \State $\bm{A}_t \sim \mathcal{N}(\bm{A}_t | \alpha_t \bm{A}_0, \sigma_t^2 \bm{I}), \ 
    \bm{x}_t \sim \mathcal{N}(\bm{x}_t | \alpha_t \bm{x}_0, \sigma_t^2 \bm{I}), \ 
    \bm{h}_t \sim \mathcal{N}(\bm{h}_t | \alpha_t \bm{h}_0, \sigma_t^2 \bm{I}) $
    \State $\bm{x}_t \gets \bm{x}_t - \overline{\bm{x}_t}$
    \State $\bm{G}_t \gets (\bm{A}_t, \bm{x}_t, \bm{h}_t)$
    \State $\hat{\bm{x}}_0 \gets \mathrm{KabschAlign}(\bm{x}_0, \bm{x}_t)$
    \State $\tilde{\bm{G}}_0 \gets (\tilde{\bm{A}}_0, \tilde{\bm{x}}_0, \tilde{\bm{h}}_0) \gets (\bm{0}, \bm{0}, \bm{0})$
    \If{$\mathrm{Uniform}(0, 1.0) > 0.5 $} 
    \State $\tilde{\bm{A}}_0, \tilde{\bm{x}}_0, \tilde{\bm{h}}_0 \gets \bm{d_{\theta}}(\bm{G}_t, \tilde{\bm{G}}_0, \log(\alpha^2_t / \sigma^2_t))$
    \State $\tilde{\bm{G}}_0 \gets \mathrm{StopGradient}(\tilde{\bm{A}}_0, \tilde{\bm{x}}_0, \tilde{\bm{h}}_0)$
    \EndIf
    \State $\bm{d_{\theta}^A}, \bm{d_{\theta}^x}, \bm{d_{\theta}^h} \gets \bm{d_{\theta}}(\bm{G}_t, \tilde{\bm{G}}_0, \log(\alpha^2_t / \sigma^2_t))$ 
    \State Minimize $\sqrt{\frac{\alpha_t}{\sigma_t}}\ [\gamma_1 ||\bm{d_{\theta}^A} - \bm{A}_0||^2_2 \  + \ 
    \gamma_2 ||\bm{d_{\theta}^x} - \hat{\bm{x}}_0||^2_2 \  + \ 
    \gamma_3 ||\bm{d_{\theta}^h} - \bm{h}_0||^2_2]$
\end{algorithmic}
\end{algorithm}

% sampling
\begin{algorithm}[t]
\setstretch{1.3}
\caption{Sampling from JODO.}
\label{alg:sampling}
\textbf{Require}: time step schedule $\{t_i\}_{i=0}^M$

\begin{algorithmic}[1]
\State $\tilde{\bm{G}}_0 \gets (\tilde{\bm{A}}_0, \tilde{\bm{x}}_0, \tilde{\bm{h}}_0) \gets (\bm{0}, \bm{0}, \bm{0})$
\State $\bm{G}_{t_0} \gets (\bm{A}_T, \bm{x}_T-\overline{\bm{x}_T}, \bm{h}_T)$, where $\bm{A}_T \sim \mathcal{N}(\bm{0}, \bm{I}), \bm{x}_T \sim \mathcal{N}(\bm{0}, \bm{I}), \bm{h}_T \sim \mathcal{N}(\bm{0}, \bm{I})$
\For{$i \gets 1$ to $M$}
    \State $t \gets t_{i-1}, s \gets t_{i}$
    \State $\tilde{\bm{G}}_0 \gets \bm{d_\theta}(\bm{G}_{t}, \tilde{\bm{G}}_0, \log(\alpha^2_t / \sigma^2_t))$
    \State $\bar{\bm{G}}_s \gets \frac{\alpha_{t|s} \sigma_s^2}{\sigma_t^2} \bm{G}_{t} + \frac{\alpha_s \sigma_{t|s}^2}{\sigma_t^2} \tilde{\bm{G}}_0$
    \State $\bm{G}_{\bm{\epsilon}} \gets (\bm{\epsilon_A}, \bm{\epsilon_x} - \overline{\bm{\epsilon_x}}, \bm{\epsilon_h})$, where $\bm{\epsilon_A} \sim \mathcal{N}(\bm{0}, \bm{I}), \bm{\epsilon_x} \sim \mathcal{N}(\bm{0}, \bm{I}), \bm{\epsilon_h} \sim \mathcal{N}(\bm{0}, \bm{I})$
    \State $\bm{G}_s \gets \bar{\bm{G}}_s + \frac{\sigma_{t|s}\sigma_s}{\sigma_t}\bm{G}_{\bm{\epsilon}}$
\EndFor
\State \Return $\bar{\bm{G}}_{t_M}$
\end{algorithmic}
\end{algorithm}

\subsection{Diffusion Graph Transformer}

The generation quality of diffusion models depends largely on the design of data prediction models.
These models need to handle three distinct components in the context of complete molecule generation: node scalar features $\bm{\mathrm{H}}$, edge scalar features $\bm{\mathrm{E}}$, and node 3D coordinates $\bm{\mathrm{P}}$.  
The three components are independently injected with noise and loss correlation gradually during the diffusion process. 
This makes it difficult for the data prediction model to recover the original molecule.
To address this challenge, we propose a Diffusion Graph Transformer (DGT) that parameterizes $\bm{d_\theta}(\bm{G}_t, \tilde{\bm{G}}_0, \log(\alpha^2_t / \sigma^2_t))$. 

DGT adopts the typical Transformer architecture \cite{vaswani2017attention} consisting of multi-head attention (MHA), layer normalization (LN), and feed-forward networks (FFN). 
It can also be considered as a geometric graph neural network that performs message passing over fully connected geometric graphs.
We first explain how DGT extracts from $\bm{G}_t=(\bm{A}_t, \bm{x}_t, \bm{h}_t)$, and incorporates the information from $\tilde{\bm{G}}_0=(\tilde{\bm{A}}_0, \tilde{\bm{x}}_0, \tilde{\bm{h}}_0)$. 
For the initial input of the first block, we project $\bm{h}_t$ and $\bm{A}_t$ to $\bm{\mathrm{H}}^1 \in \mathbb{R}^{N \times b1}$ and $\bm{\mathrm{E}}^1 \in \mathbb{R}^{N \times N \times b2}$ respectively ($b1$ and $b2$ denote the feature dimensions), while $\bm{\mathrm{P}}^1 \gets \bm{x}_t \in \mathbb{R}^{N \times 3}$.
The noise level $\log(\alpha^2_t / \sigma^2_t)$, which corresponds to the timesteps in the diffusion process, is projected through learnable sinusoidal positional embeddings as conditional features $\bm{\mathrm{C}} \in \mathbb{R}^{b3}$.  
The architecture is shown in Figure \ref{fig:pipeline}.

We design a new relational attention mechanism to propagate and update representations on geometric graphs.
For the $l$-th DGT block, we first scalarize the Euclidean variables $\bm{\mathrm{P}}^l$ to distance and encode them to $\Phi_{i,j}^l$ using the Gaussian Basis Kernel function as in \cite{shi2022benchmarking, luo2022one}. 
Then, we augment the edge features as $\bar{\bm{\mathrm{E}}}_{i,j}^l = [\bm{\mathrm{E}}_{i,j}^l; ||\bm{\mathrm{P}}_i^l - \bm{\mathrm{P}}_j^l||^2; \Phi_{i,j}^l]$. 
To preserve the computational complexity $O(N^2)$, we apply dot-product attention to scalar node features, while edge features affect both attention score weights and propagated node features.  
Given $\bm{\mathrm{Q}}$, $\bm{\mathrm{K}}$ and $\bm{\mathrm{V}}$ linearly projected from $\bm{\mathrm{H}}^l$ with dimension $d_k$, a single attention head computes as 
% follows:
\begin{equation}
\begin{aligned}
    & a_{i,j} = \frac{(\mathrm{tanh} (\phi_0 (\bar{\bm{\mathrm{E}}}_{i,j}^l)) \cdot \bm{\mathrm{Q}}_i) \bm{\mathrm{K}}_j^\top}{\sqrt{d_k}}, \ a = \mathrm{softmax}(a), \\ 
    & \mathrm{Attn}(\bm{\mathrm{H}}^l, \bar{\bm{\mathrm{E}}}^l)_{i} =  \sum_{j=0}^{N-1}a_{i,j} (\mathrm{tanh}(\phi_1(\bar{\bm{\mathrm{E}}}_{i,j}^l)) \cdot \bm{\mathrm{V}}_j)\ ,
\end{aligned}
\label{eq:rel_attn}
\end{equation}
where $\phi_0$ and $\phi_1$ are learnable projections, and $\mathrm{tanh}$ is the activation layer.
We extend Eq. (\ref{eq:rel_attn}) to multi-head attention in a standard way and denote it completely as $\mathrm{MHA}(\bm{\mathrm{H}}^l, \bm{\mathrm{E}}^l, \bm{\mathrm{P}}^l)$.
We also use multilayer perceptions (MLP) to learn adaptive scale and shift parameters and define two extra functions to incorporate conditional information such as noise level:
$\mathrm{AdaLN}(\mathrm{h}, \bm{\mathrm{C}}) = (1+\mathrm{MLP}_s(\bm{\mathrm{C}})) \cdot \mathrm{LN}(\mathrm{h}) + \mathrm{MLP}_b(\bm{\mathrm{C}})$ and $\mathrm{Scale}(\mathrm{h}, \bm{\mathrm{C}}) = \mathrm{MLP}_s'(\bm{\mathrm{C}}) \cdot \mathrm{h}$.
These layers are assembled as follows: 
\begin{equation}
  \bm{\mathrm{M}}^{l} = \mathrm{MHA}(\mathrm{AdaLN}(\bm{\mathrm{H}}^l, \bm{\mathrm{C}}), \ 
  \mathrm{AdaLN}(\bm{\mathrm{E}}^l, \bm{\mathrm{C}}), \ 
  \bm{\mathrm{P}}^l) \ , 
\end{equation}
where $\bm{\mathrm{M}}^{l} \in \mathbb{R}^{N \times b1}$ is the intermediate node representation that effectively fuses the three distinct components and serves as the bridge to update scalar and geometric features with high correlation.
We update the scalar node and edge features in parallel as
\begin{equation}
\begin{aligned}
    & {\bm{\mathrm{H}}^{l+1}}' = \mathrm{Scale}(\bm{\mathrm{M}}^{l}, \bm{\mathrm{C}}) + \bm{\mathrm{H}}^{l}, \\
    & \bm{\mathrm{H}}^{l+1} = \mathrm{Scale}(\mathrm{FFN}(\mathrm{AdaLN}({\bm{\mathrm{H}}^{l+1}}', \bm{\mathrm{C}})), \bm{\mathrm{C}}) + {\bm{\mathrm{H}}^{l+1}}', \\
    & \hat{\bm{\mathrm{E}}}_{i,j}^l = (\bm{\mathrm{M}}^l_i + \bm{\mathrm{M}}^l_j) \bm{\mathrm{W}}_1, \ {\bm{\mathrm{E}}^{l+1}}' = \mathrm{Scale}(\hat{\bm{\mathrm{E}}}^l, \bm{\mathrm{C}}) + \bm{\mathrm{E}}^{l}, \\
    & \bm{\mathrm{E}}^{l+1} = \mathrm{Scale}(\mathrm{FFN}(\mathrm{AdaLN}({\bm{\mathrm{E}}^{l+1}}', \bm{\mathrm{C}})), \bm{\mathrm{C}}) + {\bm{\mathrm{E}}^{l+1}}',
\end{aligned}
\end{equation}
where $\bm{\mathrm{W}}_1$ is a learnable matrix.
We equivariantly update the coordinates using the scalar node and edge output as
\begin{equation}
\begin{aligned}
    & \bm{\mathrm{e}}_{i,j}^{l+1} = \mathrm{AdaLN}(\bm{\mathrm{W}}_2[\bm{\mathrm{H}}_i^{l+1}, \bm{\mathrm{H}}_j^{l+1}, \bm{\mathrm{E}}_{i,j}^{l+1}, ||\bm{\mathrm{P}}_i^l - \bm{\mathrm{P}}_j^l||^2] ,\bm{\mathrm{C}}), \\
    & \bm{\mathrm{P}}_i^{l+1} = \sum_{i \neq j} \gamma^l \frac{\bm{\mathrm{P}}_i^l - \bm{\mathrm{P}}_j^l}{||\bm{\mathrm{P}}_i^l - \bm{\mathrm{P}}_j^l||^2} \mathrm{tanh}(\mathrm{MLP}(\bm{\mathrm{e}}_{i,j}^{l+1})),
\end{aligned}
\label{eq:3D_update}
\end{equation}
where $\bm{\mathrm{W}}_2$ is a projection matrix and $\gamma^l \in \mathbb{R}$ is a learnable parameter to control the scale of coordinate update.
The activation function $\mathrm{tanh}$ and $\gamma^l$ help to stabilize the training on geometric vectors.
A complete DGT consists of $L$ consecutive blocks and makes the final scalar node and edge prediction by applying extra MLPs on the concatenation of $L$ block outputs. 
The position output subtracts the CoM to stay in zero CoM subspace.
DGT satisfies the permutation-equivariant property in addition to SE(3)-equivariance, as it does not use any node ordering-dependent operations.

To take advantage of the previous estimate $\tilde{\bm{G}}_0$ for data recovery, we integrate it into the existing architecture.
We concatenate $\tilde{\bm{h}}_0$ and $\tilde{\bm{A}_0}$ with the network initial features $\bm{\mathrm{H}}^1$ and $\bm{\mathrm{E}}^1$ respectively,
and use $\tilde{\bm{x}}_0$ to augment edge features through invariant distance encoding. 
Moreover, we extract a 2D adjacency matrix $\mathrm{A^{2D}}$ and a spatial distance cut-off adjacency matrix $\mathrm{A^{3D}}$ from $\tilde{\bm{G}}_0$, which captures more accurate graph connectivity and spatial arrangement.
By adding $\mathrm{A^{2D}}$ and $\mathrm{A^{3D}}$ as additional heads of attention weights in Eq. (\ref{eq:rel_attn}), we conduct hybrid propagation on geometric graphs.
In Eq. (\ref{eq:3D_update}), we generalize the coordinate update to the multi-head version with aggregation on three adjacency matrix types. 
Thus, we inject self-conditioning information into DGT and preserve its equivariant property.

\subsection{Complete Molecule Generation Process}

Using the optimized data prediction model $\bm{d_\theta}$, we construct the generative diffusion processes via the parameterized reverse-time SDE in Eq. (\ref{eq:reverse_SDE}). 
Various methods can be applied for molecule generation from the SDE, such as the Euler-Maruyama method, ancestral sampling, etc.
We adopt convenient ancestral sampling combined with the data prediction model and self-conditioning to generate complete high-quality molecules, as shown in Figure \ref{fig:pipeline} and Algorithm \ref{alg:sampling}.
For more details on the noise schedule parameters, refer to \cite{VDM2021}.
Intuitively, we sample noisy random data from the prior distribution and iteratively transform it towards realistic data.     
The number of atoms $N$ is sampled from the categorical distribution $p(N)$ computed in the training set before the generative process.

\subsection{Model Variants}

Our diffusion model can be easily extended for conditional generation $\bm{G} \sim p(\bm{G}|c)$ with desired properties $c$.
In practice, we add the representation of conditional properties and the noise level embeddings in DGT.
Through commonly used $\mathrm{AdaLN}(\cdot)$ and $\mathrm{Scale}(\cdot)$, conditional information effectively controls the generation process of three distinct parts of molecules.
Other conditional signals, such as text, can also be integrated into our flexible model in a similar way, facilitating potential applications of language-guided molecule generation and editing.
By removing the 3D components, our model denoted as JODO-2D, can support molecular graph generation and other graph tasks.

%% file: experiments.tex
\section{Experiments}

This section presents the results of JODO on two molecule datasets with both 2D and 3D information, demonstrating its superior molecule generation quality. It also reports the performance of our model on conditional molecule generation with targeted quantum properties and unconditional 2D molecular graph generation. More experimental details are in the Appendix.

\subsection{Joint 2D and 3D Molecule Generation}

\subsubsection{Experimental Setup}

We train and evaluate models on two molecule datasets, QM9 \cite{qm9} and GEOM-Drugs \cite{axelrod2022geom}, which contain 2D bonding graphs and 3D conformations.

QM9 is a widely used molecule dataset that includes bonding information, atom coordinates, and molecular properties for approximately $130K$ small molecules with up to $9$ heavy atoms (up to $29$ atoms including hydrogen). The molecules in QM9 have several types of covalent bonds, such as single, double, and triple bonds. We split the dataset into the training/validation/test partitions of $100K/18K/13K$ samples.

Compared to QM9 with small molecules, GEOM-Drugs is a larger-scale dataset that contains molecules with up to $181$ atoms and an average of $44.4$ atoms. 
GEOM-Drugs provides multiple conformations for each molecule with corresponding energies, and we retain one stable conformation with the lowest energy to construct the dataset.
There are $16$ atom types in this dataset. In addition to single, double, and triple bonds, we explicitly include aromatic bonds.
The training/validation/test split ratio is $8:1:1$.

Models are trained to unconditionally generate complete molecules with 3-dimensional coordinates, atom types (including hydrogen), edge types, and formal charges.
We then sample $10K$ new molecules for evaluation. 

\subsubsection{Evaluation Metrics}

Our goal is to generate chemically valid and complete molecules, model accurate molecular distribution, and obtain well-aligned 2D topology and 3D geometries.
Evaluating the joint generation of molecules is still an open problem. 
We carefully set up our evaluation pipeline to reflect generation quality, but more unified metrics and downstream-specific metrics are still urgently needed.
We evaluate our model from three perspectives to comprehensively check the quality of molecule generation:

(1) \textit{2D molecular graph metrics}.

The benchmark evaluation of molecule generation with connectivity-based description has been extensively studied \cite{du2022molgensurvey}.
We first collect the largest connected component from the generated molecules without post-hoc chemical valency correction and define validity using the RDKit \cite{RDKit} molecular structure parser, which checks the valency of the atoms and the consistency of the aromatic ring bonds.
We report a more challenging metric, the fraction of valid and complete molecules in which all atoms are connected (\textbf{V\&C}), since we train on single connected molecules instead of fragmented ones.
We also report common metrics considering hydrogen, \ie the fraction of valid and unique molecules (\textbf{V\&N}) and the fraction of valid, unique and novel molecules that are not present in the training set (\textbf{V\&U\&N}).
\cite{HoogeboomEDM22} argues that validity could be tricked by reducing the number of bonds and proposes stability metrics, \ie
\textbf{Atom stable} (the fraction of atoms that have precisely the right valency) and \textbf{Mol stable} (the fraction of generated molecules for which all atoms are stable). 
Notably, we take formal charges into consideration for the statistic of the allowed number of bonds per atom.

We then convert all valid molecules to SMILES strings and build distribution learning metrics primarily on the MOSES benchmark \cite{polykovskiy20MOSES}.
\textbf{Fr{\'{e}}chet ChemNet Distance (FCD)} measures the distance between the test set and the generated set with the activation of the penultimate layer of ChemNet. Lower FCD values indicate more similarity between the two distributions.
\textbf{Similarity to the nearest neighbor (SNN)} calculates an average Tanimoto similarity between the fingerprints of a generated molecule and its closest molecule in the test set.
\textbf{Fragment similarity (Frag)} compares the distributions of BRICS fragments \cite{degen2008fragment} in the generated and test sets, and \textbf{Scaffold similarity (Scaf)} compares the frequencies of Bemis-Murcko scaffolds \cite{bemis1996scaffold} between them.

(2) \textit{3D geometry metrics}.

To evaluate the quality of the generated 3D geometry, \cite{garciaE-NF21, HoogeboomEDM22} compute the distances between all pairs of atoms and use a simple lookup table of typical distances in chemistry to determine the bonds and their orders.
Stability metrics are utilized to show the quality of these molecular graphs constructed based on rules.
Higher-quality 3D atom spatial arrangements are expected to result in more typical atomic distances, and thus more stable molecular graphs could be constructed.
We report these metrics to make a fair comparison with previous work, but there are still two concerns about the stability metrics.
First, stability metrics fail to describe the conformation quality in the GEOM-Drugs dataset, which has more complicated and atypical spatial arrangements.
Second, stability metrics could be tricked by predicting more common atom pairs that have a typical distance, without considering the overall atom-type and edge-type distribution in a molecule. 
Therefore, we additionally report the complementary FCD metric as a reference for the rule-based constructed molecule quality evaluation, which focuses more on global molecule similarity.

(3) \textit{Substructure geometry alignment metrics}.
The alignment between the generated 2D molecular graphs and the 3D conformations is a key factor in joint generation.
Obtaining accurate and stable ground truth conformations based on density functional theory as in \cite{axelrod2022geom} is too time-consuming for a large-scale evaluation. 
We instead evaluate the distribution distance of the common substructure geometries between the generated samples and the test set as in \cite{luoG-sphere22, peng2022pocket2mol}.
The misalignment of local geometries indicates that the model may fail to capture stable conformations, showing more irregular behavior. 
Specifically, we select the $8$ most frequent types of bonds, bond pairs, and bond triples, compute the Maximum Mean Discrepancy (MMD) \cite{gretton12MMD} distances of the bond length (\textbf{Bond}), bond angle (\textbf{Angle}), and dihedral angle (\textbf{Dihedral}) distributions separately, and report their mean MMD distances.

% QM9_H
\begin{table*}[t]

\centering
\caption{Performance on the QM9 dataset with explicit hydrogen atoms.}
\label{tab:QM9_H_results}

\begin{subtable}{0.95\textwidth}
\centering
\renewcommand\arraystretch{1.1}
\resizebox{\textwidth}{!}{%
\begin{tabular}{lccccccccc}
\hline
\textbf{\textit{Metric-2D}} & Atom stable $\uparrow$ & Mol stable $\uparrow$ & FCD $\downarrow$ & V\&C $\uparrow$ & V\&U $\uparrow$ & V\&U\&N $\uparrow$ & SNN $\uparrow$ & Frag $\uparrow$ & Scaf $\uparrow$ \\ \hline
\textit{Train} & \textit{99.9 \%} & \textit{98.8 \%} & \textit{0.063} & \textit{98.9 \%} & \textit{98.9 \%} & \textit{0.0 \%}& \textit{0.490} & \textit{0.992} & \textit{0.946} \\
CDGS \cite{huangCDGS23} & 99.7 \% & 95.1 \% & 0.798 & 95.1 \% & 93.6 \% & \textbf{89.8} \% & 0.493 & 0.973 & 0.784 \\
JODO (ours) & \textbf{99.9} \% & \textbf{98.8} \% & \textbf{0.138} & \textbf{99.0} \% & \textbf{96.0} \% & 89.5 \% & \textbf{0.522} & \textbf{0.986} & \textbf{0.934} \\  \hline
\end{tabular}
}
\end{subtable}

\begin{subtable}{0.95\textwidth}
\centering
\renewcommand\arraystretch{1.1}
\resizebox{\textwidth}{!}{%
\begin{tabular}{lcccclccc}
% \hline
\textbf{\textit{Metric-3D}} & Atom stable $\uparrow$ & Mol stable $\uparrow$ & FCD $\downarrow$ &  & \textbf{\textit{Metric-Align}} & Bond $\downarrow$ & Angle $\downarrow$ & Dihedral $\downarrow$ \\ \cline{1-4} \cline{6-9} 
\textit{Train} & \textit{99.4 \%} & \textit{95.3 \%} & \textit{0.877} &  & \textit{Train} & \textit{5.44e-4} & \textit{4.65e-4} & \textit{1.78e-4} \\
E-NF \cite{garciaE-NF21} & 84.7 \% & 4.5 \% & 4.452 &  & E-NF \cite{garciaE-NF21} & 0.6165 & 0.4203 & 0.0056 \\
G-SchNet \cite{gebauerG-sch19} & 95.7 \% & 68.1 \% & 2.386 &  & G-SchNet \cite{gebauerG-sch19} & 0.3622 & 0.0727 & 0.0042 \\
G-SphereNet \cite{luoG-sphere22} & 67.2 \% & 13.4 \% & 6.659 &  & G-SphereNet \cite{luoG-sphere22} & 0.1511 & 0.3537 & 0.0129 \\
EDM \cite{HoogeboomEDM22} & 98.6 \% & 81.7 \% & 1.285 &  & EDM \cite{HoogeboomEDM22} & \textbf{0.1303} & 0.0182 & 6.64e-4 \\
MDM \cite{huangMDM22} & \textbf{99.2} \% & 89.6 \% & 4.861 &  & MDM \cite{huangMDM22} & 0.2735 & 0.0660 & 0.0239 \\
JODO (ours) & \textbf{99.2} \% & \textbf{93.4} \% & \textbf{0.885} &  & JODO (ours) & 0.1475 & \textbf{0.0121} & \textbf{6.29e-4} \\ 
\hline
\end{tabular}%
}
\end{subtable}

\end{table*}

% Geom Drug 
\begin{table*}[t]
\centering
\caption{Performance on the GEOM-Drugs dataset.}
\label{tab:geom_drug_results}

\begin{subtable}{0.95\textwidth}
\centering
\renewcommand\arraystretch{1.1}
\resizebox{\textwidth}{!}{%
\begin{tabular}{lccccccccc}
\hline
\textbf{\textit{Metric-2D}} & Atom stable $\uparrow$ & Mol stable $\uparrow$ & FCD $\downarrow$ & V\&C $\uparrow$ & V\&U $\uparrow$ & V\&U\&N $\uparrow$ & SNN $\uparrow$ & Frag $\uparrow$ & Scaf $\uparrow$ \\ \hline
\textit{Train} & \textit{100.0 \%} & \textit{100.0 \%} & \textit{0.251} & \textit{100.0 \%} & \textit{100.0 \%} & \textit{0.0 \%} & \textit{0.585} & \textit{0.999} & 0.584 \\
CDGS \cite{huangCDGS23} & 99.1 \% & 70.6 \% & 22.051 & 28.5 \% & 28.5 \% & 28.5 \% & 0.262 & 0.789 & 0.022 \\
JODO (ours) & \textbf{100.0} \% & \textbf{98.1} \% & \textbf{2.523} & \textbf{87.4} \% & \textbf{90.5} \% & \textbf{90.2} \% & \textbf{0.417} & \textbf{0.993} & \textbf{0.483} \\ \hline
\end{tabular}}
\end{subtable}

\begin{subtable}{0.95\textwidth}
\centering
\renewcommand\arraystretch{1.1}
\resizebox{\textwidth}{!}{%
\begin{tabular}{lcccclccc}
\textbf{\textit{Metric-3D}} & Atom stable $\uparrow$ & Mol stable $\uparrow$ & FCD $\downarrow$ & & \textbf{\textit{Metric-Align}} & Bond $\downarrow$ & Angle $\downarrow$ & Dihedral $\downarrow$ \\ \cline{1-4} \cline{6-9} 
\textit{Train} & \textit{86.1 \%} & \textit{2.8 \%} & \textit{13.733} & & \textit{Train} & \textit{1.56e-4} & \textit{1.81e-4} & \textit{1.56e-4} \\
EDM \cite{HoogeboomEDM22} & 83.1 \% & 0.2 \% & 31.290 & & EDM \cite{HoogeboomEDM22} & 0.4286 & 0.4959 & 0.0146 \\
% MDM \cite{huangMDM22} & xx.x \% & x.x \% & xx.xxx &  & MDM \cite{huangMDM22} & x.xxxx & x.xxxx & x.xxxx \\
JODO (ours) & \textbf{84.5} \% & \textbf{1.0} \% & \textbf{19.993} & & JODO (ours) & \textbf{0.0849} & \textbf{0.0115} & \textbf{6.68e-4} \\ \hline
\end{tabular}%
}
\end{subtable}

\end{table*}

% visualization
\begin{figure*}[!htbp]
\centering
\includegraphics[width=0.9\textwidth]{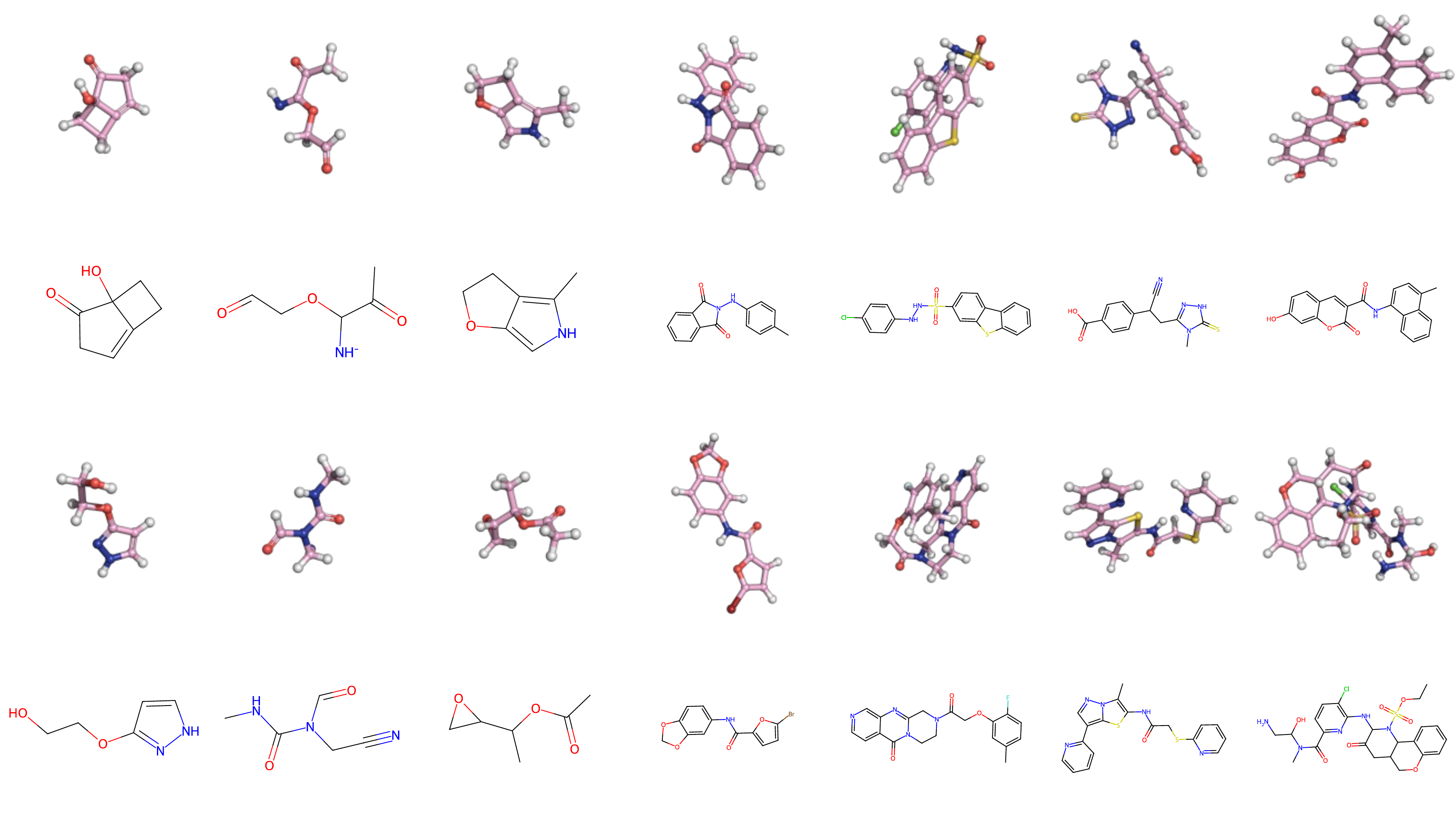}
\caption{Visualization of molecules generated by JODO trained on QM9 (left three columns) and GEOM-Drugs (right four columns).
The 2D molecular graphs are shown below their corresponding 3D geometries.}
\label{fig:mol_vis}
\end{figure*}

% Distribution Analysis
\begin{figure*}[t]
    \centering
    
    \begin{subfigure}{\columnwidth}
    \centering
    \includegraphics[width=\textwidth]{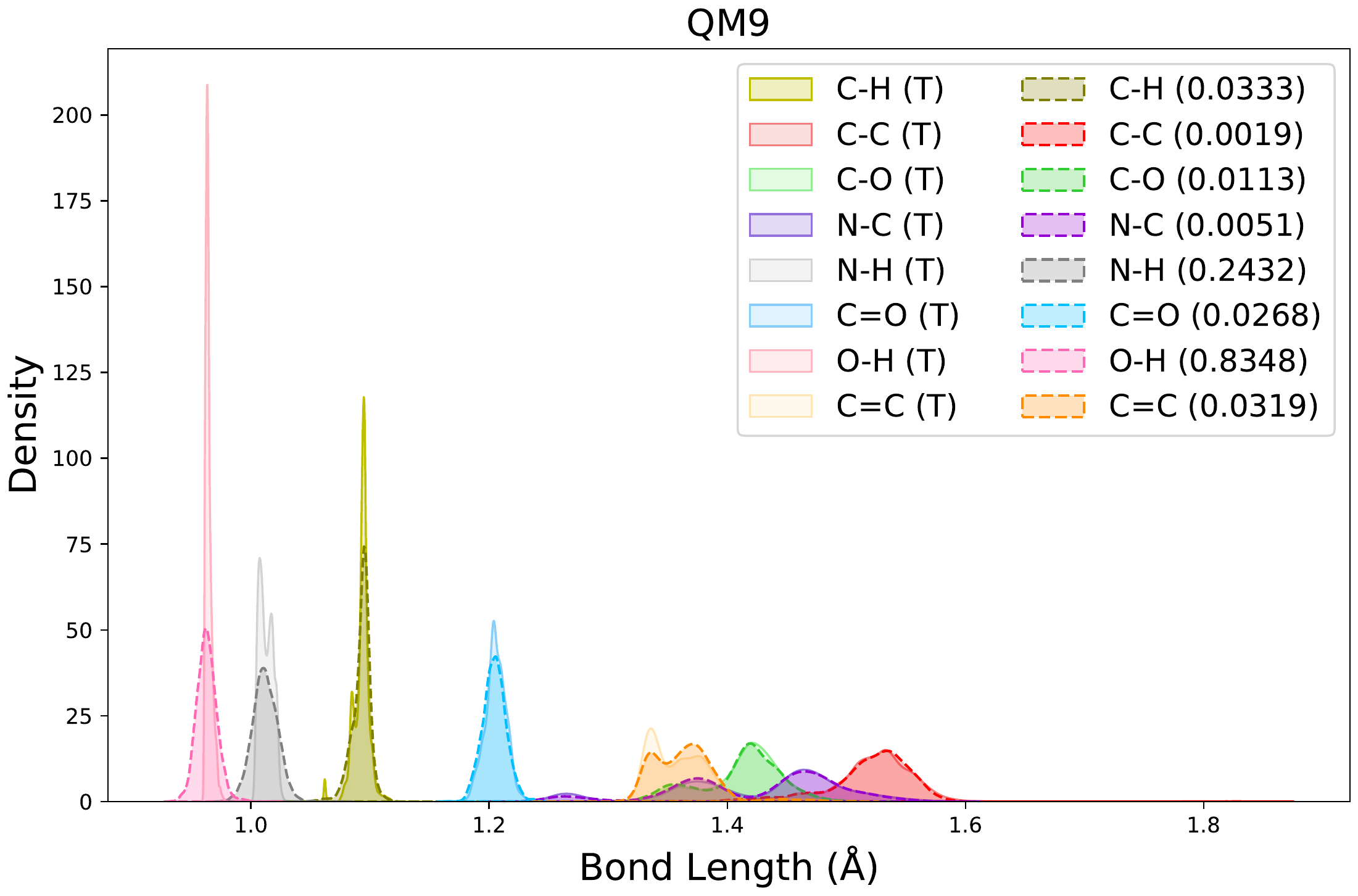}
    \end{subfigure}
    % \hfill
    \begin{subfigure}{\columnwidth}
    \centering
    \includegraphics[width=\textwidth]{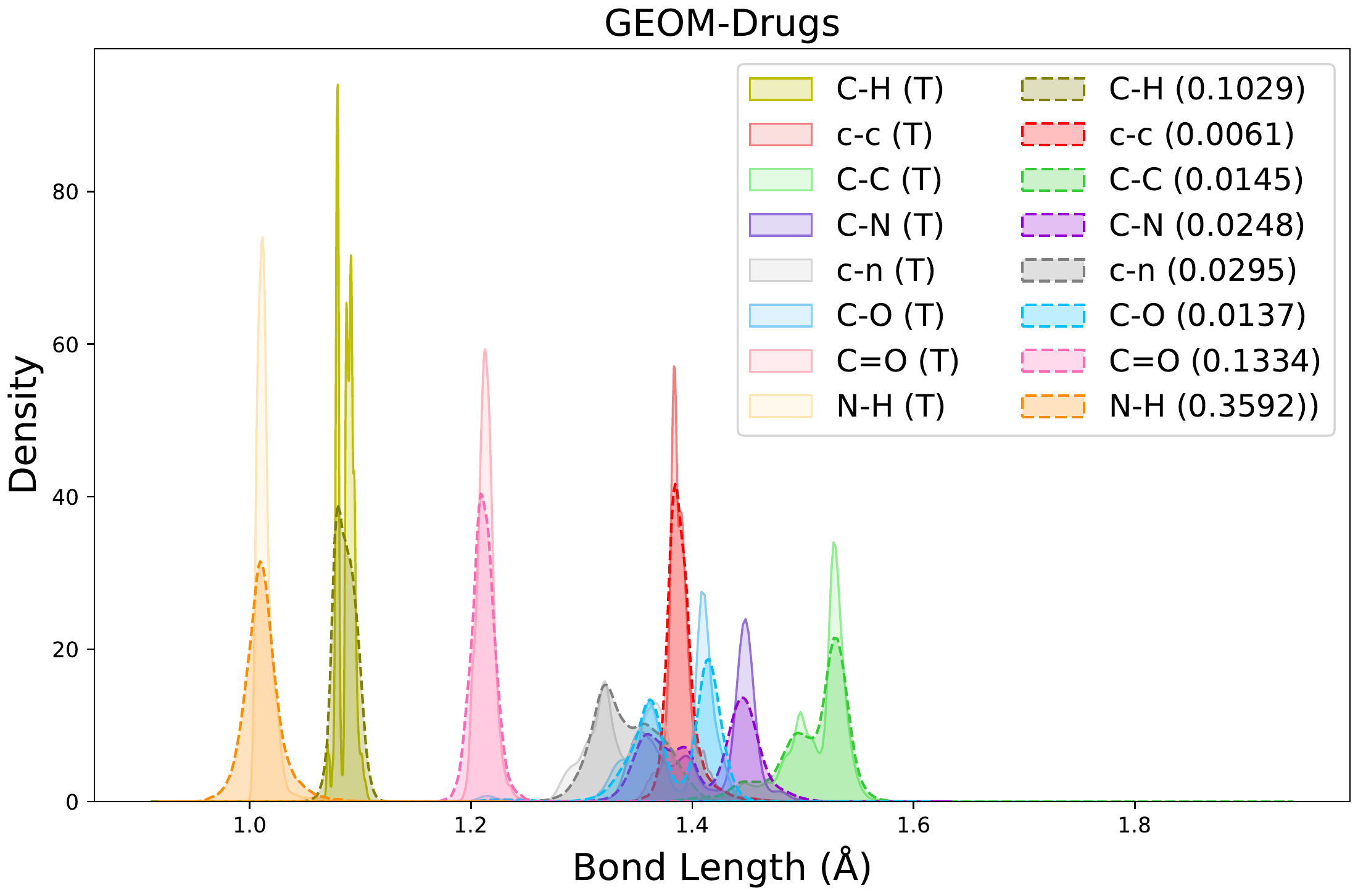}
    \end{subfigure}

    \hfill

    \begin{subfigure}{0.33\columnwidth}
    \centering
    \includegraphics[width=\textwidth]{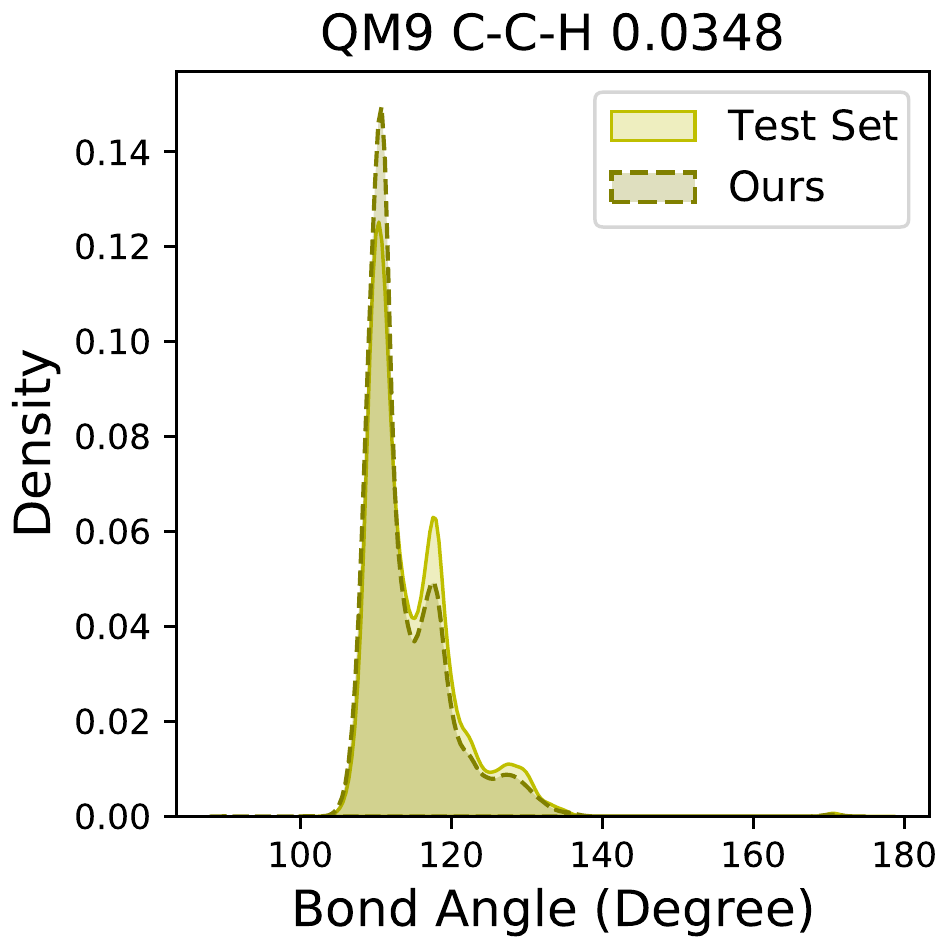}
    \end{subfigure}
    \begin{subfigure}{0.33\columnwidth}
    \centering
    \includegraphics[width=\textwidth]{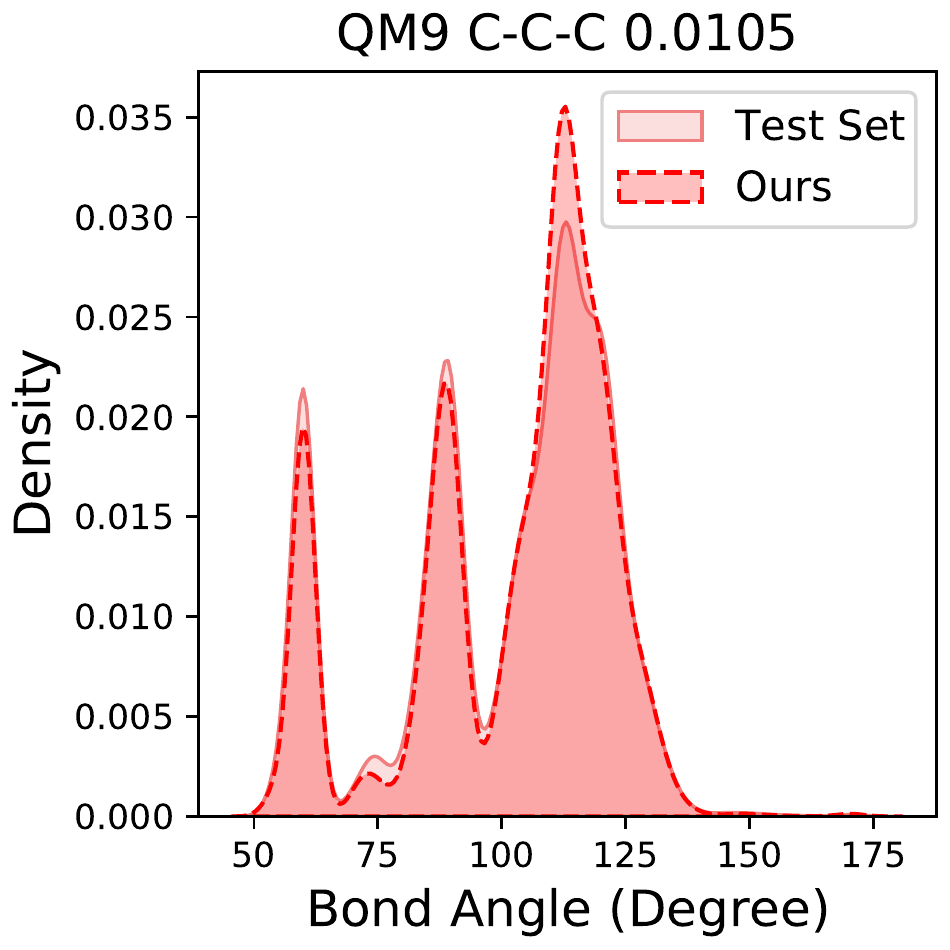}
    \end{subfigure}
    \begin{subfigure}{0.33\columnwidth}
    \centering
    \includegraphics[width=\textwidth]{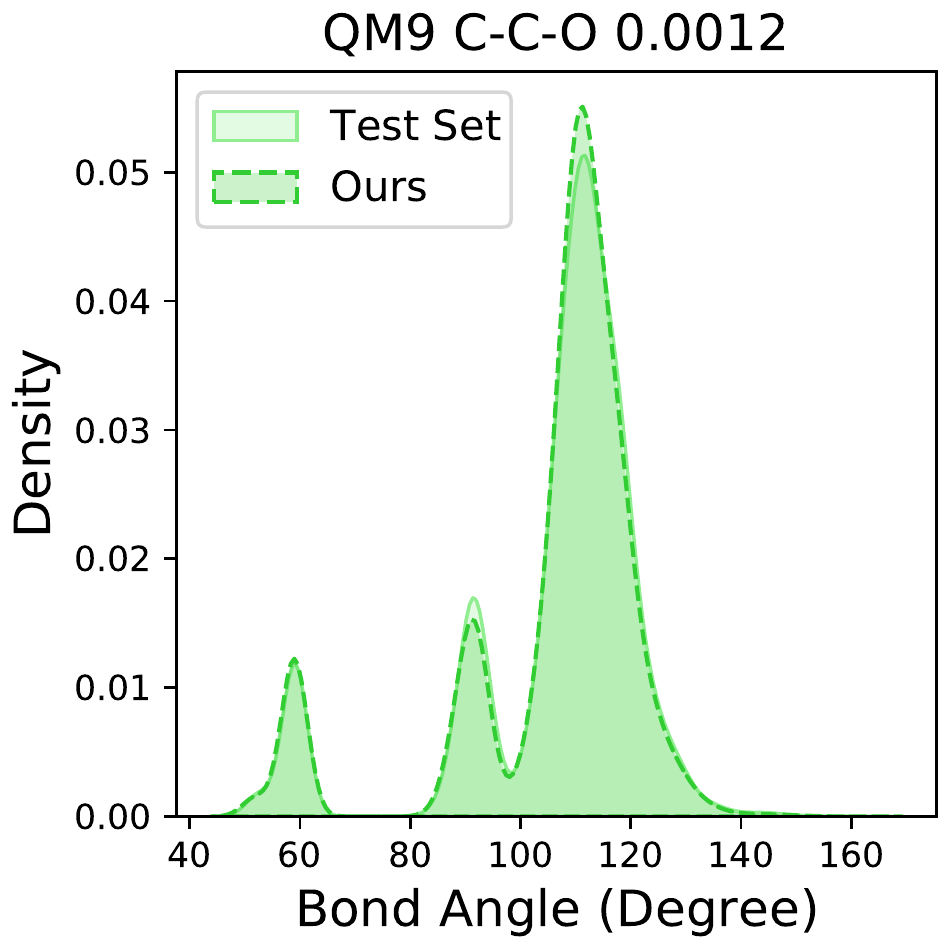}
    \end{subfigure}
    \begin{subfigure}{0.33\columnwidth}
    \centering
    \includegraphics[width=\textwidth]{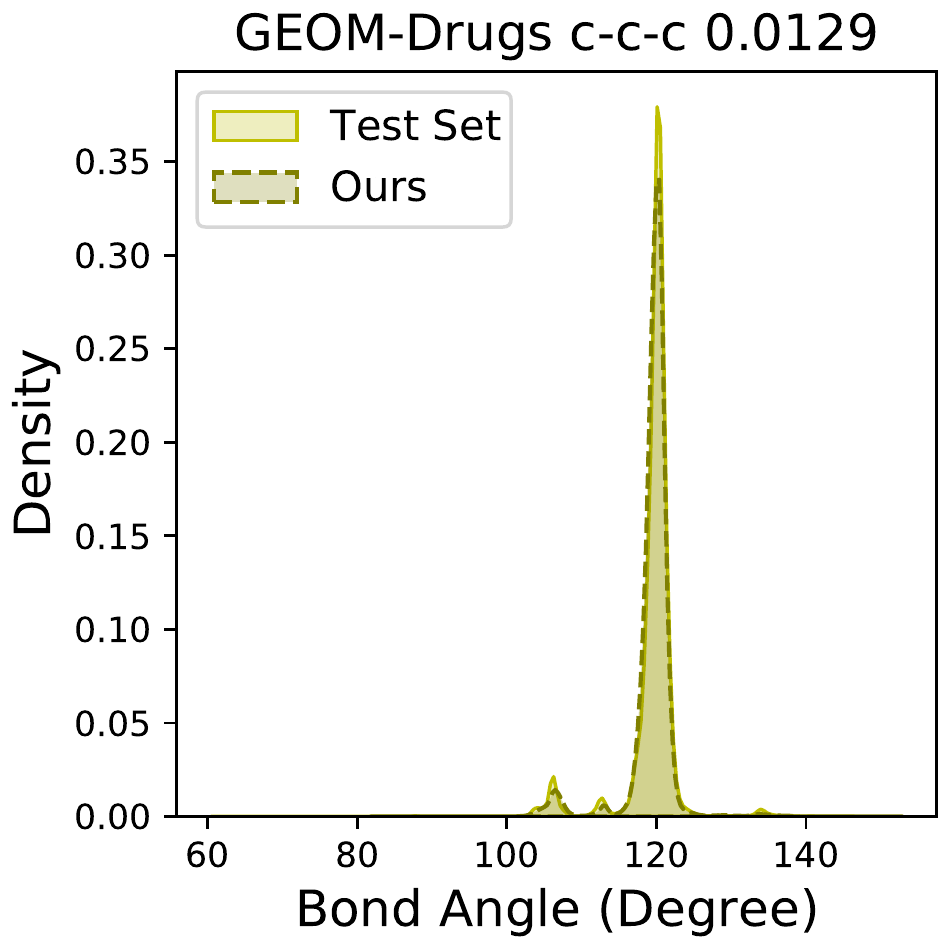}
    \end{subfigure}
    \begin{subfigure}{0.33\columnwidth}
    \centering
    \includegraphics[width=\textwidth]{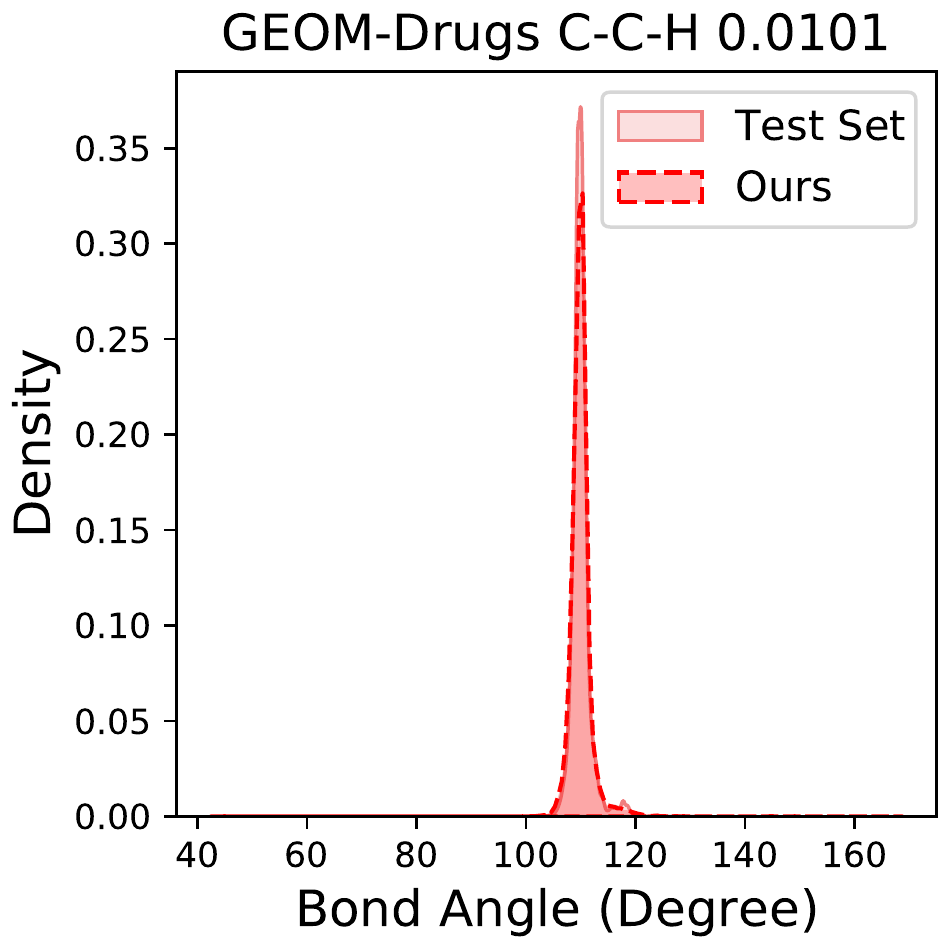}
    \end{subfigure}
    \begin{subfigure}{0.33\columnwidth}
    \centering
    \includegraphics[width=\textwidth]{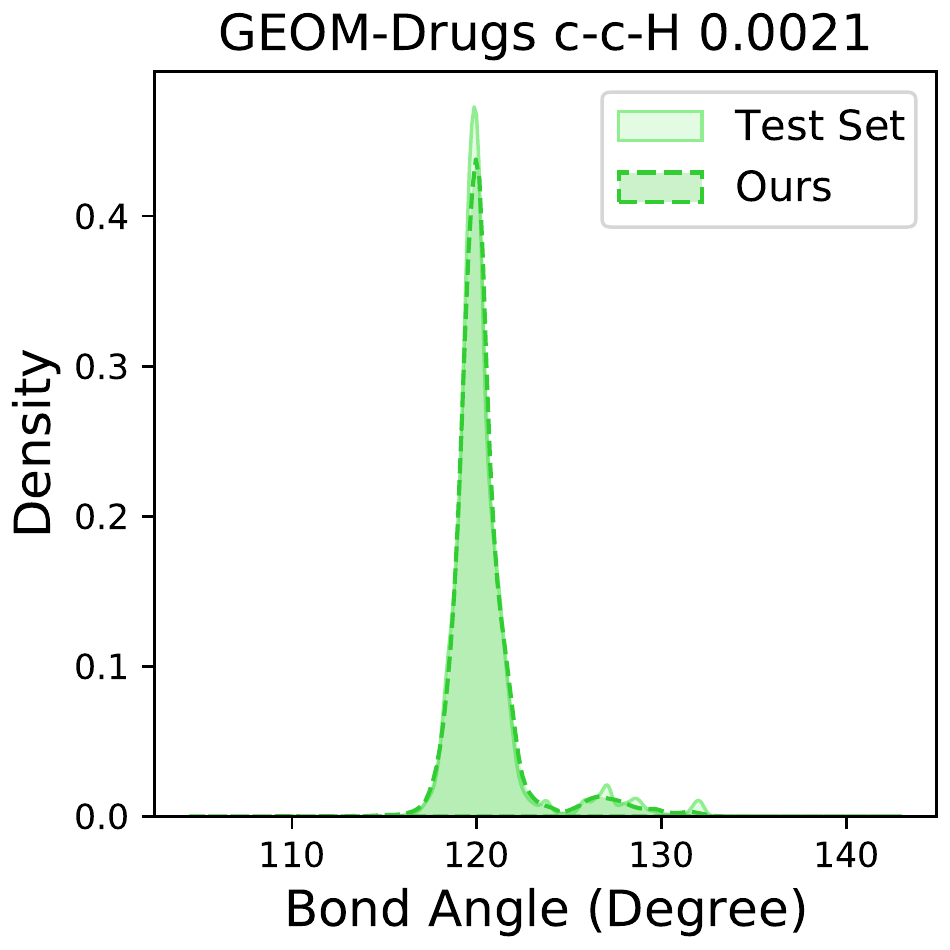}
    \end{subfigure}

    \hfill

    \begin{subfigure}{0.33\columnwidth}
    \centering
    \includegraphics[width=\textwidth]{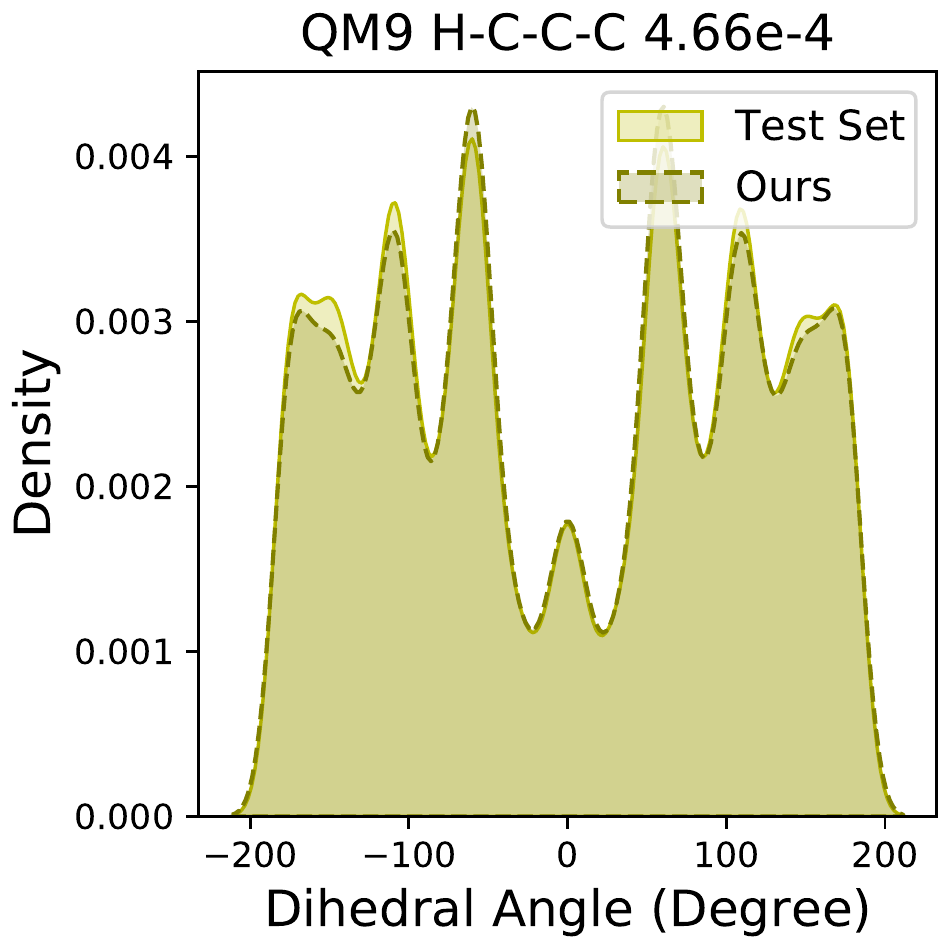}
    \end{subfigure}
    \begin{subfigure}{0.33\columnwidth}
    \centering
    \includegraphics[width=\textwidth]{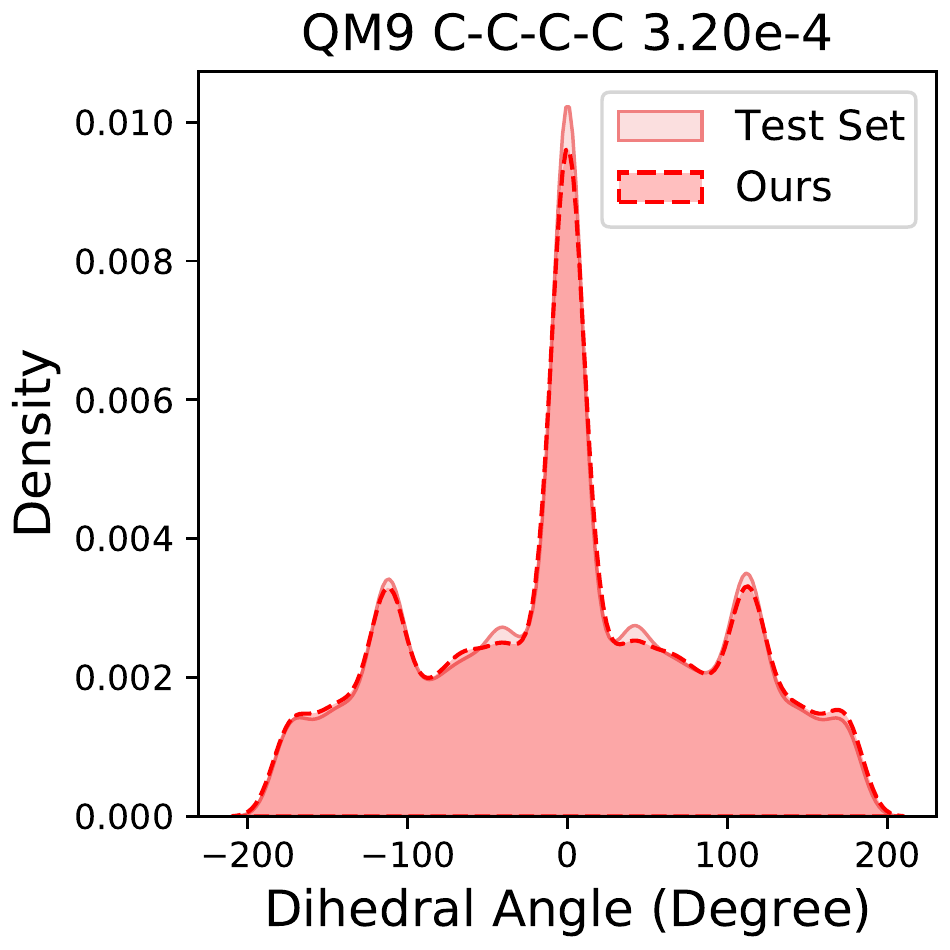}
    \end{subfigure}
    \begin{subfigure}{0.33\columnwidth}
    \centering
    \includegraphics[width=\textwidth]{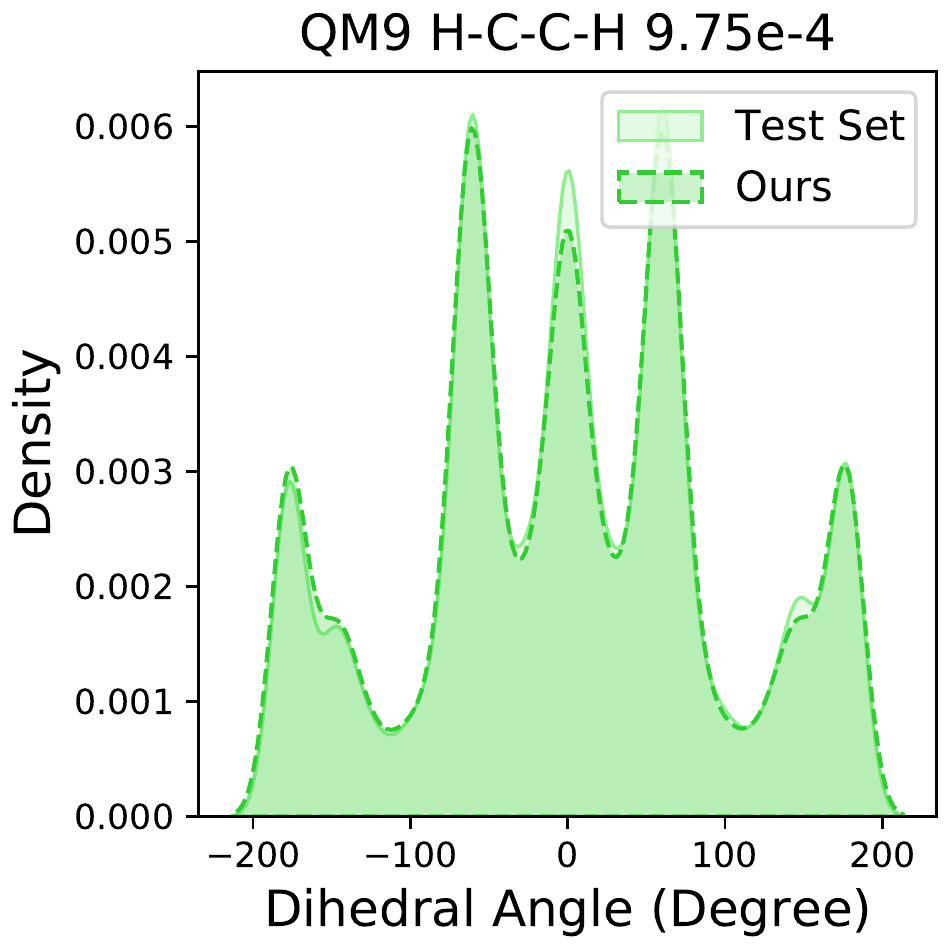}
    \end{subfigure}
    \begin{subfigure}{0.33\columnwidth}
    \centering
    \includegraphics[width=\textwidth]{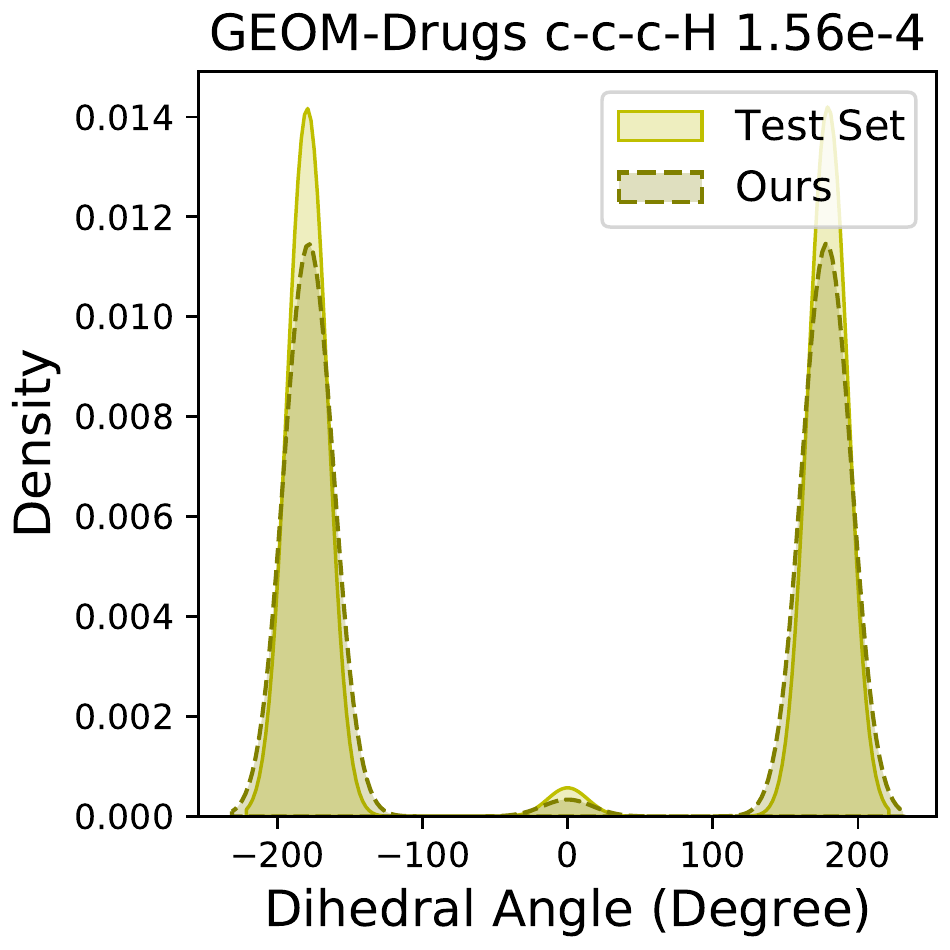}
    \end{subfigure}
    \begin{subfigure}{0.33\columnwidth}
    \centering
    \includegraphics[width=\textwidth]{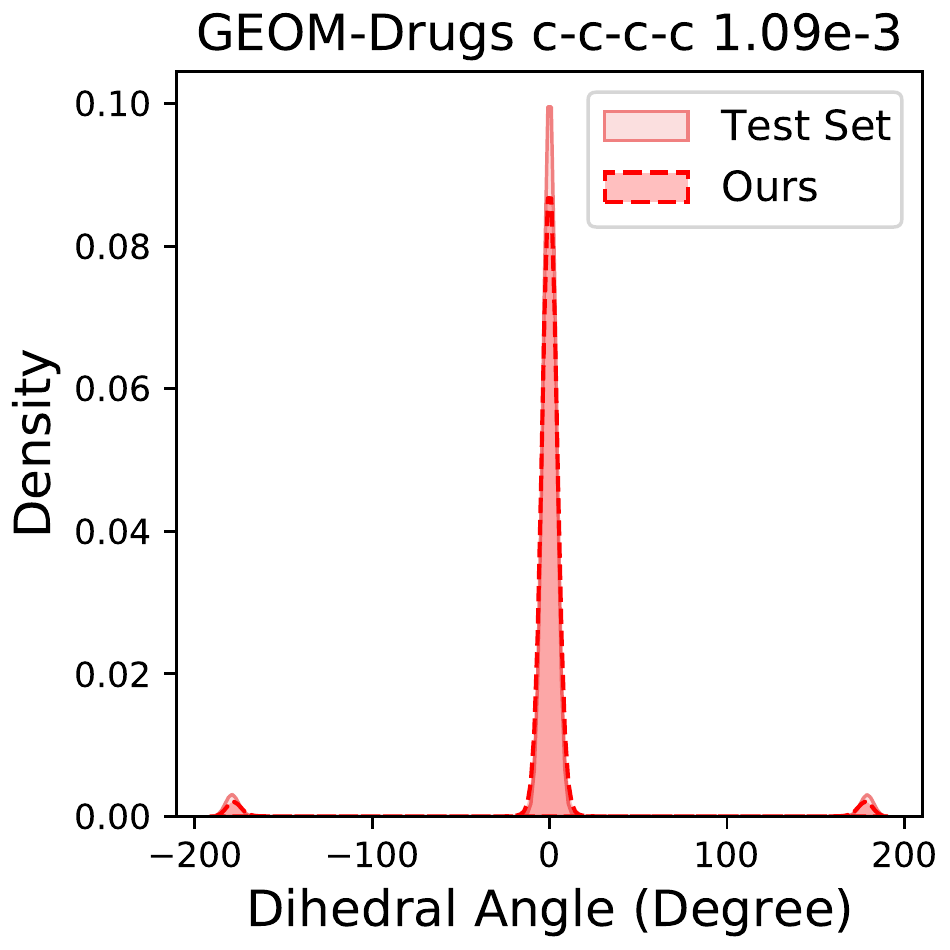}
    \end{subfigure}
    \begin{subfigure}{0.33\columnwidth}
    \centering
    \includegraphics[width=\textwidth]{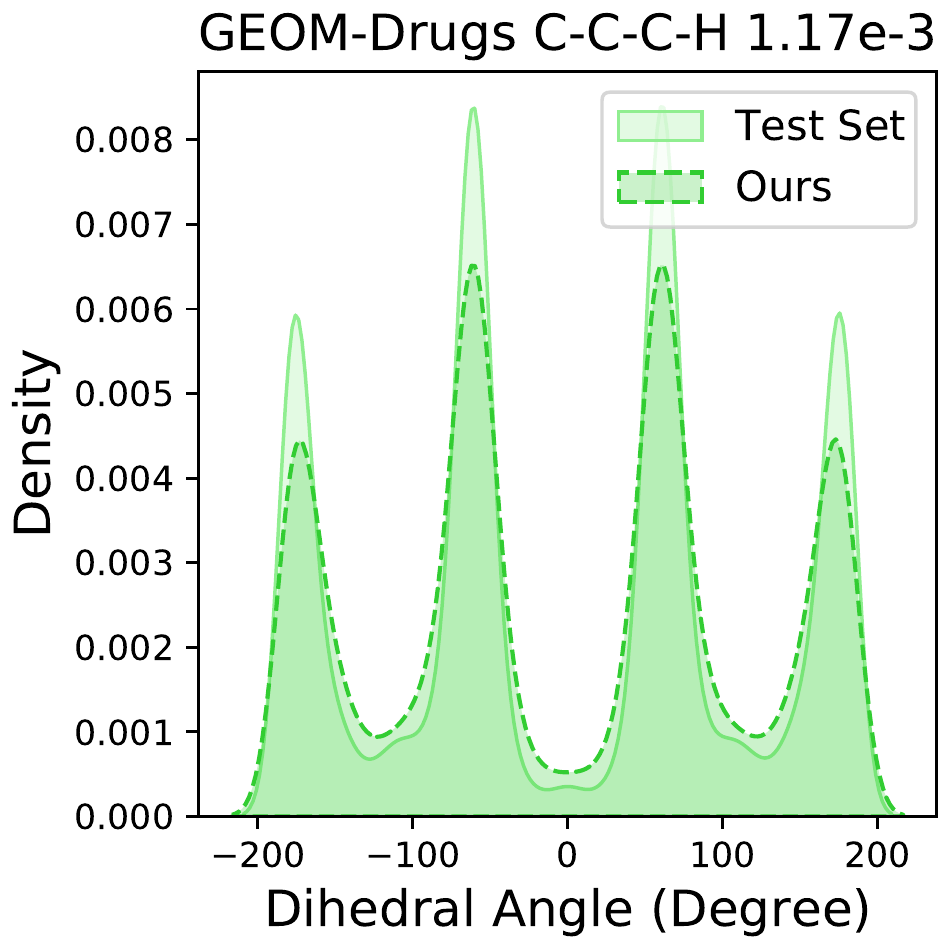}
    \end{subfigure}

    \caption{Distribution comparison of bond lengths, bond angles, and dihedral angles between test set molecules and JODO generated molecules.
    MMD distances are also reported (lower values are better).
    Better view by zooming in.
    }
    \label{fig:sub_geo_dist}
\end{figure*}

\subsubsection{Baselines}
We compare JODO with existing 2D molecular graph generative models and 3D equivariant generative models.
Among the 2D graph models, we select CDGS \cite{huangCDGS23} as the representative model, which achieves great performance in the complicated Zinc250k dataset \cite{zinc250k} in a permutation-invariant one-shot generation form. 
For the 3D models, we consider G-schNet \cite{gebauerG-sch19} and G-SphereNet \cite{luoG-sphere22}, which are autoregressive models that arrange the atom positions sequentially.
We also include Equivariant Normalizing Flows (E-NF) \cite{garciaE-NF21}, a continuous-time normalizing flow model that is independent of unnatural atom orderings, and E(3) Equivariant Diffusion Model (EDM) \cite{HoogeboomEDM22} and Molecular Diffusion Model (MDM) \cite{huangMDM22}, which are diffusion-based models that iteratively generate 3D coordinates and atom types.
We run the publicly available code of baselines and generate $10K$ samples for each model.
Some baselines fail to scale to the GEOM-Drugs dataset for reasonable results and are therefore omitted.
Additionally, we sample $10K$ molecules from the training set three times and report average results as the performance upper bound.

\subsubsection{Generation Quality}

The results of the QM9 dataset and the GEOM-Drugs dataset are presented in Table \ref{tab:QM9_H_results} and Table \ref{tab:geom_drug_results}, respectively. 
Some visualizations of the molecules generated by the proposed model are shown in Figure \ref{fig:mol_vis}.
The proposed method is run three times, and the mean performance is reported.

Our proposed JODO outperforms 2D and 3D baselines remarkably from three evaluation perspectives on both datasets.
For the QM9 dataset including hydrogen atoms, by taking advantage of the 2D and 3D joint modeling and the diffusion model design, JODO learns accurate molecule distribution and generates high-fidelity molecules. It achieves near-optimal performance in 2D metrics and significant improvement in 3D geometry metrics.
The GEOM-Drugs dataset poses a great challenge for one-shot generative models because of the large number of atoms per molecule.
JODO surpasses the previous state-of-the-art one-shot diffusion-based CDGS \cite{huangCDGS23} by a large margin in 2D metrics, especially FCD and V\&C, indicating that JODO can generate more valid and connected drug-sized molecules.
Although the stability metrics of the 3D evaluation on the GEOM-Drugs dataset are likely to be tricked, better models still tend to have lower FCD values and closer stability ratios to the training set. Therefore, our model generates conformations that are more reasonable than EDM \cite{HoogeboomEDM22}.
For the alignment between bond-connective molecular graphs and 3D geometries on two datasets, JODO exhibits excellent MMD performance, consistent with the distributions of common substructure geometries in the test set. 

A notable observation is that even for the samples in the training set, the FCD values of the molecules constructed from the 3D coordinates based on the typical distance lookup table are obviously lower than those calculated from the topological molecular descriptors.
This implies a detrimental distribution shift in the chemical space.
Therefore, the efficient rule-based bond post-processing method used in \cite{garciaE-NF21, HoogeboomEDM22, huangMDM22} struggles to handle larger molecules,
which also motivates us to develop end-to-end joint 2D and 3D generative models to directly generate complete molecules.

\subsubsection{Further Analysis}

% To further examine the performance of our model in generating well-aligned 2D topologies and 3D geometries, we display the distribution of the most common bond lengths, bond angles, and dihedral angles in the test set and generated samples in Figure  \ref{fig:sub_geo_dist}, rather than reporting the average MMD values.

We further examine the performance of our model in generating well-aligned 2D topologies and 3D geometries by comparing the distributions of the most common bond lengths, bond angles and dihedral angles in the test set and the generated samples (Figure \ref{fig:sub_geo_dist}), instead of only reporting the mean MMD values.
Atom types and bond types are combined sequentially to form substructure identifiers (lowercase letters are used for atom types linked by aromatic bonds).
The MMD distances of the substructure distributions are also included in the figure.
It can be observed that different types of bond lengths show differences in the distance distribution, and our model captures this distribution difference and fits the distribution shape of the bond lengths well.
The performance gap mainly comes from some distribution shapes with high peaks where the bond length is more stable in the small value range.
For bond angles and dihedral angles, our model excellently represents their distributions, even those with multiple peaks.
The excellent alignment indicates that our model generates stable local geometries.
We provide more distribution alignment figures of bond angles and dihedral angles in Appendix.

\begin{figure}[t]
    \centering

    \begin{subfigure}{0.49\columnwidth}
        \centering
        \includegraphics[width=\columnwidth]{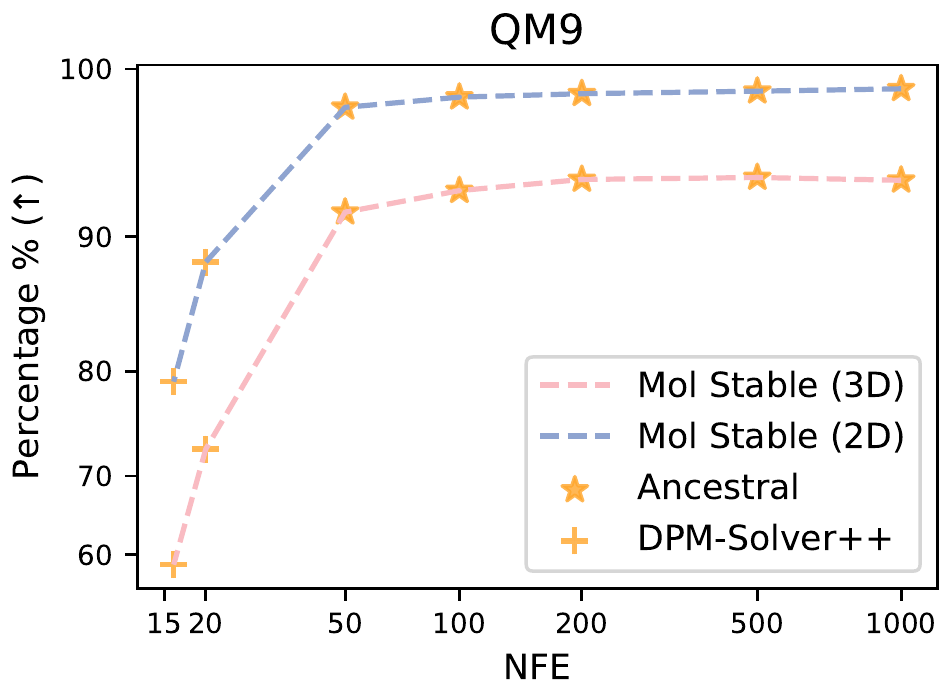}
    \end{subfigure}
    \begin{subfigure}{0.49\columnwidth}
        \centering
        \includegraphics[width=\columnwidth]{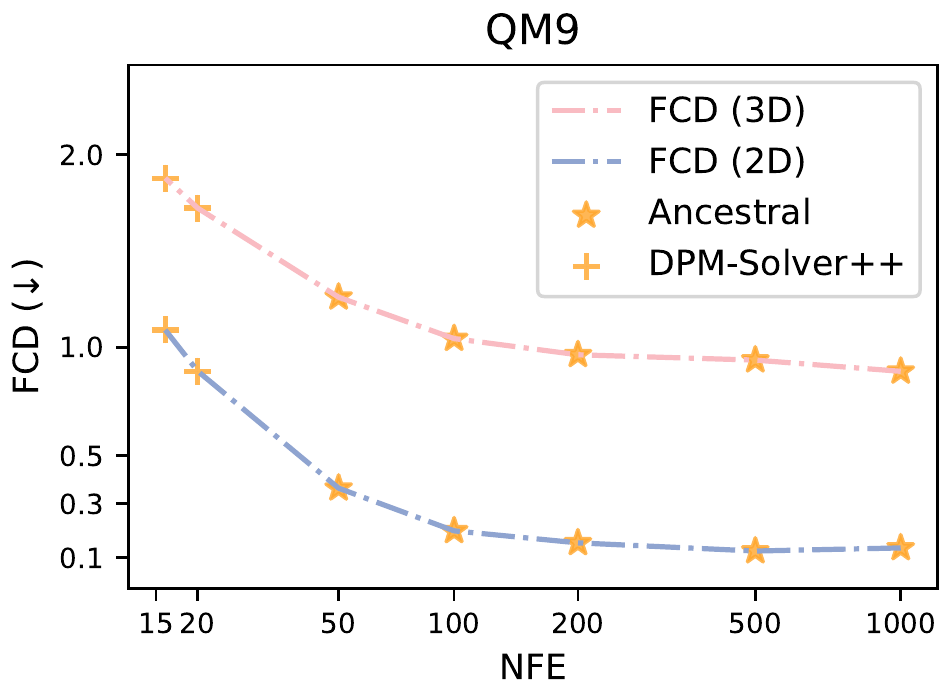}
    \end{subfigure}
    
    \hfill

    \begin{subfigure}{0.49\columnwidth}
        \centering
        \includegraphics[width=\columnwidth]{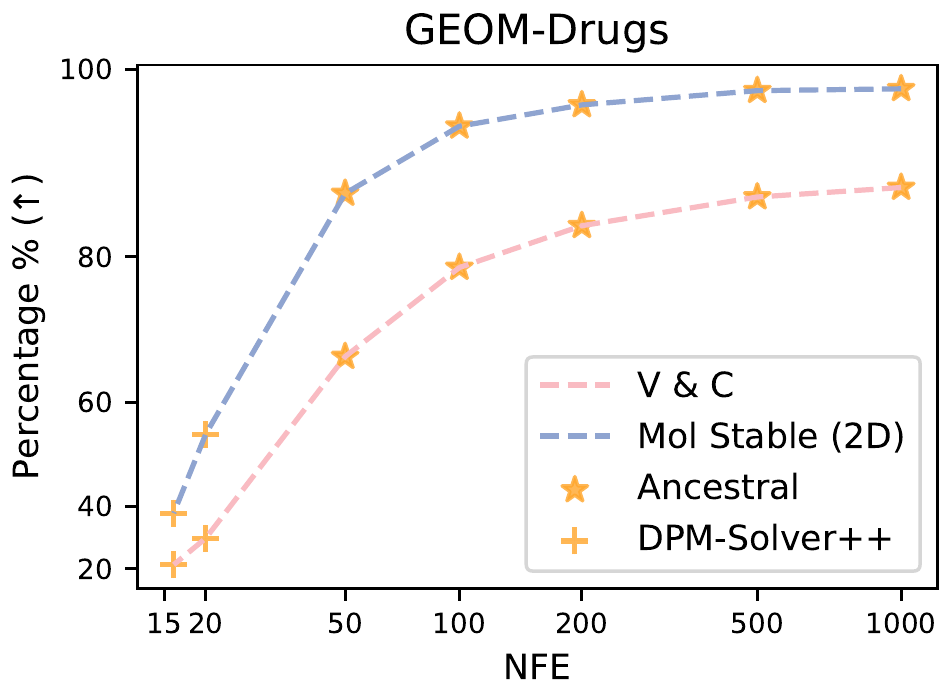}
    \end{subfigure}
    \begin{subfigure}{0.49\columnwidth}
        \centering
        \includegraphics[width=\columnwidth]{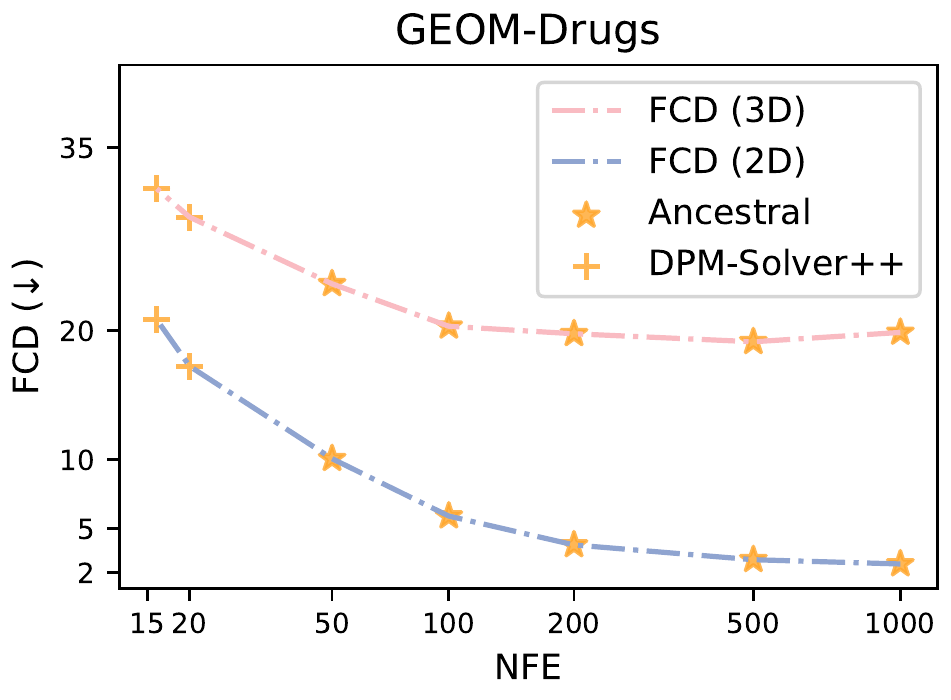}
    \end{subfigure}
    
    \caption{Molecule sampling results with different numbers of function evaluations (NFE).}
    \label{fig:fast_sampling}
\end{figure}

Diffusion-based models can trade off sample quality and computational cost by choosing the number of iteration steps (a.k.a., number of function evaluations (NFE)) in sampling.
This allows for fast sampling, which benefits applications such as virtual screening by generating more molecules in a reasonable amount of time. 
Figure \ref{fig:fast_sampling} shows how the sample quality of our model varies with different NFEs on two datasets.
Using ancestral sampling, our model generates high-fidelity molecules between $50$ NFE and $1000$ NFE. 
Even with 50 NFE, our model outperforms the previous baselines in Table \ref{tab:QM9_H_results} and Table \ref{tab:geom_drug_results}. 
CDGS \cite{huangCDGS23} successfully employs ODE solvers such as DPM-Solvers \cite{DPMS22} for fast molecular graph sampling.
Since our model preserves continuity in the data space, we exploit a hybrid sampling method to explore the few-step generation ($15$ NFE to $20$ NFE), with DPM-Solvers++ \cite{lu2022dpm++} for 2D graph sampling and ancestral sampling for 3D coordinates. 
Although some reasonable molecules are generated with limited steps, fast samplers on equivariant diffusion models are desired for drug-size molecule generation. 

% fast sampling

\subsubsection{Ablation Studies}

We conduct ablation studies on the GEOM-Drugs dataset to verify the effectiveness of our framework component and report the results in Table \ref{tab:abl_geom}.
We also perform the same study on the QM9 dataset, which is smaller and simpler than GEOM-Drugs, and present the results in Appendix.

We first vary the hidden feature dimension and the number of layers to increase the model parameters. 
The larger model achieves better generation quality overall. 
Compared with the base version model, we observe from Table \ref{tab:abl_geom} that:
(1) Removing the 3D geometry modeling reduces the validity and completeness of the generated molecules and worsens the distribution learning with lower FCD values, even though the model (\textit{2D only}) can maintain high stability ratios. 
The model (\textit{3D only}) also performs poorly in the 3D generation without the support of 2D molecular graphs. 
These results confirm that joint learning in 2D and 3D is complementary and beneficial for generating complete drug-sized molecules.
(2) We further investigate the effect of the design of the attention mechanism and implement two variants: \textit{Attn-add} projects edge representations as node-node attention bias, and \textit{Attn-multi} only multiplies the projected edge representations to node-node attention weights.
These two variants lag behind the base model in performance. 
Our relational attention design can capture node and edge correlations more effectively and improve quality.
(3) Removing self-conditioning (\textit{w/d self-cond}) impairs our model’s ability to learn molecule distributions and generate complete drug-sized molecules. 
Using the self-conditioning node and edge features as the model input (\textit{self-cond input}) without adding additional adjacency matrices in message passing, yields similar results except for worse FCD.  
(4) Applying the noise level conditioning (\textit{w/d noise-cond}) also enhances the overall generation quality.

% abl
\begin{table}[t]
\caption{Ablation studies on the GEOM-Drugs dataset.
FCD-3D: FCD values in 3D metrics; Mol-S-2D: Mol stable in 2D metrics; FCD-2D: FCD values in 2D metrics.}
\label{tab:abl_geom}
\resizebox{\columnwidth}{!}{%
\renewcommand\arraystretch{1.3}
\begin{tabular}{lcccc}
\hline
Ablation & \multicolumn{4}{c}{Geom-Drugs} \\ \hline
 & FCD-3D$\downarrow$ & Mol-S-2D$\uparrow$ & FCD-2D$\downarrow$ & V\&C$\uparrow$ \\ \hline
Train & 13.733 & 100.0\% & 0.251 & 100.0\% \\ \hline
Large (315.9MB) & 19.993 & 98.1\% & 2.523 & 87.4\% \\
Medium (140.6MB) & 19.162 & 96.7\% & 2.534 & 84.5\% \\
Base (21.7MB) & 22.645 & 92.1\% & 4.049 & 70.0\% \\ \hline
2D only & - & 98.9\% & 7.156 & 54.4\% \\
3D only & 29.875 & - & - & - \\ \hline
Attn-add & 23.687 & 68.7\% & 5.219 & 61.6\% \\
Attn-multi & 24.713 & 90.8\% & 4.745 & 64.3\% \\ \hline
w/d self-cond & 26.964 & 84.9\% & 13.822 & 23.4\% \\
self-cond input & 22.705 & 93.5\% & 4.291 & 71.5\% \\ \hline
w/d noise-cond & 23.568 & 91.6\% & 4.581 & 71.8\% \\ \hline
\end{tabular}%
}
\end{table}

% conditional generation table
\begin{table*}[t]
\centering
\caption{Performance in aligning the generated molecules with the conditional single target quantum property. The best results are highlighted in bold.}
\label{tab:cond_single_results}
\renewcommand\arraystretch{1.1}
\resizebox{0.95\textwidth}{!}{%
\begin{tabular}{llllllll}
\hline
Method                    & MAE $\downarrow$          &  & Method                        & MAE $\downarrow$             &  & Method                        & MAE $\downarrow$             \\ \cline{1-2} \cline{4-5} \cline{7-8} 
\multicolumn{2}{c}{$C_v \ (\frac{cal}{mol}\mathrm{K})$} &  & \multicolumn{2}{c}{$\mu \ (\mathrm{D})$}                        &  & \multicolumn{2}{c}{$\alpha \ (\mathrm{Bohr}^3)$}                \\ \cline{1-2} \cline{4-5} \cline{7-8} 
U-bound                   & 6.879 $\pm$ 0.015         &  & U-bound                       & 1.613 $\pm$ 0.003            &  & U-bound                       & 8.98 $\pm$ 0.02              \\
\#Atoms                   & 1.971                     &  & \#Atoms                       & 1.053                        &  & \#Atoms                       & 3.86                         \\
Conditional EDM           & 1.065 $\pm$ 0.010         &  & Conditional EDM               & 1.123 $\pm$ 0.013            &  & Conditional EDM               & 2.78 $\pm$ 0.04              \\
EEGSDE ($s=10$)           & 0.941 $\pm$ 0.005         &  & EEGSDE ($s=2$)                & 0.777 $\pm$ 0.007            &  & EEGSDE ($s=3$)                & 2.50 $\pm$ 0.02              \\
Conditional JODO          & \textbf{0.581} $\pm$ 0.001&  & Conditional JODO              & \textbf{0.628} $\pm$ 0.003   &  & Conditional JODO              & \textbf{1.42} $\pm$ 0.01     \\
\textit{L-bound}          & \textit{0.040}            &  & \textit{L-bound}              & \textit{0.043}               &  & \textit{L-bound}              & \textit{0.09}                 \\ \cline{1-2} \cline{4-5} \cline{7-8} 
\multicolumn{2}{c}{$\Delta \epsilon \ (\mathrm{meV})$}  &  & \multicolumn{2}{c}{$\epsilon_{\mathrm{HOMO}} \ (\mathrm{meV})$} &  & \multicolumn{2}{c}{$\epsilon_{\mathrm{LUMO}} \ (\mathrm{meV})$} \\ \cline{1-2} \cline{4-5} \cline{7-8} 
U-bound                   & 1464 $\pm$ 4              &  & U-bound                       & 645 $\pm$ 41                 &  & U-bound                       & 1457 $\pm$ 5                 \\
\#Atoms                   & 866                       &  & \#Atoms                       & 426                          &  & \#Atoms                       & 813                          \\
Conditional EDM           & 671 $\pm$ 5               &  & Conditional EDM               & 371 $\pm$ 2                  &  & Conditional EDM               & 601 $\pm$ 7                  \\
EEGSDE ($s=3$)            & 487 $\pm$ 3               &  & EEGSDE ($s=1$)                & 302 $\pm$ 2                  &  & EEGSDE ($s=3$)                & 447 $\pm$ 6                  \\
Conditional JODO          & \textbf{335} $\pm$ 3      &  & Conditional JODO              & \textbf{226} $\pm$ 1        &  & Conditional JODO              & \textbf{256} $\pm$ 1         \\
\textit{L-bound}          & \textit{65}               &  & \textit{L-bound}              & \textit{39}                  &  & \textit{L-bound}              & \textit{36}                   \\ \hline
\end{tabular}%
}
\end{table*}

\begin{table}[!htbp]
\centering
\caption{Performance in aligning the generated molecules with multiple target quantum properties.} 
% The best results are highlighted in bold.}
\label{tab:cond_multi_results}
\renewcommand\arraystretch{1.1}
\resizebox{\columnwidth}{!}{%
\begin{tabular}{lll}
\hline
Method                      & MAE1 $\downarrow$   & MAE2 $\downarrow$   \\ \hline
\multicolumn{3}{c}{$C_v (\frac{cal}{mol}\mathrm{K}), \ \  \mu(\mathrm{D})$} \\ \hline
Conditional EDM             & 1.097 $\pm$ 0.007   & 1.156 $\pm$ 0.011   \\
EEGSDE ($s_1=10, s_2=1$)    & 0.981 $\pm$ 0.008   & 0.912 $\pm$ 0.006   \\
Conditional JODO            & \textbf{0.634} $\pm$ 0.002   & \textbf{0.716} $\pm$ 0.006   \\ \hline
\multicolumn{3}{c}{$\Delta\epsilon(\mathrm{meV}), \ \  \mu(\mathrm{D})$}      \\ \hline
Conditional EDM             & 683 $\pm$ 1         & 1.130 $\pm$ 0.007   \\
EEGSDE ($s_1=s_2=1$)        & 563 $\pm$ 3         & 0.866 $\pm$ 0.003   \\
Conditional JODO            & \textbf{350} $\pm$ 4         & \textbf{0.752} $\pm$ 0.006   \\ \hline
\multicolumn{3}{c}{$\alpha(\mathrm{Bohr}^3), \ \  \mu(\mathrm{D})$}           \\ \hline
Conditional EDM             & 2.76 $\pm$ 0.01     & 1.158 $\pm$ 0.002   \\
EEGSDE ($s_1=s_2=1.5$)      & 2.61 $\pm$ 0.01     & 0.855 $\pm$ 0.007   \\
Conditional JODO            & \textbf{1.52} $\pm$ 0.01     & \textbf{0.717} $\pm$ 0.006   \\ \hline
\end{tabular}%
}
\end{table}

\subsection{Conditional Molecule Generation with Desired Quantum Properties}

In this section, we apply our conditional JODO for inverse molecule design, which aims to explore the chemical space and generate new molecules with desired properties for potential drug discovery applications. 
We perform conditional generation based on single-property conditioning or multiple-property conditioning.

\subsubsection{Setup}
We conduct conditional generation experiments on the QM9 dataset. 
Following \cite{HoogeboomEDM22, baoEEGSDE22}, we use the same training/validation/test splits and further divide the training sets into two equal non-overlapping $D_a, D_b$ sets with $50K$ samples each.
We consider six quantum properties as the conditional information: heat capacity $C_v$, dipole moment $\mu$, polarizability $\alpha$, highest occupied molecular orbital energy $\epsilon_{\mathrm{HOMO}}$, lowest unoccupied molecular orbital energy $\epsilon_{\mathrm{LUMO}}$, and HOMO-LUMO gap $\Delta \epsilon$. 
To obtain the properties of generated molecules conveniently, we train a property prediction network $\phi_c$ \cite{satorras21EGNN} on the first part $D_a$ as in previous work.
The generative models are trained on the second part $D_b$ to avoid leakage.
We generate $10K$ samples for each evaluation and report the Mean Average Error (MAE) between the given conditional property values and the properties of the generated samples predicted by the pretrained property prediction network.
We report the MAE of $\phi_c$ evaluated on the molecules of $D_b$ as the performance bound (\textit{L-bound}) for reference.
The smaller the gap between the generative model and the \textit{L-bound}, while maintaining the generative quality of the molecules themselves, the better the ability to exploit conditional information.

\subsubsection{Baselines}
We adopt two naive baselines from \cite{HoogeboomEDM22} that are agnostic to the given properties.
The 'U-bound' baseline randomly shuffles the property labels of molecules in $D_b$ and reports their MAE using $\phi_c$. 
The ‘\#Atoms’ baseline predicts molecular properties in $D_b$ based only on the number of atoms in molecules.
We also compare our model with two diffusion-based conditional generative models.
Conditional EDM \cite{HoogeboomEDM22} directly concatenates the given property values with atom features in the noise prediction model of EDM. 
EEGSDE \cite{baoEEGSDE22} trains extra time-dependent property prediction models as the energy function to guide the generation process towards the desired space.

\subsubsection{Results}
We report the performance of single target quantum property conditional generation in Table \ref{tab:cond_single_results} and multiple target quantum properties conditional generation in Table \ref{tab:cond_multi_results}.
Based on our powerful diffusion model, Conditional JODO outperforms previous EDM-based models for the six single quantum property condition settings.
Compared to EEGSDE, our model reduces MAE by more than $40\%$ in properties $\alpha$, $\epsilon_\mathrm{LUMO}$, and more than $30\%$ in properties $C_v$, $\Delta \epsilon$, without extra energy guidance.
When targeting multiple properties, Conditional JODO consistently outperforms baselines.
In addition to aligning well between the conditional information and generated molecules, our model maintains the generation quality of molecules, which is reported in Appendix.
As we preserve the continuity for the diffusion model design, we can flexibly plug in the widely used classifier or the classifier-free guidance \cite{dhariwal2021diffusion, ho2022classifier-free} to support a more general inverse molecular design.

\subsection{2D Molecular Graph Generation}

We simplify our model as a one-shot 2D molecular graph generative model by removing the 3D geometry modeling part.
We conduct 2D molecular graph generation experiments to further investigate the ability of JODO to model the permutation-invariant molecular graph distribution.

\subsubsection{Setup}
We use two molecule datasets, ZINC250k \cite{zinc250k} and MOSES \cite{polykovskiy20MOSES}, to train and evaluate models.
Following \cite{JoLH22GDSS, huangCDGS23}, we kekulize all molecules in the ZINC250k dataset using RDKit, where we remove hydrogen atoms and replace aromatic bonds.
JODO-2D generates atom types, formal charges, and edge types in the molecules. 
For the ZINC250k dataset, we report the validity ratio without valency checking, FCD, and scaffold similarity.
For the MOSES dataset, we follow its original split and construct molecular graphs from SMILES strings as in \cite{vignacDigress22}.
We report FCD and SNN that are computed on the test set with separate scaffolds. 
We compute FCD and SNN on the test set with separate scaffolds and report them.
The Filters metric is the fraction of generated molecules that pass the dataset construction filters.
We generate 10K molecules for evaluation by default.

\subsubsection{Baselines}
We compare JODO-2D with several autoregressive and one-shot molecular graph generative models on the ZINC250k dataset. 
GraphAF \cite{ShiGraphaf20} and GraphDF \cite{LuoGraphdf21} are autoregressive models based on normalizing flow models.
MoFlow \cite{zangMoflow20} and GraphCNF \cite{lippeGraphcnf21} are two other flow-based models that do not depend on node orderings.
EDP-GNN \cite{niuEDP-GNN20} is an early attempt to apply a score-based model for graph generation.
GDSS \cite{JoLH22GDSS} and CDGS \cite{huangCDGS23} are representative works that utilize diffusion-based models with continuous-time stochastic differential equations for molecular graph generation, while DiGress \cite{vignacDigress22} uses a discrete denoising diffusion model.
For the MOSES dataset, we follow \cite{vignacDigress22} and compare JODO-2D with SMILES-based VAE from MOSES \cite{polykovskiy20MOSES}, fragment-based JT-VAE \cite{Jin18JT-VAE}, autoregressive GraphINVENT \cite{mercado21graphinvent} and one-shot DiGress \cite{vignacDigress22}.

\subsubsection{Results}
The performance on the Zinc250k dataset is shown in Table \ref{tab:zinc_results}. 
JODO-2D achieves state-of-the-art performance as a permutation-invariant one-shot graph generative model.
We attribute our 2D model’s superior performance to the formal charge modeling and the new data prediction model design, which enable JODO-2D to learn the underlying molecule distribution more faithfully than existing diffusion-based models. 
Table \ref{tab:moses_results} reports the results of the MOSES dataset.
Without extra graph structural and positional encoding, JODO-2D is comparable to or better than diffusion-based DiGress in most metrics. 
However, it still lags behind other methods based on efficient molecular representation, such as SMILES and fragments, on this million-scale dataset.
Enhancing the perception of ring structures and common fragments in the data prediction network of diffusion-based models may further improve their distribution learning ability and efficiency.

\begin{table}[t]
\centering
\caption{Generation quality on the ZINC250k dataset.}
\label{tab:zinc_results}

\renewcommand\arraystretch{1.1}
\resizebox{0.93\columnwidth}{!}{%
\begin{tabular}{lccc}
\hline
Method   & \begin{tabular}[c]{@{}c@{}}VALID \\ w/o check\end{tabular} $\uparrow$ & FCD $\downarrow$ & Scaf. $\uparrow$ \\ \hline
\textit{Train}    & \textit{100.00 \%}           & \textit{0.195} & \textit{0.610}      \\ \hline
MoFlow \cite{zangMoflow20}  & 63.11 \%                              & 20.931           & 0.013              \\
GraphAF \cite{ShiGraphaf20} & 68.47 \%                             & 16.023           & 0.067              \\
GraphDF \cite{LuoGraphdf21} & 90.61 \%                             & 33.546           & 0.000              \\
GraphCNF \cite{lippeGraphcnf21} & 96.35 \%                             & 13.532           & 0.032              \\ \hline
EDP-GNN \cite{niuEDP-GNN20} & 82.97 \%                             & 16.737           & 0.000              \\
GDSS \cite{JoLH22GDSS} & 97.01 \%                             & 14.656           & 0.047              \\
DiGress \cite{vignacDigress22} & 92.10 \%                             & 3.597            & 0.355              \\
CDGS \cite{huangCDGS23} & 98.13 \%                             & 2.069            & 0.515            \\ \hline
JODO-2D (ours)  & \textbf{99.91} \%                    & \textbf{0.472}   & \textbf{0.605}             \\ \hline
\end{tabular}%
}
\end{table}

\begin{table}[t]
\centering
\caption{Generation quality on the MOSES dataset.}
\label{tab:moses_results}
\renewcommand\arraystretch{1.4}
\resizebox{\columnwidth}{!}{%
\begin{tabular}{lccccc}
\hline
Method              & VAE    & JT-VAE   & GraphINVENT & DiGress  & JODO-2D  \\ \hline
Class              & SMILES & Frag.    & Autoreg.    & One-shot & One-shot \\ \hline
Valid $\uparrow$     & 97.7   & 100.0      & 96.4        & 85.7     & 88.9     \\
Unique $\uparrow$  & 99.8   & 100.0      & 99.8        & 100.0      & 100.0     \\
Novel $\uparrow$   & 69.5   & 99.9     & -           & 95.0     & 91.0     \\
Filters $\uparrow$ & 99.7   & 97.8     & 95.0        & 97.1     & 98.7     \\
FCD $\downarrow$   & 0.57   & 1.00     & 1.22        & 1.19     & 1.14     \\
SNN $\uparrow$     & 0.58   & 0.53     & 0.54        & 0.52     & 0.55     \\ \hline
\end{tabular}%
}
\end{table}

%% file: appendix.tex
\appendix

\subsection{Ancestral Sampling}
We provide some definitions of the coefficients used in the ancestral sampling as
\begin{equation}
    \begin{aligned}
    \alpha_{t|s} & = \alpha_t / \alpha_s \ , \\
    \sigma_{t|s}^2 & = \sigma_t^2 - \alpha_{t|s}^2 \sigma_s^2 \ .    
    \end{aligned}
\end{equation}
For detailed derivation, refer to \cite{VDM2021}.

\subsection{Experimental Details}

We first summarize the details of the molecule datasets in Table \ref{tab:mol_dataset}, including the number of molecules, the range of the number of atoms in a molecule,  and the number of atom types and bond types.  

\begin{table}[!htbp]
\centering
\caption{Molecule dataset information.}
\label{tab:mol_dataset}
\renewcommand\arraystretch{1.5}
\resizebox{\columnwidth}{!}{%
\begin{tabular}{ccccc}
\hline
Dataset  & N-molecule & N-atom  & N-atom type & N-bond type \\ \hline
QM9 & 130,831 & $3 \leq|V|\leq 29$ & 5 & 3 \\
GEOM-Drugs & 304,294 & $3 \leq|V|\leq 181$ & 16 & 4 \\
ZINC250k & 249,455             & $6\leq|V|\leq38$ & 9                    & 3                  \\
MOSES    & 1,936,962             & $8 \leq|V|\leq 27$  & 7                & 4                   \\ \hline
\end{tabular}%
}
\end{table}

We slightly modify the Diffusion Graph Transformer (DGT) architecture for different molecule datasets.
For the QM9 dataset with explicit hydrogen atoms, we stack $8$ DGT blocks with $16$ attention heads, $256$ node hidden feature channels and $64$ edge hidden feature channels. 
The feed-forward network (FFN) has a dimension expansion ratio of $2$.
The model architecture remains the same in conditional generation except for extra MLPs to encode given properties.
For the GEOM-Drugs dataset, we create three variants with different numbers of parameters to explore the performance limit of the current model design.
The base variant has $6$ blocks with $128$ node channels and $32$ edge channels, the medium variant has $10$ blocks with $256$ node channels and $64$ edge channels, and the large variant has $384$ node channels and $96$ edge channels.

We train all our models with the Adam optimizer and a constant learning rate of $2e^{-4}$.
The unconditional generation on QM9 takes $1.5$M iterations for training  with a batch size of $64$, on GEOM-Drugs it takes from $1.5$M to $1.75$M iterations with batch size $16$.
The conditional generation on QM9 takes longer to converge, with $2$M iterations for single property conditioning and $2.5$M iterations for multiple property conditioning. 
For the conditional generation on QM9, we train models longer to achieve convergence, with $2$M for single property conditioning and $2.5$M for multiple property conditioning.
We apply the gradient clip technique during training.
The models on QM9 are trained on a single RTX 3090 GPU, and the models on GEOM-Drugs are trained on one or two A100-40GB GPUs. 

\subsection{More Result Analysis}

\begin{figure*}[b]
    \centering
    
    \begin{subfigure}{0.19\textwidth}
    \centering
    \includegraphics[width=\textwidth]{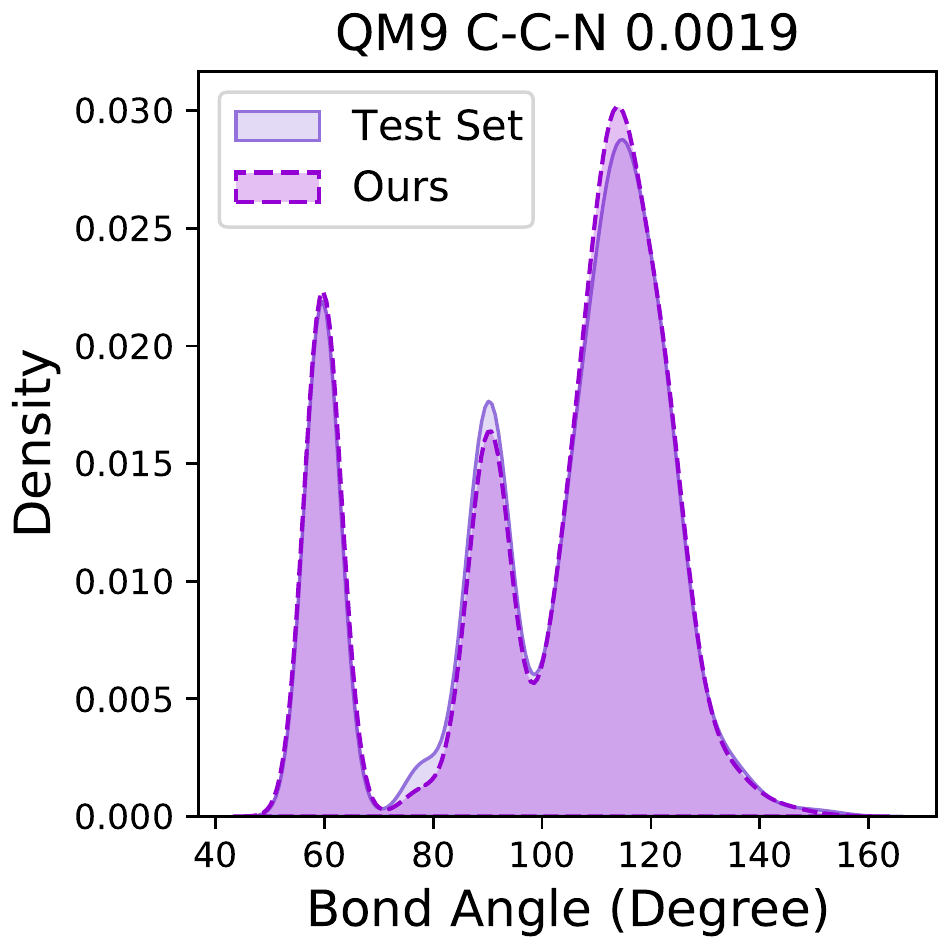}
    \end{subfigure}
    \begin{subfigure}{0.19\textwidth}
    \centering
    \includegraphics[width=\textwidth]{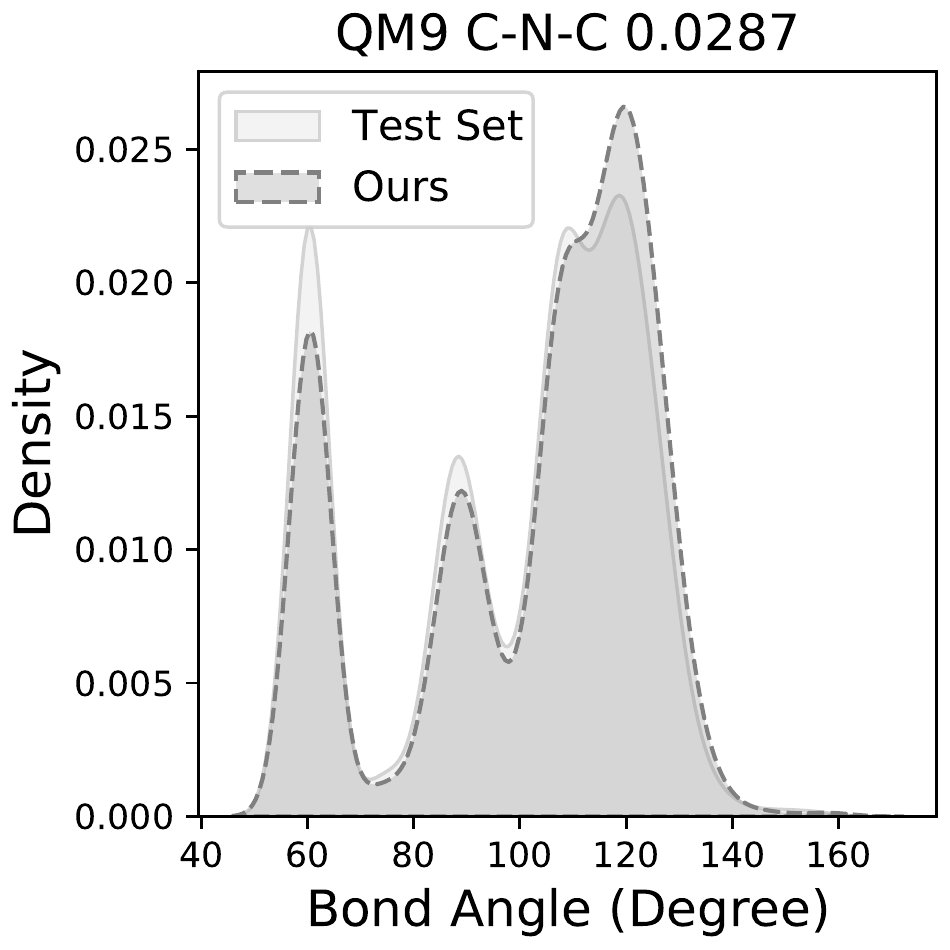}
    \end{subfigure}
    \begin{subfigure}{0.19\textwidth}
    \centering
    \includegraphics[width=\textwidth]{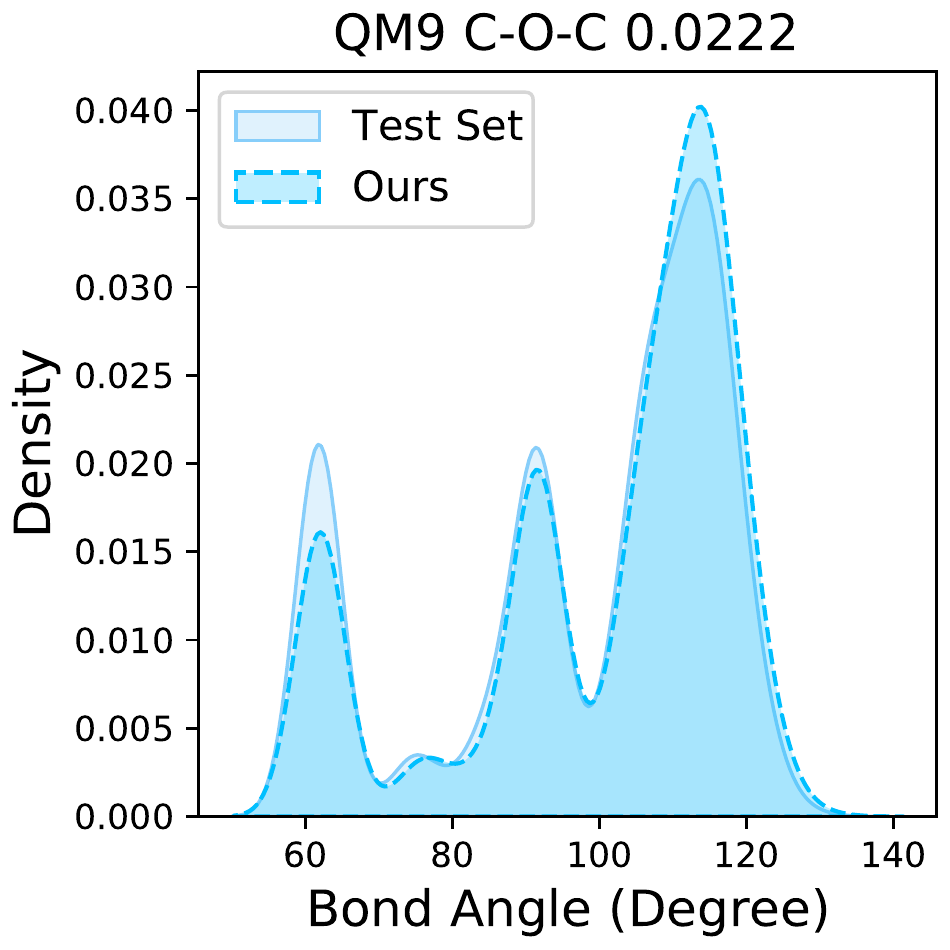}
    \end{subfigure}
    \begin{subfigure}{0.19\textwidth}
    \centering
    \includegraphics[width=\textwidth]{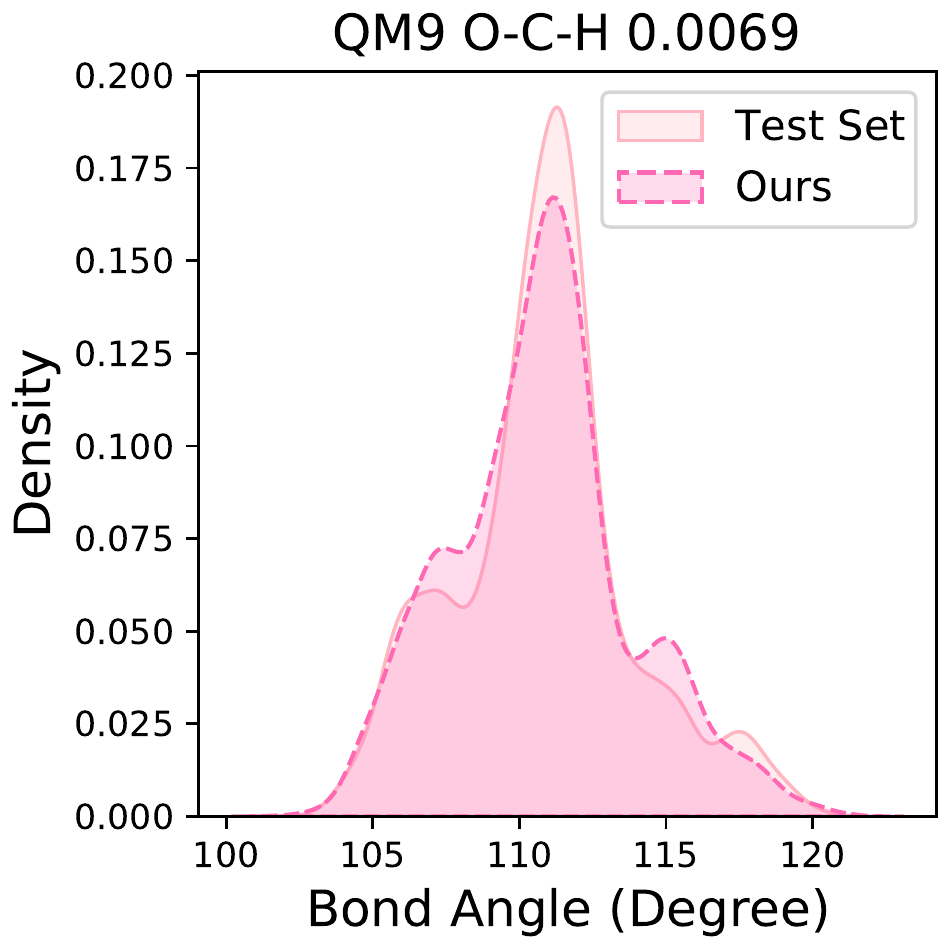}
    \end{subfigure}
    \begin{subfigure}{0.19\textwidth}
    \centering
    \includegraphics[width=\textwidth]{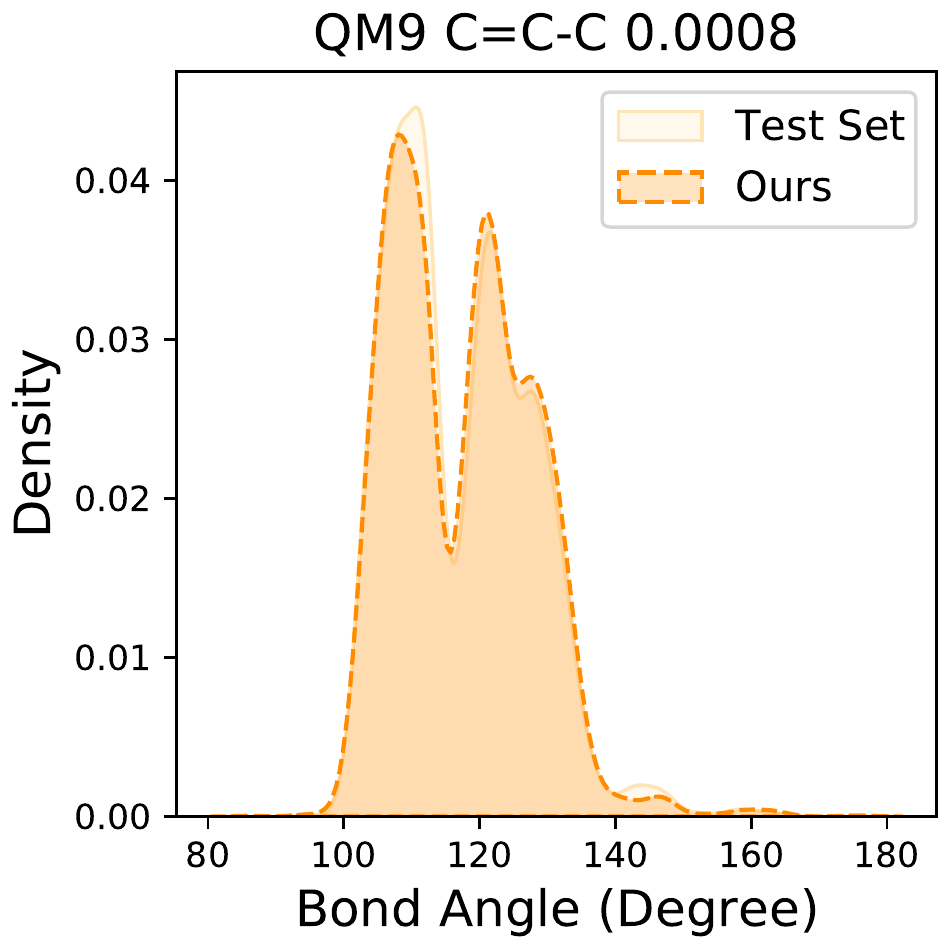}
    \end{subfigure}

    \hfill

    \begin{subfigure}{0.19\textwidth}
    \centering
    \includegraphics[width=\textwidth]{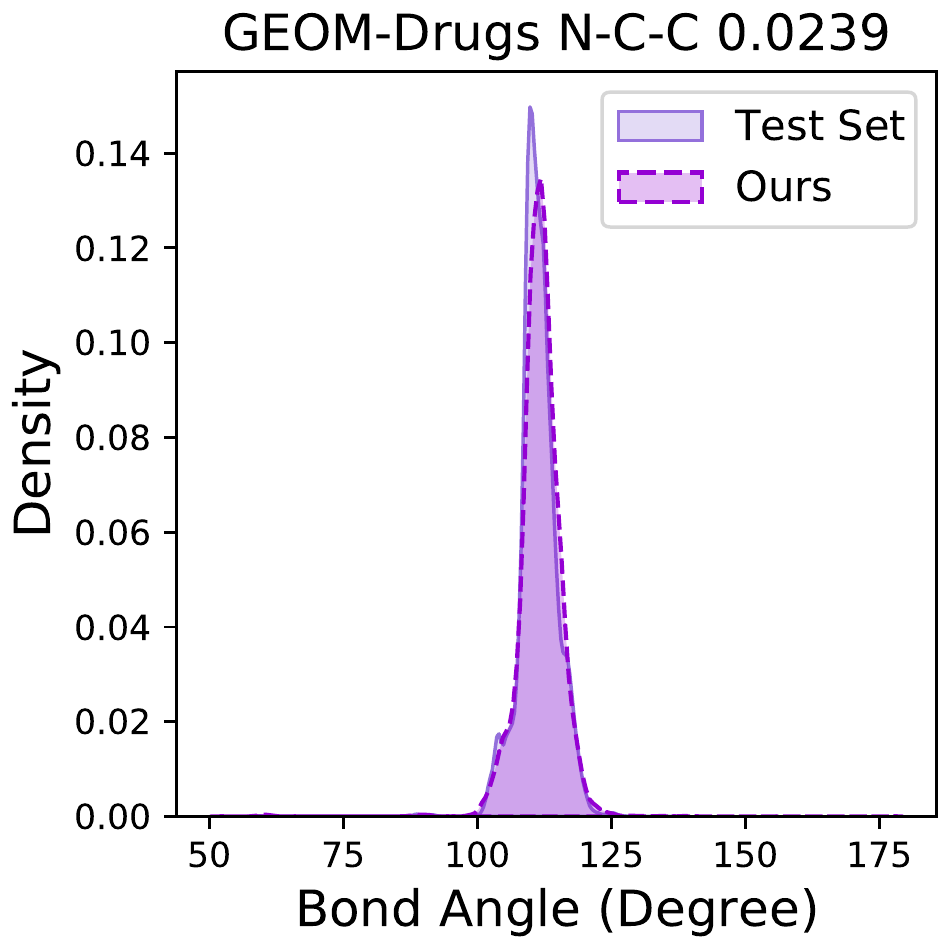}
    \end{subfigure}
    \begin{subfigure}{0.19\textwidth}
    \centering
    \includegraphics[width=\textwidth]{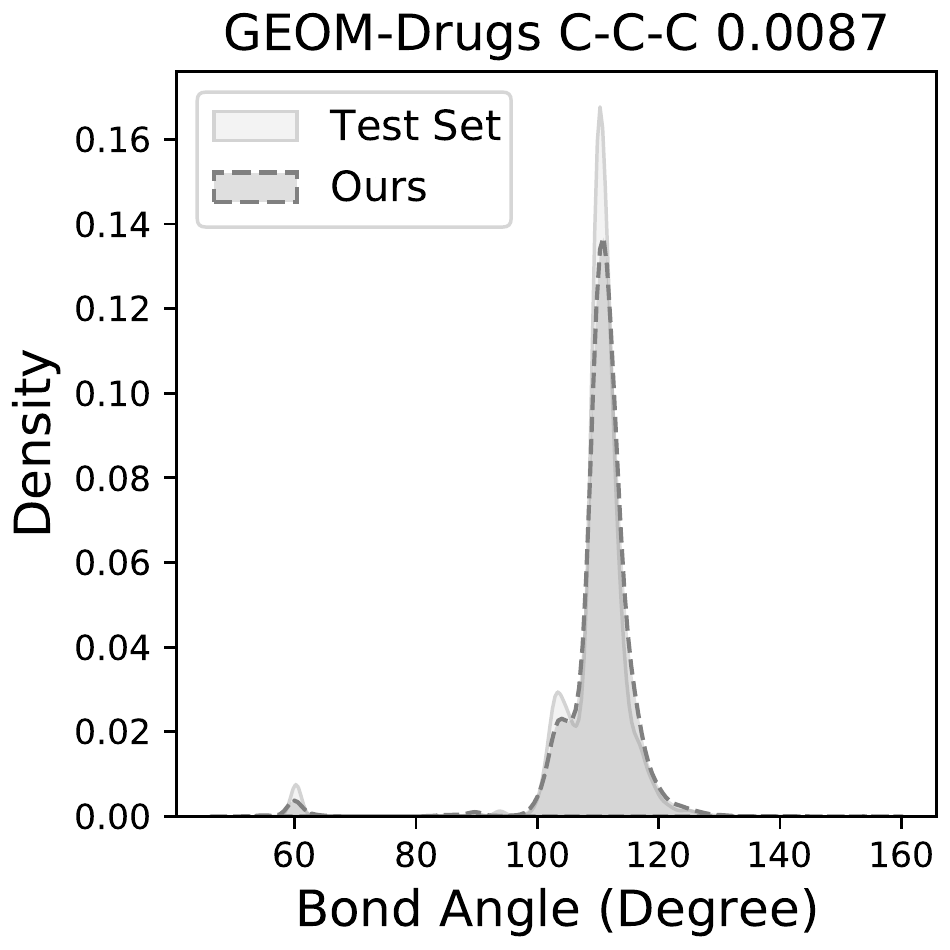}
    \end{subfigure}
    \begin{subfigure}{0.19\textwidth}
    \centering
    \includegraphics[width=\textwidth]{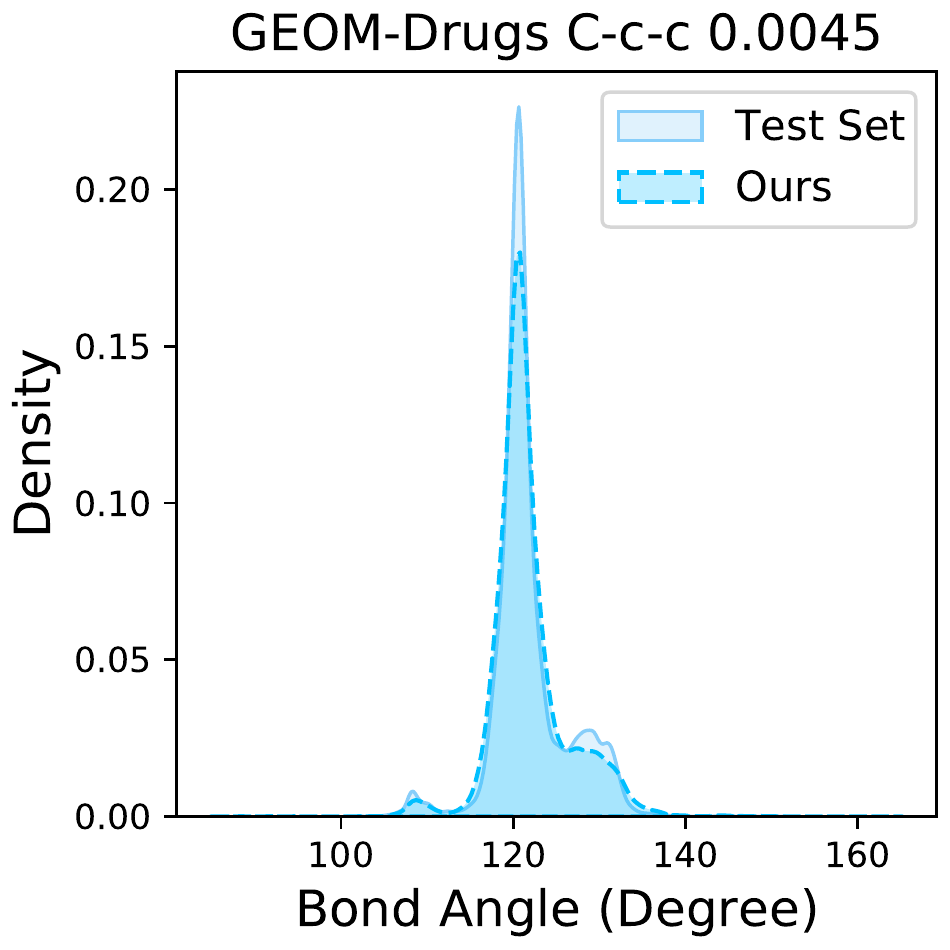}
    \end{subfigure}
    \begin{subfigure}{0.19\textwidth}
    \centering
    \includegraphics[width=\textwidth]{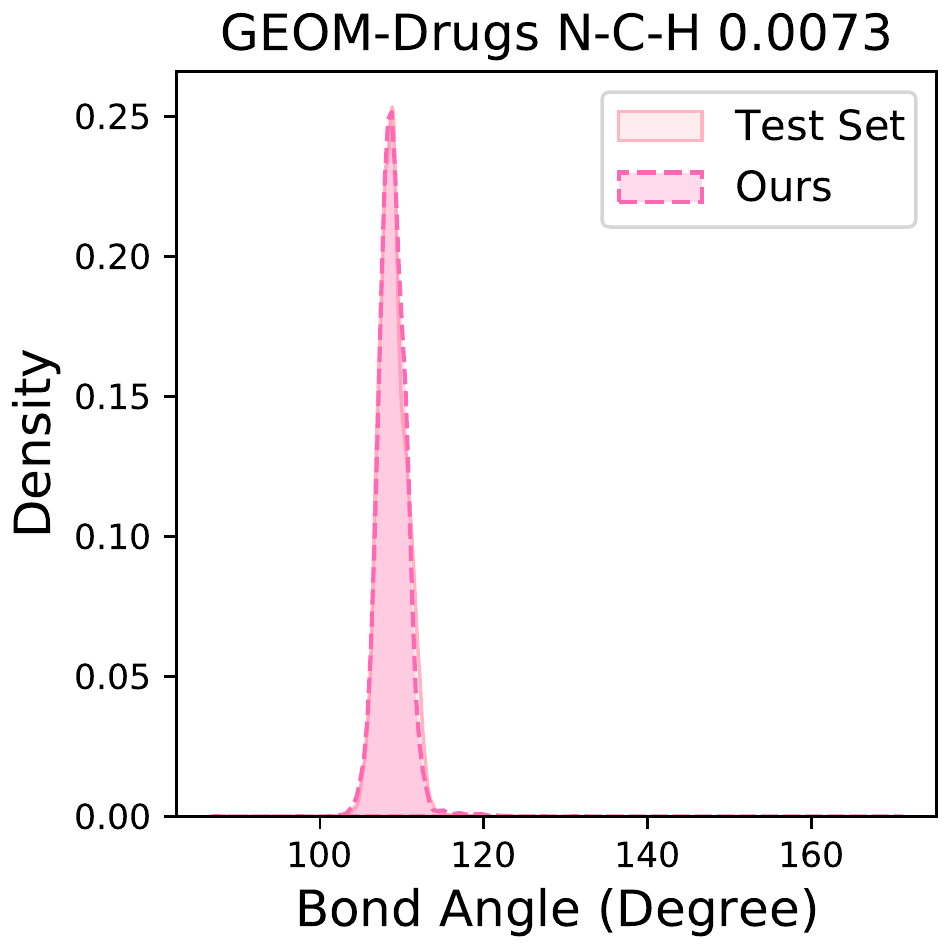}
    \end{subfigure}
    \begin{subfigure}{0.19\textwidth}
    \centering
    \includegraphics[width=\textwidth]{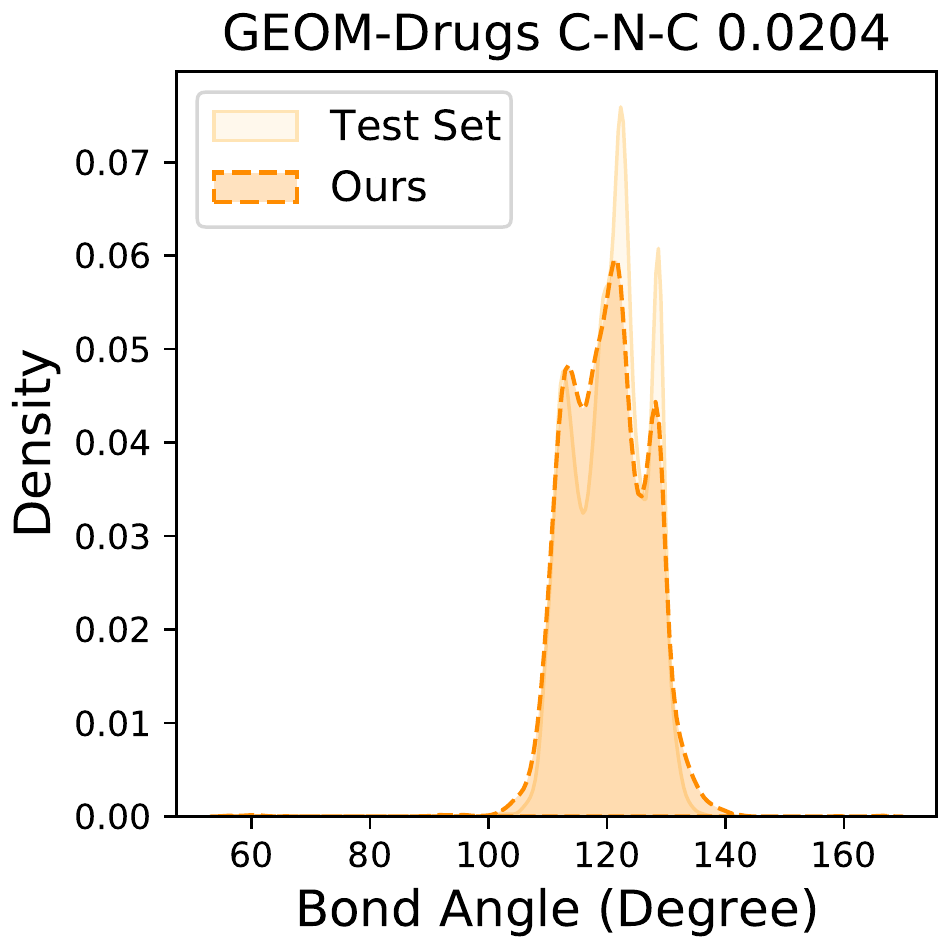}
    \end{subfigure}

    \hfill

    \begin{subfigure}{0.19\textwidth}
    \centering
    \includegraphics[width=\textwidth]{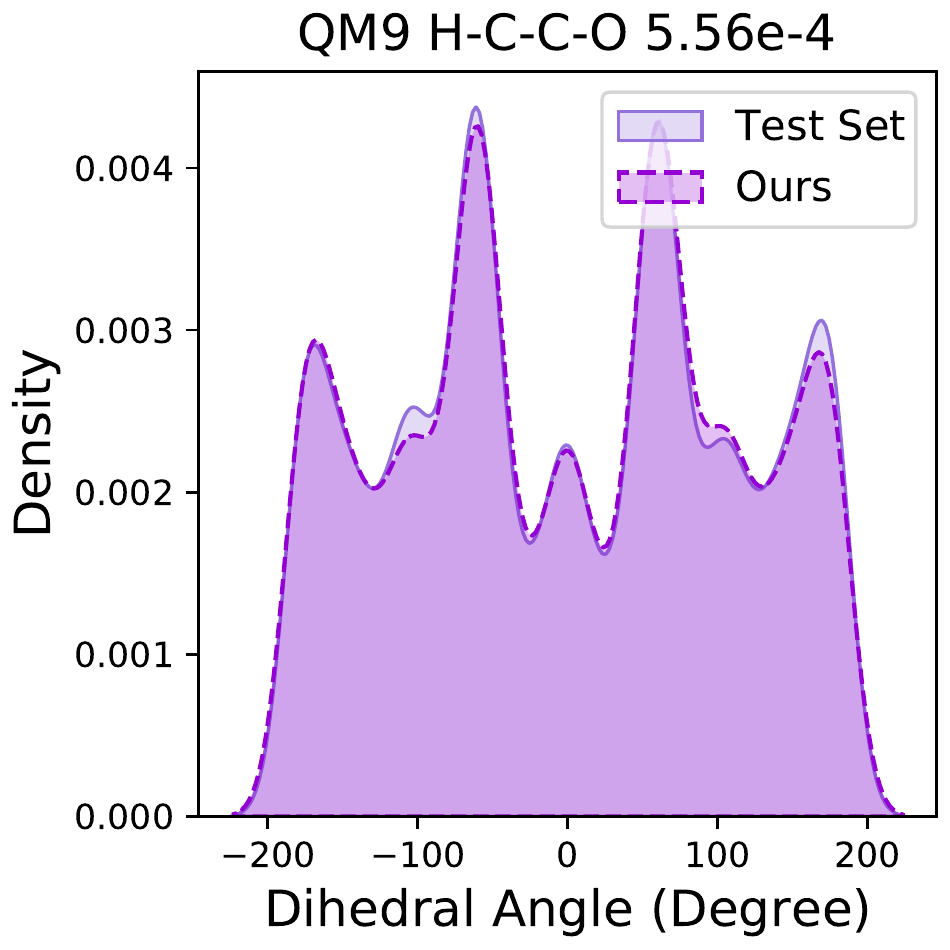}
    \end{subfigure}
    \begin{subfigure}{0.19\textwidth}
    \centering
    \includegraphics[width=\textwidth]{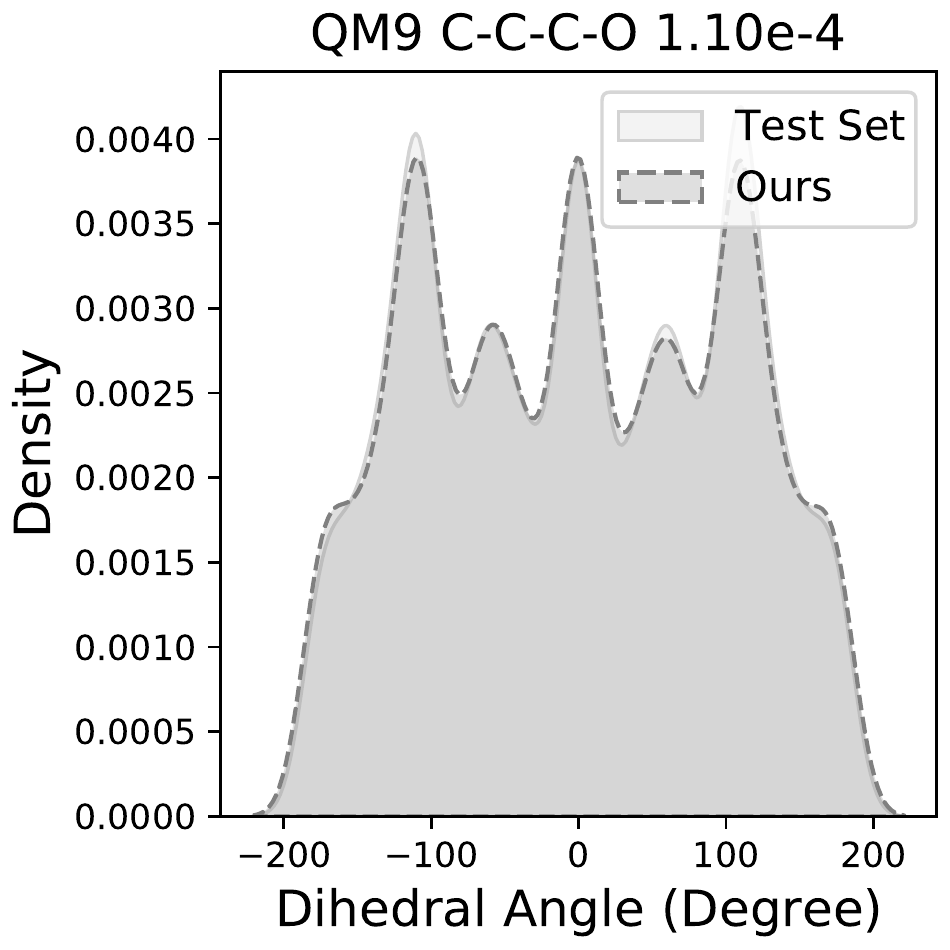}
    \end{subfigure}
    \begin{subfigure}{0.19\textwidth}
    \centering
    \includegraphics[width=\textwidth]{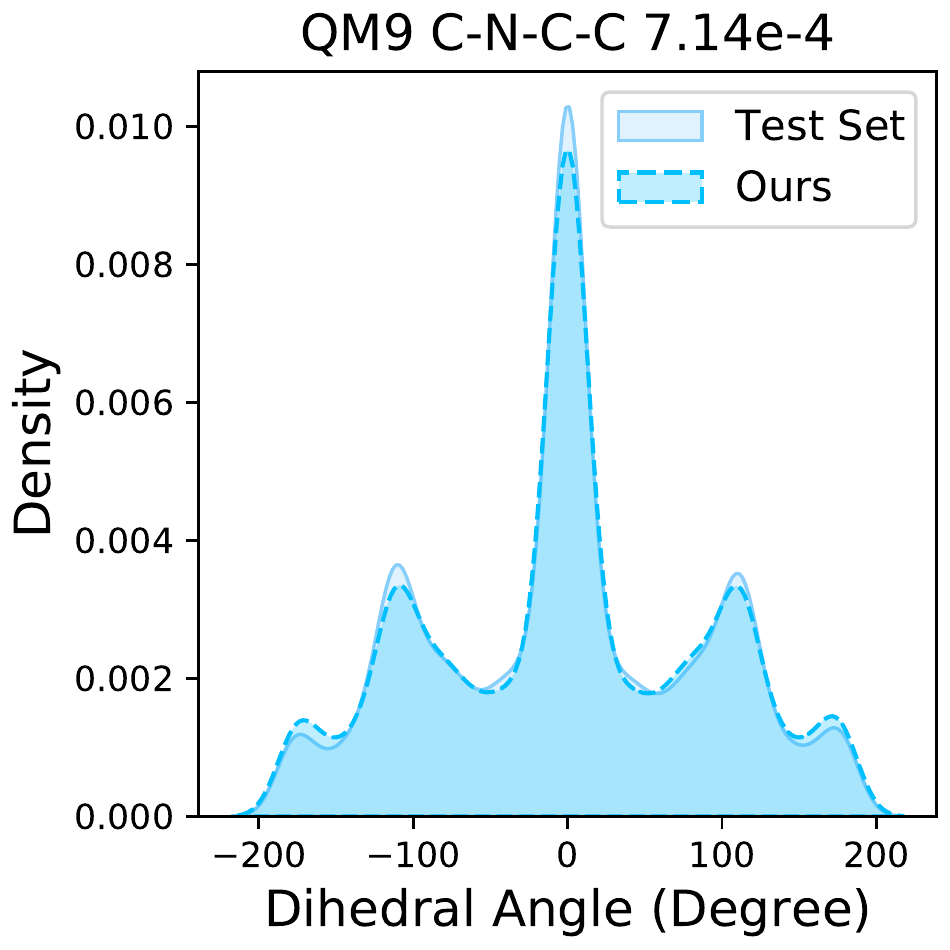}
    \end{subfigure}
    \begin{subfigure}{0.19\textwidth}
    \centering
    \includegraphics[width=\textwidth]{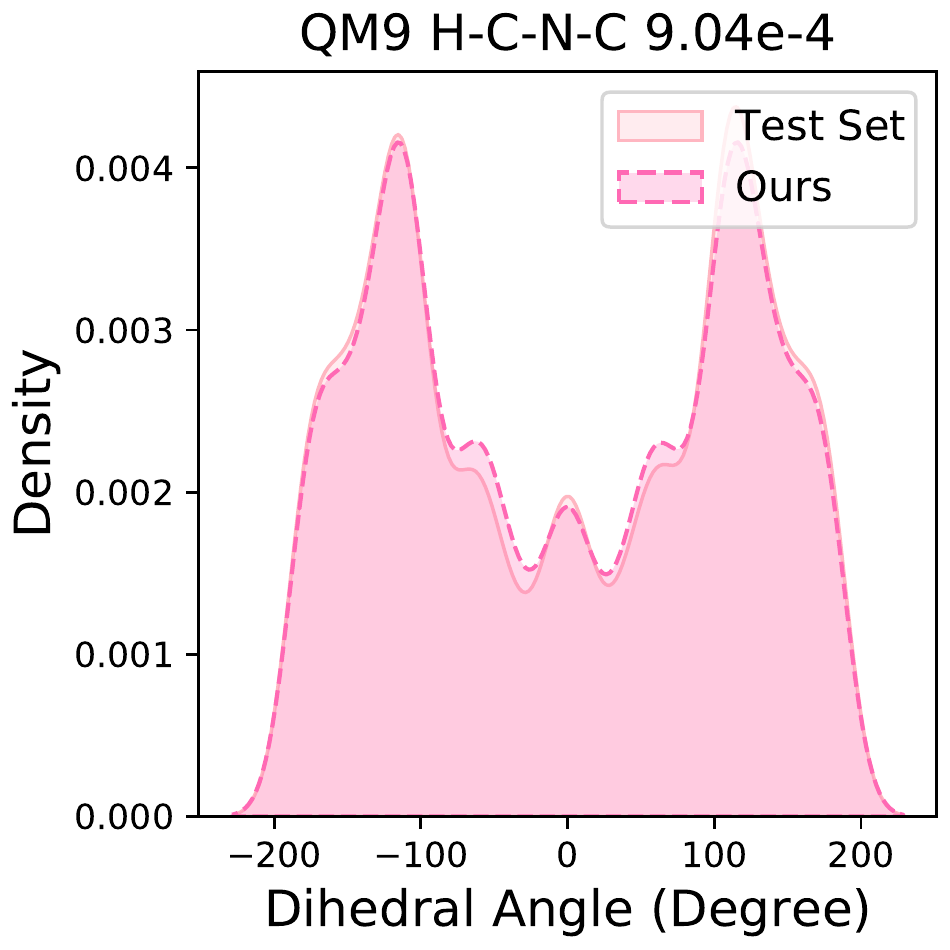}
    \end{subfigure}
    \begin{subfigure}{0.19\textwidth}
    \centering
    \includegraphics[width=\textwidth]{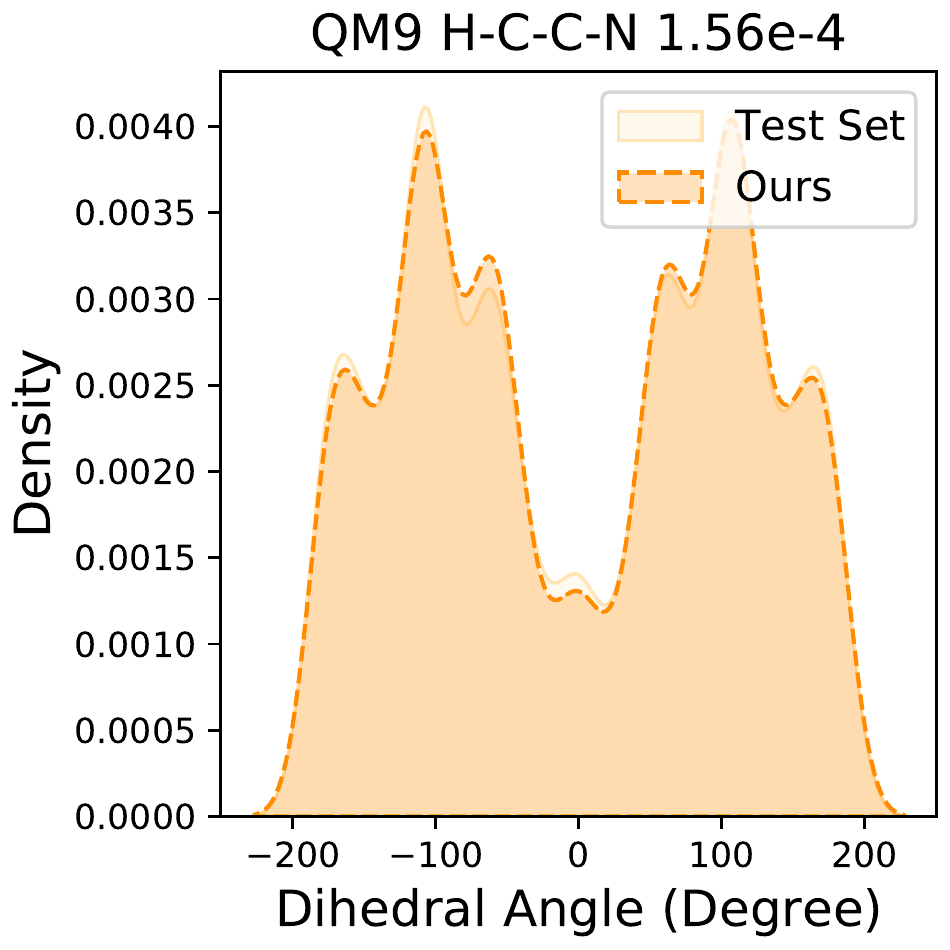}
    \end{subfigure}    

    \hfill

    \begin{subfigure}{0.19\textwidth}
    \centering
    \includegraphics[width=\textwidth]{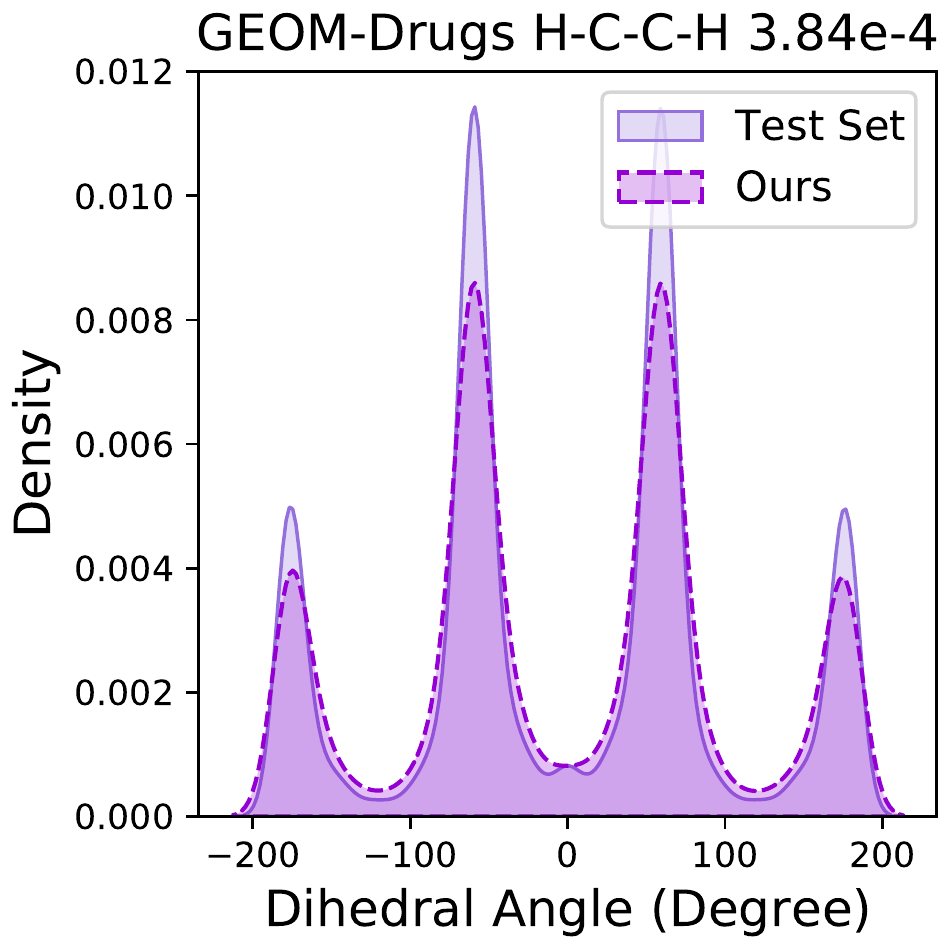}
    \end{subfigure}
    \begin{subfigure}{0.19\textwidth}
    \centering
    \includegraphics[width=\textwidth]{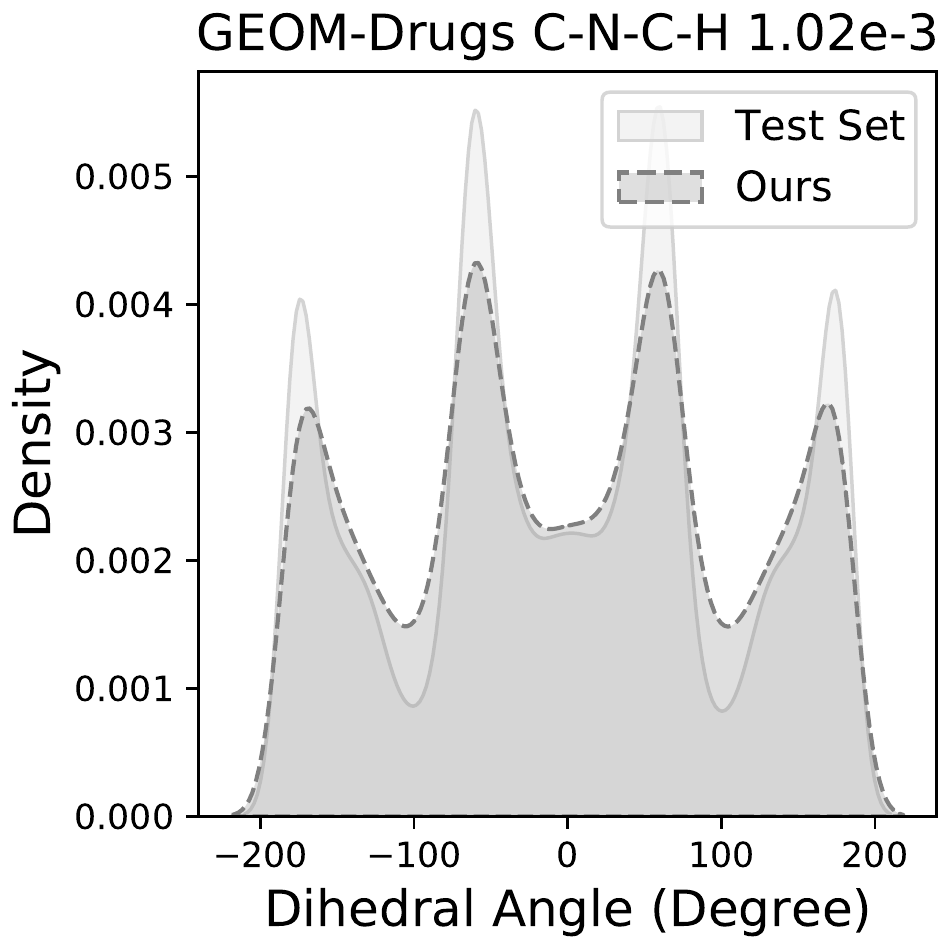}
    \end{subfigure}
    \begin{subfigure}{0.19\textwidth}
    \centering
    \includegraphics[width=\textwidth]{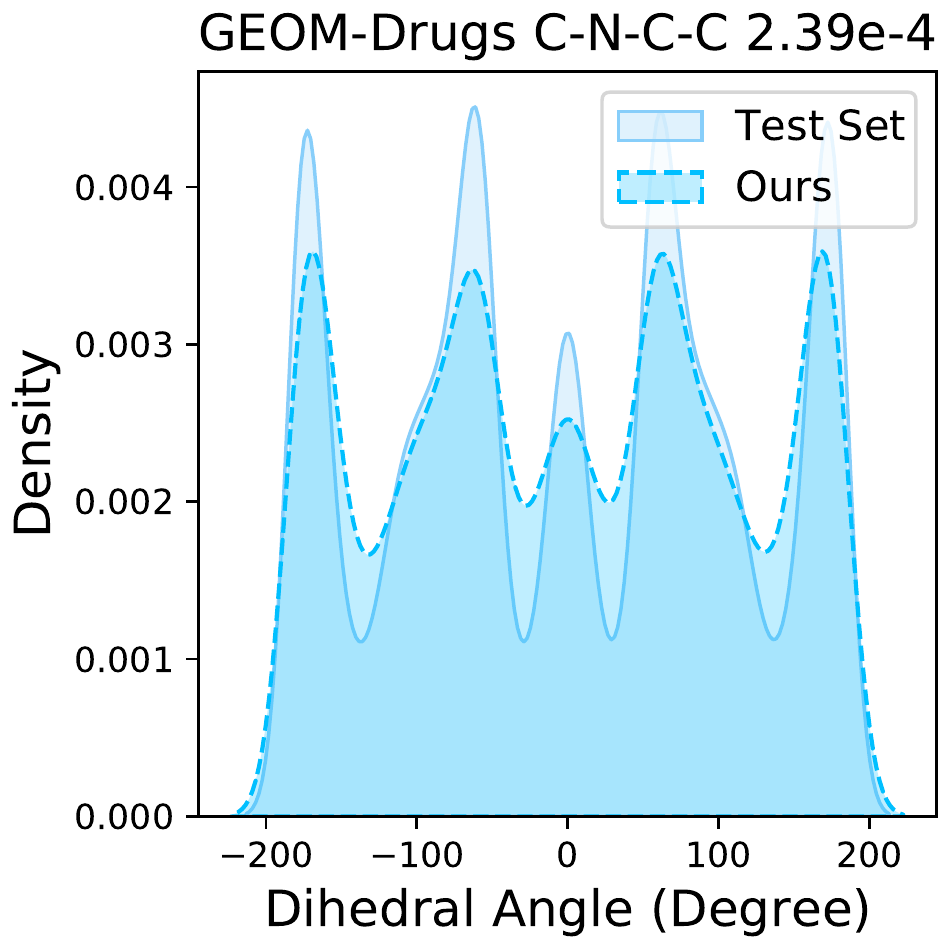}
    \end{subfigure}
    \begin{subfigure}{0.19\textwidth}
    \centering
    \includegraphics[width=\textwidth]{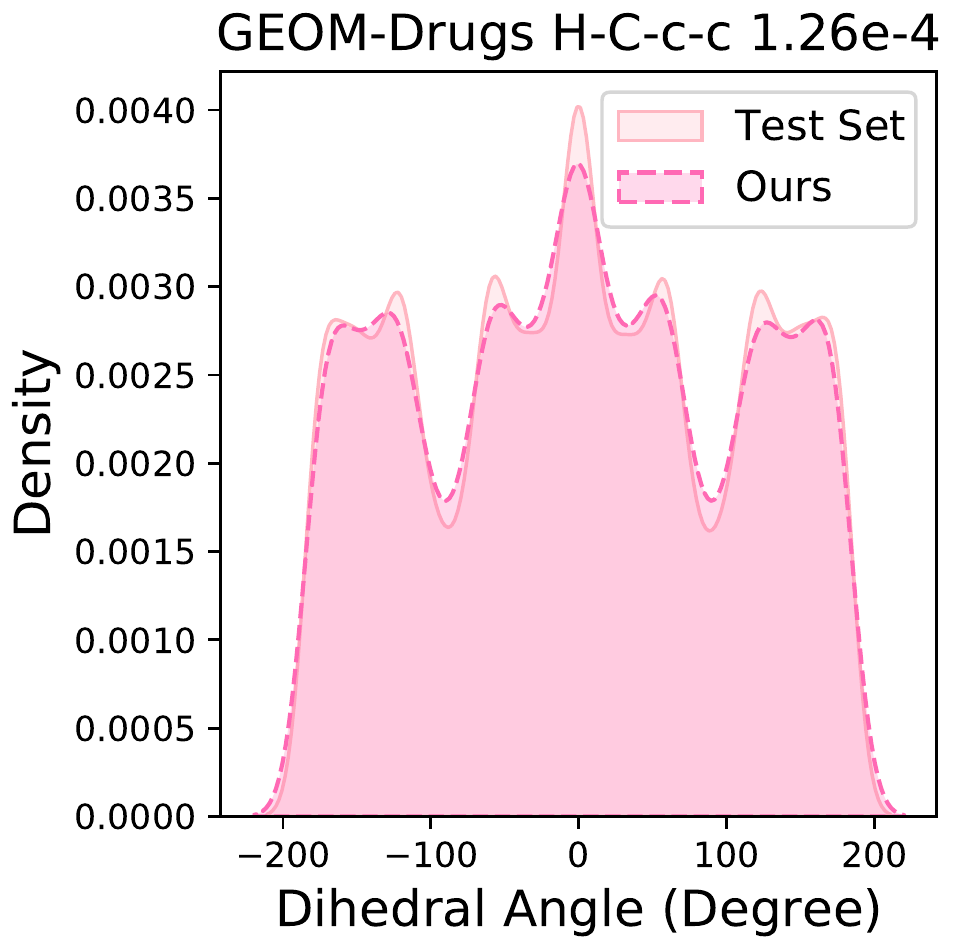}
    \end{subfigure}
    \begin{subfigure}{0.19\textwidth}
    \centering
    \includegraphics[width=\textwidth]{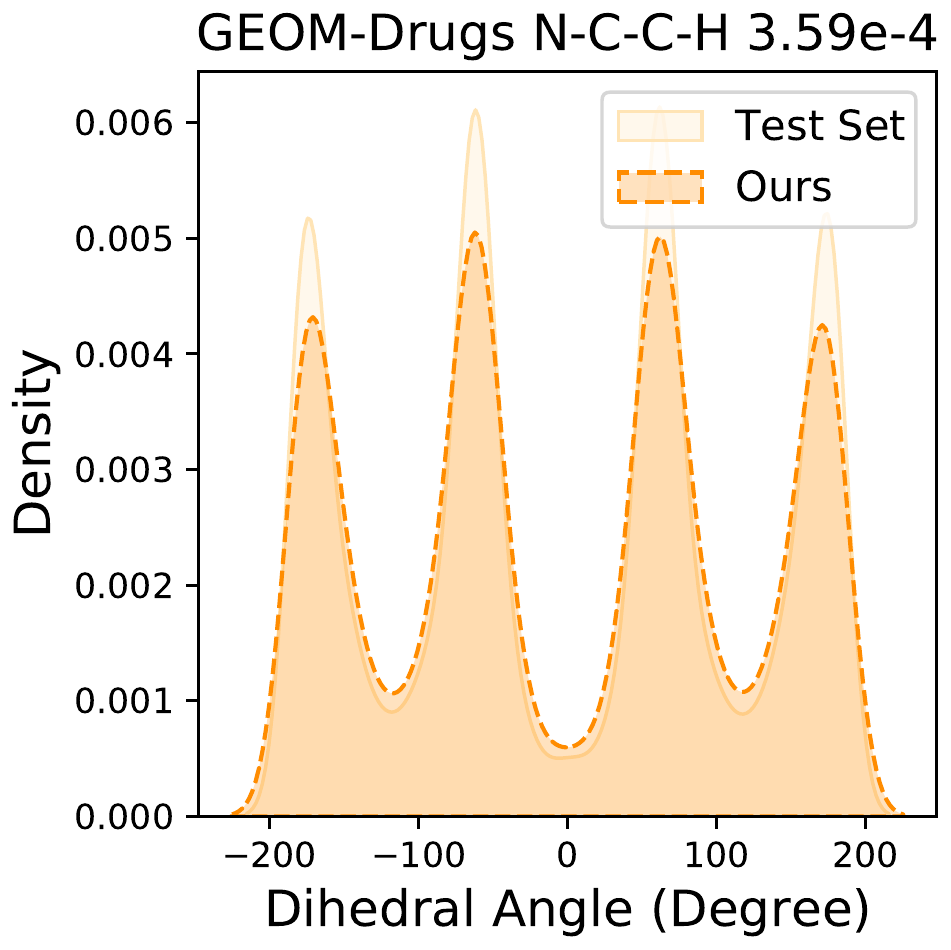}
    \end{subfigure} 

    \caption{Extra distribution comparison of bond angles and dihedral angles between test set molecules and molecules generated from JODO. The MMD distances and corresponding substructures are reported in the titles.}
    \label{fig:more_dist}
\end{figure*}

In this section, we present more experimental results to evaluate the generation quality of JODO. 
Figure \ref{fig:more_dist} shows that JODO matches the test set distribution well for more frequent bond angles and dihedral angles.
This indicates that JODO generates coherent molecular graphs and 3D geometries instead of random spatial configurations.
Extra ablation studies in the QM9 dataset are reported in Table \ref{tab:abl_qm9}.
We obtain similar conclusions as on the GEOM-Drug dataset, but the performance difference is smaller because QM9 has simpler molecules and is easier to learn.

\begin{figure*}[t]
    \centering
    \begin{subfigure}{\textwidth}
        \centering
        \includegraphics[width=\textwidth]{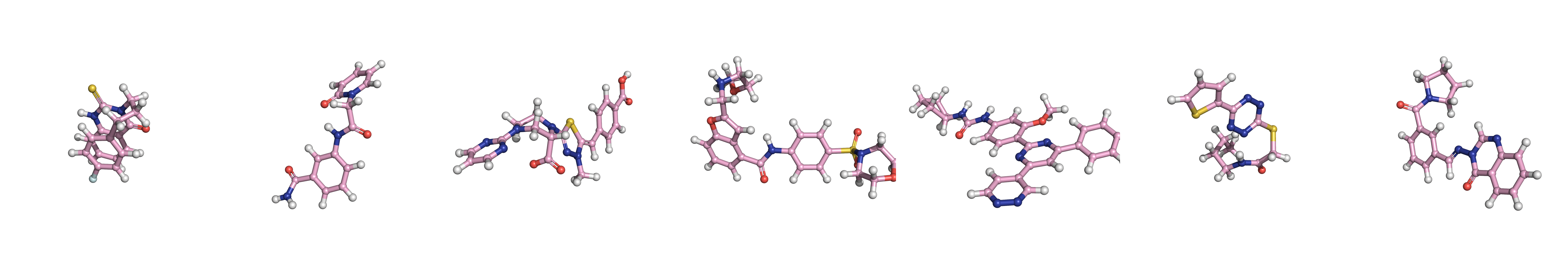}
    \end{subfigure}
    \begin{subfigure}{\textwidth}
        \centering
        \includegraphics[width=\textwidth]{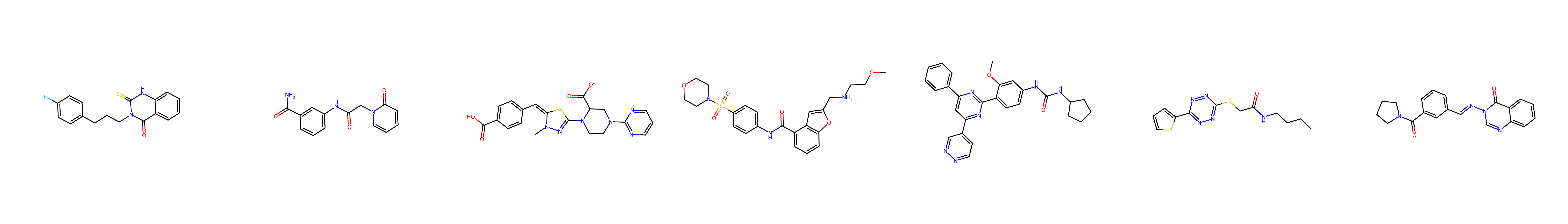}
    \end{subfigure}
    
    \begin{subfigure}{\textwidth}
        \centering
        \includegraphics[width=\textwidth]{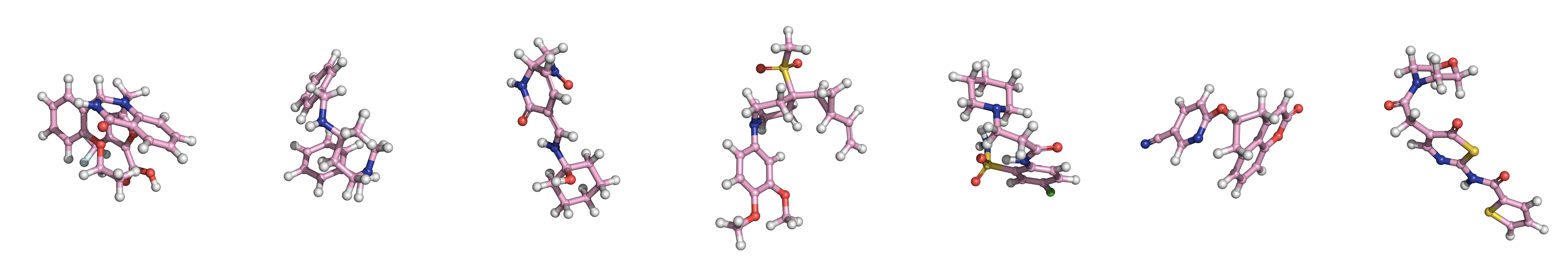}
    \end{subfigure}
    \begin{subfigure}{\textwidth}
        \centering
        \includegraphics[width=\textwidth]{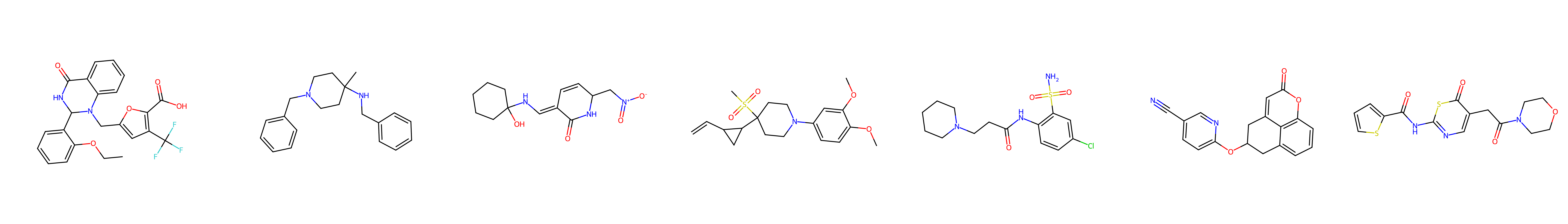}
    \end{subfigure}
    \begin{subfigure}{\textwidth}
        \centering
        \includegraphics[width=\textwidth]{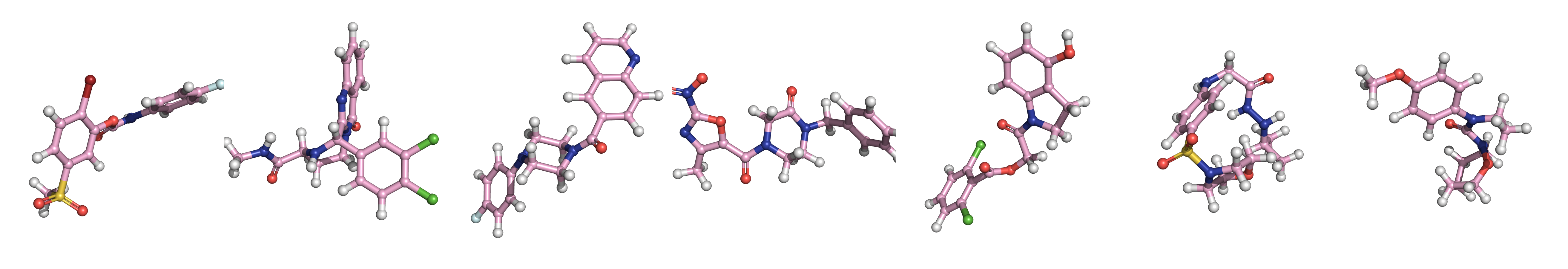}
    \end{subfigure}
    \begin{subfigure}{\textwidth}
        \centering
        \includegraphics[width=\textwidth]{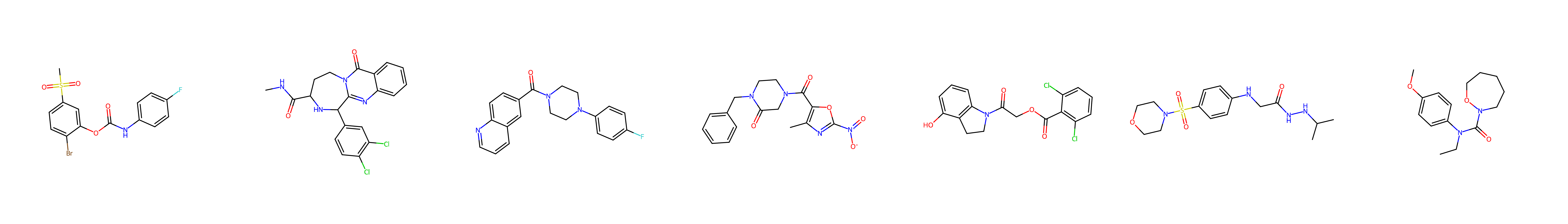}
    \end{subfigure}
    \caption{Visualization of random samples generated by JODO trained on GEOM-Drugs.}
    \label{fig:more_geom_vis}
\end{figure*}

\begin{figure*}[t]
\centering
    \begin{subfigure}{\textwidth}
        \centering
        \includegraphics[width=\textwidth]{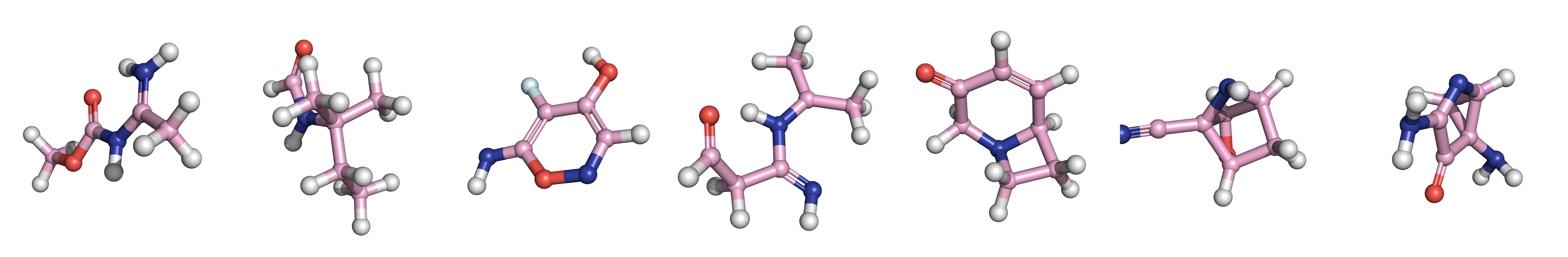}
    \end{subfigure}
    \begin{subfigure}{\textwidth}
        \centering
        \includegraphics[width=\textwidth]{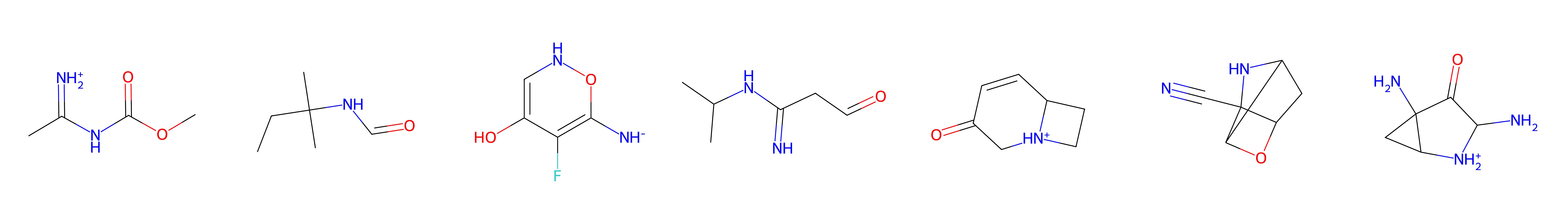}
    \end{subfigure}

    \begin{subfigure}{\textwidth}
        \centering
        \includegraphics[width=\textwidth]{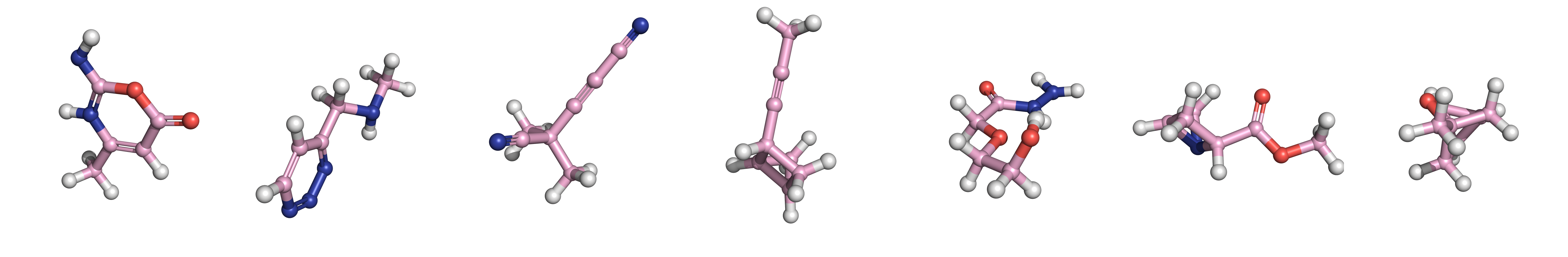}
    \end{subfigure}
    \begin{subfigure}{\textwidth}
        \centering
        \includegraphics[width=\textwidth]{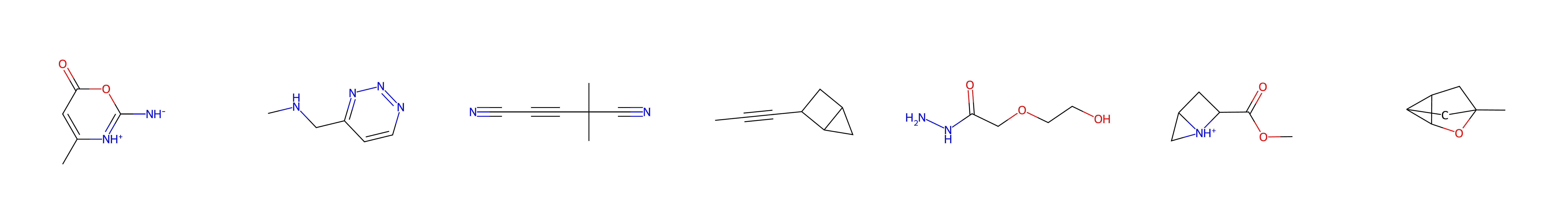}
    \end{subfigure}

    \begin{subfigure}{\textwidth}
        \centering
        \includegraphics[width=\textwidth]{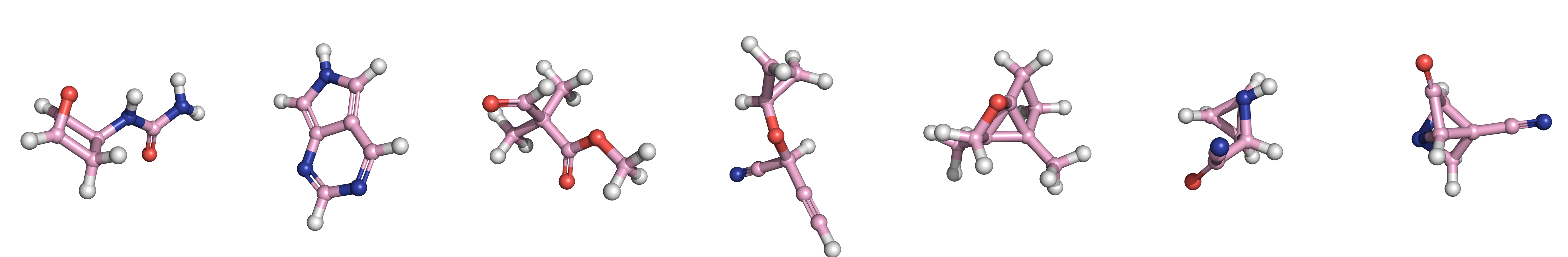}
    \end{subfigure}
    \begin{subfigure}{\textwidth}
        \centering
        \includegraphics[width=\textwidth]{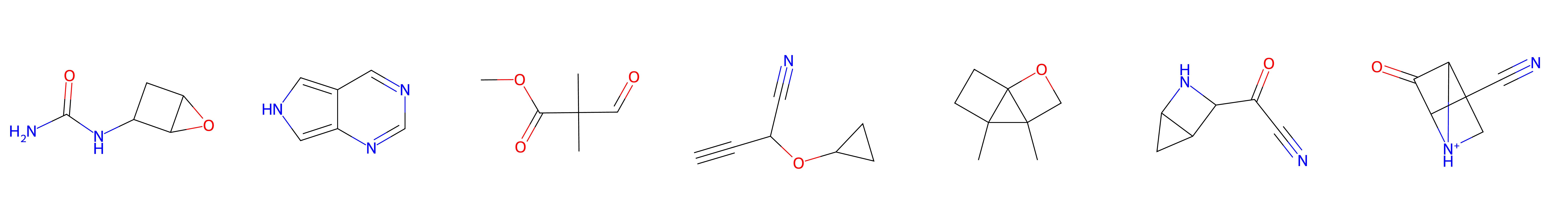}
    \end{subfigure}
    
    \caption{Visualization of random samples generated by JODO trained on QM9 with explicit hydrogen atoms.}
    \label{fig:more_qm9_vis}
\end{figure*}

\begin{figure*}
    \centering
    \includegraphics[width=\textwidth]{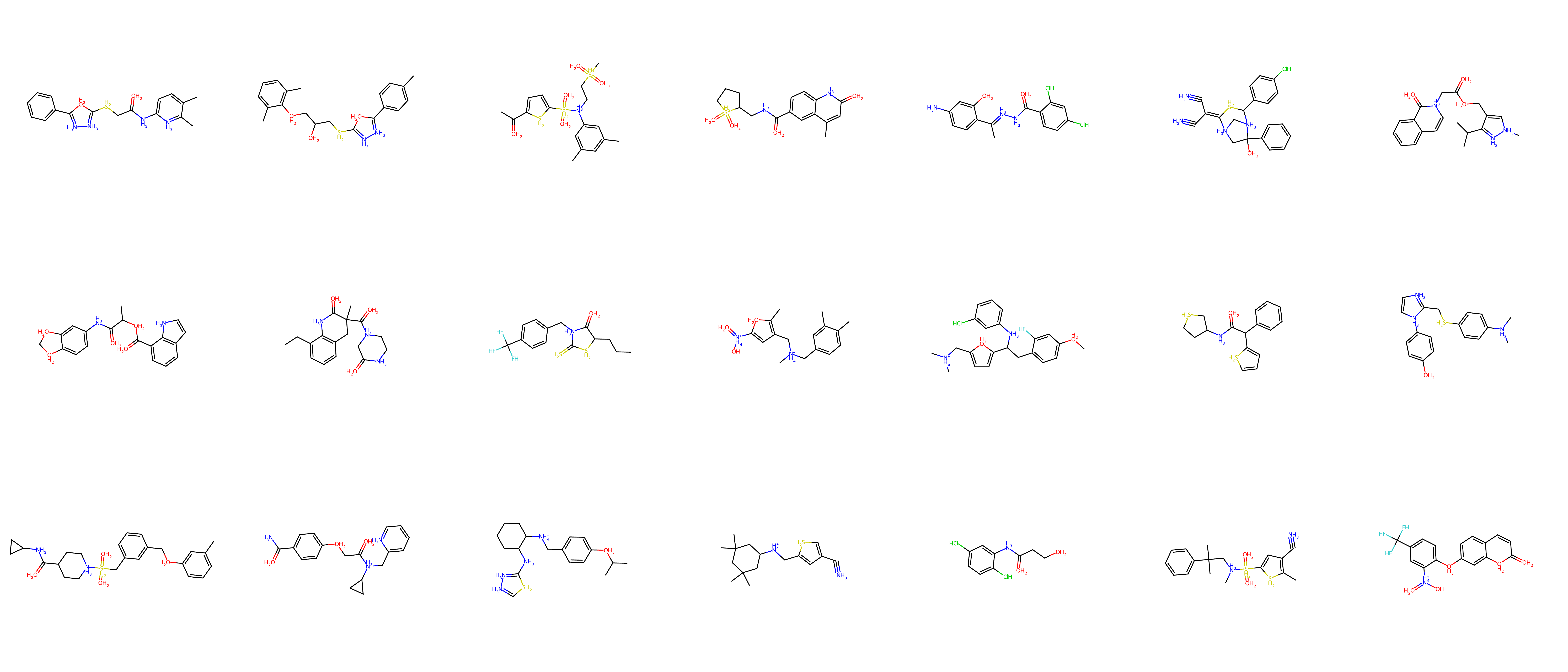}
    \caption{Visualization of random samples generated by JODO-2D trained on the ZINC250k dataset.}
    \label{fig:vis_zinc}
\end{figure*}

\begin{figure*}
    \centering
    \includegraphics[width=\textwidth]{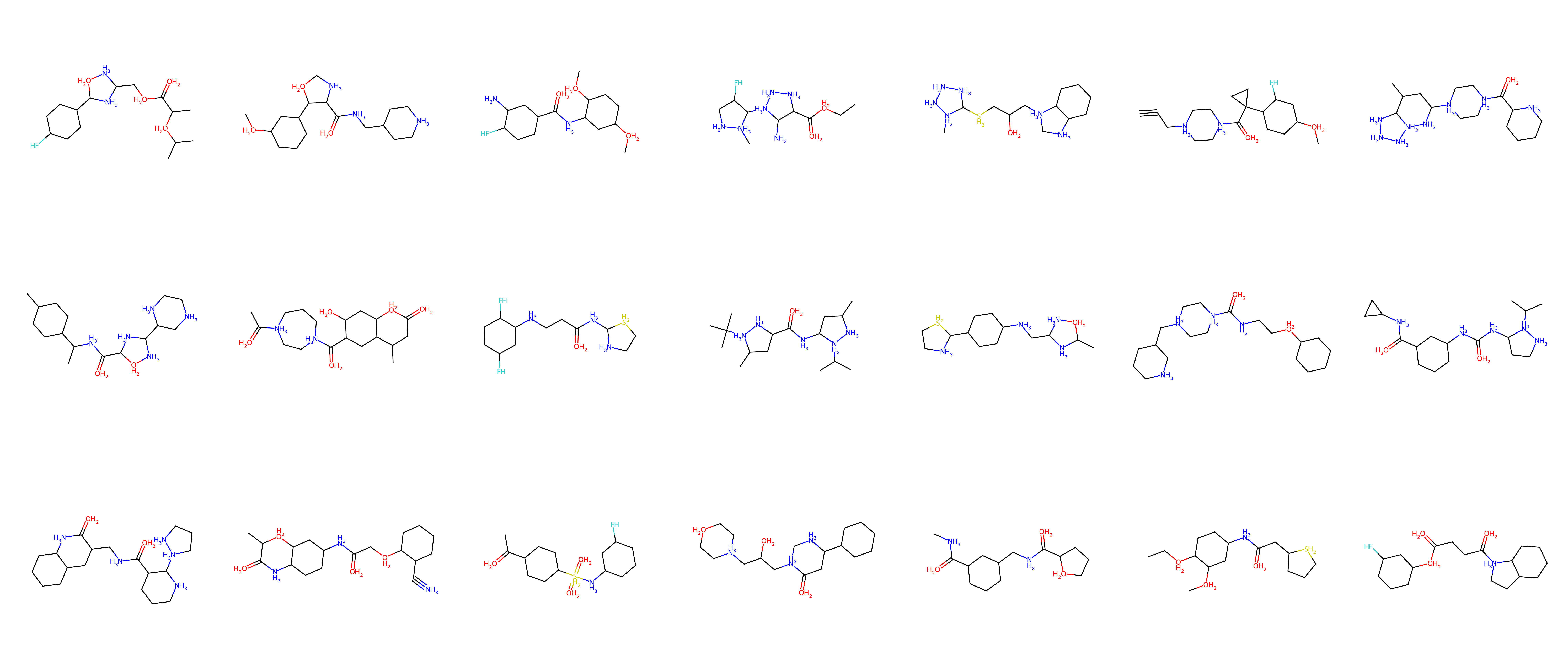}
    \caption{Visualization of random samples generated by JODO-2D trained on the MOSES dataset.}
    \label{fig:vis_mose}
\end{figure*}

More visualizations of complete molecules sampled from JODO on GEOM-Drugs and QM9 are shown in Figure \ref{fig:more_geom_vis} and Figure \ref{fig:more_qm9_vis} respectively.
These samples are randomly selected.
Some conformers may not show their 3D geometries clearly due to the viewing direction setting. 
JODO generates high-fidelity molecules with realistic 2D and 3D descriptors for various molecule weights.
Especially in the complex GEOM-Drugs, JODO not only avoids generating molecules with disconnected components but also preserves the stable geometric planes of aromatic ring structures.
% In particular, on complicated GEOM-Drugs, JODO is not only less likely to generate molecules with disconnected components but also able to generate stable geometric planes of aromatic ring structures.
Furthermore, Figure \ref{fig:vis_zinc} and Figure \ref{fig:vis_mose} show the molecular graphs generated by JODO-2D trained on ZINC250k and MOSES datasets.
Our model generates complete and valid molecular graphs in one shot.

Table \ref{tab:extra_cond} also provides additional results of molecule quality in conditional generation targeted at quantum properties. We conclude that our conditional model is faithful to the given information and achieves superior general quality in stability and distribution learning.

\begin{table}[!htbp]
\caption{Ablation studies on the QM9 dataset.}
\label{tab:abl_qm9}
\renewcommand\arraystretch{1.5}
\resizebox{\columnwidth}{!}{%
\begin{tabular}{lccccc}
\hline
Ablation & \multicolumn{5}{c}{QM9} \\ \hline
 & Mol-S-3D$\uparrow$ & FCD-3D$\downarrow$ & Mol-S-2D$\uparrow$ & FCD-2D$\downarrow$ & V\&C$\uparrow$ \\ \hline
Train & 95.3\% & 0.877 & 98.8\% & 0.063 & 98.9\% \\ \hline
Base Model & 93.4\% & 0.885 & 98.8\% & 0.138 & 99.0\% \\ \hline
2D only & - & - & 98.0\% & 0.104 & 98.0\% \\
3D only & 79.3\% & 0.961 & - & - & - \\ \hline
Attn-add & 92.0\% & 0.846 & 98.4\% & 0.162 & 98.5\% \\
Attn-multi & 92.2\% & 0.929 & 98.7\% & 0.134 & 98.7\% \\ \hline
w/d self-cond & 90.8\% & 1.031 & 98.2\% & 0.218 & 98.4\% \\
self-cond input & 93.4\% & 0.884 & 98.8\% & 0.152 & 99.0\% \\ \hline
w/d time-cond & 92.0\% & 0.861 & 98.8\% & 0.151 & 98.9\% \\ \hline
\end{tabular}%
}
\end{table}

\begin{table}[!htbp]
\caption{Additional results of molecule quality in quantum property conditional generation.}
\label{tab:extra_cond}
\renewcommand\arraystretch{1.2}
\resizebox{\columnwidth}{!}{%
\begin{tabular}{lcccc}
\hline
Condition & Mol-S-3D & FCD-3D & Mol-S-2D & FCD-2D \\ \hline
$C_v$ & 91.75\% & 0.878 & 98.29\% & 0.141 \\
$\mu$ & 93.86\% & 0.850 & 98.49\% & 0.110 \\
$\alpha$ & 93.07\% & 0.917 & 98.68\% & 0.180 \\
$\Delta \epsilon$ & 94.03\% & 0.867 & 98.69\% & 0.105 \\
$\epsilon_{\mathrm{HOMO}}$ & 94.02\% & 0.855 & 98.80\% & 0.131 \\
$\epsilon_{\mathrm{LUMO}}$ & 92.30\% & 0.889 & 98.02\% & 0.106 \\
$C_v, \mu$ & 92.97\% & 0.934 & 98.38\% & 0.161 \\
$\Delta \epsilon, \mu$ & 93.35\% & 0.937 & 98.41\% & 0.172 \\
$\alpha, \mu$ & 94.24\% & 0.906 & 98.61\% & 0.162 \\ \hline
\end{tabular}%
}
\end{table}

%% file: main.bbl
\begin{thebibliography}{10}
\providecommand{\url}[1]{#1}
\csname url@samestyle\endcsname
\providecommand{\newblock}{\relax}
\providecommand{\bibinfo}[2]{#2}
\providecommand{\BIBentrySTDinterwordspacing}{\spaceskip=0pt\relax}
\providecommand{\BIBentryALTinterwordstretchfactor}{4}
\providecommand{\BIBentryALTinterwordspacing}{\spaceskip=\fontdimen2\font plus
\BIBentryALTinterwordstretchfactor\fontdimen3\font minus
  \fontdimen4\font\relax}
\providecommand{\BIBforeignlanguage}[2]{{%
\expandafter\ifx\csname l@#1\endcsname\relax
\typeout{** WARNING: IEEEtran.bst: No hyphenation pattern has been}%
\typeout{** loaded for the language `#1'. Using the pattern for}%
\typeout{** the default language instead.}%
\else
\language=\csname l@#1\endcsname
\fi
#2}}
\providecommand{\BIBdecl}{\relax}
\BIBdecl

\bibitem{jumper2021highly}
J.~Jumper, R.~Evans, A.~Pritzel, T.~Green, M.~Figurnov, O.~Ronneberger,
  K.~Tunyasuvunakool, R.~Bates, A.~{\v{Z}}{\'\i}dek, A.~Potapenko
  \emph{et~al.}, ``Highly accurate protein structure prediction with
  alphafold,'' \emph{Nature}, vol. 596, no. 7873, pp. 583--589, 2021.

\bibitem{zhavoronkov2019deep}
A.~Zhavoronkov, Y.~A. Ivanenkov, A.~Aliper, M.~S. Veselov, V.~A. Aladinskiy,
  A.~V. Aladinskaya, V.~A. Terentiev, D.~A. Polykovskiy, M.~D. Kuznetsov,
  A.~Asadulaev \emph{et~al.}, ``Deep learning enables rapid identification of
  potent ddr1 kinase inhibitors,'' \emph{Nature biotechnology}, vol.~37, no.~9,
  pp. 1038--1040, 2019.

\bibitem{butler2018machine}
K.~T. Butler, D.~W. Davies, H.~Cartwright, O.~Isayev, and A.~Walsh, ``Machine
  learning for molecular and materials science,'' \emph{Nature}, vol. 559, no.
  7715, pp. 547--555, 2018.

\bibitem{polykovskiy20MOSES}
D.~Polykovskiy, A.~Zhebrak, B.~Sanchez-Lengeling, S.~Golovanov, O.~Tatanov,
  S.~Belyaev, R.~Kurbanov, A.~Artamonov, V.~Aladinskiy, M.~Veselov
  \emph{et~al.}, ``Molecular sets (moses): a benchmarking platform for
  molecular generation models,'' \emph{Frontiers in pharmacology}, vol.~11, p.
  565644, 2020.

\bibitem{du2022molgensurvey}
Y.~Du, T.~Fu, J.~Sun, and S.~Liu, ``Molgensurvey: A systematic survey in
  machine learning models for molecule design,'' \emph{arXiv preprint
  arXiv:2203.14500}, 2022.

\bibitem{Liu18CGVAE}
Q.~Liu, M.~Allamanis, M.~Brockschmidt, and A.~L. Gaunt, ``Constrained graph
  variational autoencoders for molecule design,'' in \emph{NeurIPS 2018}, 2018,
  pp. 7806--7815.

\bibitem{LuoGraphdf21}
Y.~Luo, K.~Yan, and S.~Ji, ``Graphdf: {A} discrete flow model for molecular
  graph generation,'' in \emph{ICML}, 2021, pp. 7192--7203.

\bibitem{huangCDGS23}
H.~Huang, L.~Sun, B.~Du, and W.~Lv, ``Conditional diffusion based on discrete
  graph structures for molecular graph generation,'' in \emph{AAAI}, 2023.

\bibitem{JoLH22GDSS}
J.~Jo, S.~Lee, and S.~J. Hwang, ``Score-based generative modeling of graphs via
  the system of stochastic differential equations,'' in \emph{{ICML}}, 2022,
  pp. 10\,362--10\,383.

\bibitem{vignacDigress22}
C.~Vignac, I.~Krawczuk, A.~Siraudin, B.~Wang, V.~Cevher, and P.~Frossard,
  ``Digress: Discrete denoising diffusion for graph generation,'' \emph{arXiv
  preprint arXiv:2209.14734}, 2022.

\bibitem{anderson2003SBDD}
A.~C. Anderson, ``The process of structure-based drug design,'' \emph{Chemistry
  \& biology}, vol.~10, no.~9, pp. 787--797, 2003.

\bibitem{gebauerG-sch19}
N.~Gebauer, M.~Gastegger, and K.~Sch{\"u}tt, ``Symmetry-adapted generation of
  3d point sets for the targeted discovery of molecules,'' \emph{Advances in
  neural information processing systems}, vol.~32, 2019.

\bibitem{simm2020symmetry}
G.~N. Simm, R.~Pinsler, G.~Cs{\'a}nyi, and J.~M. Hern{\'a}ndez-Lobato,
  ``Symmetry-aware actor-critic for 3d molecular design,'' \emph{arXiv preprint
  arXiv:2011.12747}, 2020.

\bibitem{simm20RL}
G.~Simm, R.~Pinsler, and J.~M. Hern{\'a}ndez-Lobato, ``Reinforcement learning
  for molecular design guided by quantum mechanics,'' in \emph{ICML}.\hskip 1em
  plus 0.5em minus 0.4em\relax PMLR, 2020, pp. 8959--8969.

\bibitem{gebauer2022inverse}
N.~W. Gebauer, M.~Gastegger, S.~S. Hessmann, K.-R. M{\"u}ller, and K.~T.
  Sch{\"u}tt, ``Inverse design of 3d molecular structures with conditional
  generative neural networks,'' \emph{Nature communications}, vol.~13, no.~1,
  p. 973, 2022.

\bibitem{garciaE-NF21}
V.~Garcia~Satorras, E.~Hoogeboom, F.~Fuchs, I.~Posner, and M.~Welling, ``E (n)
  equivariant normalizing flows,'' \emph{NeurIPS}, vol.~34, pp. 4181--4192,
  2021.

\bibitem{luoG-sphere22}
Y.~Luo and S.~Ji, ``An autoregressive flow model for 3d molecular geometry
  generation from scratch,'' in \emph{ICLR}, 2022.

\bibitem{HoogeboomEDM22}
E.~Hoogeboom, V.~G. Satorras, C.~Vignac, and M.~Welling, ``Equivariant
  diffusion for molecule generation in 3d,'' in \emph{ICML}, 2022, pp.
  8867--8887.

\bibitem{wu22diffusionbridge}
L.~Wu, C.~Gong, X.~Liu, M.~Ye, and Q.~Liu, ``Diffusion-based molecule
  generation with informative prior bridges,'' \emph{arXiv preprint
  arXiv:2209.00865}, 2022.

\bibitem{baoEEGSDE22}
F.~Bao, M.~Zhao, Z.~Hao, P.~Li, C.~Li, and J.~Zhu, ``Equivariant energy-guided
  sde for inverse molecular design,'' \emph{arXiv preprint arXiv:2209.15408},
  2022.

\bibitem{axelrod2022geom}
S.~Axelrod and R.~Gomez-Bombarelli, ``Geom, energy-annotated molecular
  conformations for property prediction and molecular generation,''
  \emph{Scientific Data}, vol.~9, no.~1, p. 185, 2022.

\bibitem{nesterov20203dmolnet}
V.~Nesterov, M.~Wieser, and V.~Roth, ``3dmolnet: a generative network for
  molecular structures,'' \emph{arXiv preprint arXiv:2010.06477}, 2020.

\bibitem{roney22GEN3D}
J.~P. Roney, P.~Maragakis, P.~Skopp, and D.~E. Shaw, ``Generating realistic 3d
  molecules with an equivariant conditional likelihood model,'' 2022.

\bibitem{sohl2015deep}
J.~Sohl-Dickstein, E.~Weiss, N.~Maheswaranathan, and S.~Ganguli, ``Deep
  unsupervised learning using nonequilibrium thermodynamics,'' in \emph{ICML},
  2015, pp. 2256--2265.

\bibitem{ho2020denoising}
J.~Ho, A.~Jain, and P.~Abbeel, ``Denoising diffusion probabilistic models,''
  \emph{NeurIPS}, vol.~33, pp. 6840--6851, 2020.

\bibitem{song2021score}
Y.~Song, J.~Sohl{-}Dickstein, D.~P. Kingma, A.~Kumar, S.~Ermon, and B.~Poole,
  ``Score-based generative modeling through stochastic differential
  equations,'' in \emph{{ICLR}}, 2021.

\bibitem{huangGraphgdp22}
H.~Huang, L.~Sun, B.~Du, Y.~Fu, and W.~Lv, ``Graphgdp: Generative diffusion
  processes for permutation invariant graph generation,'' in \emph{ICDM}, 2022,
  pp. 201--210.

\bibitem{chen2022analog}
T.~Chen, R.~Zhang, and G.~Hinton, ``Analog bits: Generating discrete data using
  diffusion models with self-conditioning,'' \emph{arXiv preprint
  arXiv:2208.04202}, 2022.

\bibitem{dieleman2022continuous}
S.~Dieleman, L.~Sartran, A.~Roshannai, N.~Savinov, Y.~Ganin, P.~H. Richemond,
  A.~Doucet, R.~Strudel, C.~Dyer, C.~Durkan \emph{et~al.}, ``Continuous
  diffusion for categorical data,'' \emph{arXiv preprint arXiv:2211.15089},
  2022.

\bibitem{satorras21EGNN}
V.~G. Satorras, E.~Hoogeboom, and M.~Welling, ``E (n) equivariant graph neural
  networks,'' in \emph{ICML}.\hskip 1em plus 0.5em minus 0.4em\relax PMLR,
  2021, pp. 9323--9332.

\bibitem{perez2018film}
E.~Perez, F.~Strub, H.~De~Vries, V.~Dumoulin, and A.~Courville, ``Film: Visual
  reasoning with a general conditioning layer,'' in \emph{Proceedings of the
  AAAI Conference on Artificial Intelligence}, vol.~32, no.~1, 2018.

\bibitem{dhariwal2021diffusion}
P.~Dhariwal and A.~Nichol, ``Diffusion models beat gans on image synthesis,''
  \emph{Advances in Neural Information Processing Systems}, vol.~34, pp.
  8780--8794, 2021.

\bibitem{peebles2022scalable}
W.~Peebles and S.~Xie, ``Scalable diffusion models with transformers,''
  \emph{arXiv preprint arXiv:2212.09748}, 2022.

\bibitem{Jin18JT-VAE}
W.~Jin, R.~Barzilay, and T.~S. Jaakkola, ``Junction tree variational
  autoencoder for molecular graph generation,'' in \emph{ICML}, J.~G. Dy and
  A.~Krause, Eds., 2018, pp. 2328--2337.

\bibitem{Youpolicy18}
J.~You, B.~Liu, Z.~Ying, V.~S. Pande, and J.~Leskovec, ``Graph convolutional
  policy network for goal-directed molecular graph generation,'' in
  \emph{NeurIPS}, 2018, pp. 6412--6422.

\bibitem{ShiGraphaf20}
C.~Shi, M.~Xu, Z.~Zhu, W.~Zhang, M.~Zhang, and J.~Tang, ``Graphaf: a flow-based
  autoregressive model for molecular graph generation,'' in \emph{{ICLR}},
  2020.

\bibitem{zangMoflow20}
C.~Zang and F.~Wang, ``Moflow: an invertible flow model for generating
  molecular graphs,'' in \emph{SIGKDD}, 2020, pp. 617--626.

\bibitem{lippeGraphcnf21}
P.~Lippe and E.~Gavves, ``Categorical normalizing flows via continuous
  transformations,'' in \emph{{ICLR}}, 2021.

\bibitem{simonovsky2018graphvae}
M.~Simonovsky and N.~Komodakis, ``Graphvae: Towards generation of small graphs
  using variational autoencoders,'' in \emph{ICANN}.\hskip 1em plus 0.5em minus
  0.4em\relax Springer, 2018, pp. 412--422.

\bibitem{de2018molgan}
N.~De~Cao and T.~Kipf, ``Molgan: An implicit generative model for small
  molecular graphs,'' \emph{arXiv preprint arXiv:1805.11973}, 2018.

\bibitem{DEFactor18}
R.~Assouel, M.~Ahmed, M.~H.~S. Segler, A.~Saffari, and Y.~Bengio, ``Defactor:
  Differentiable edge factorization-based probabilistic graph generation,''
  \emph{arXiv preprint arXiv: 1811.09766}, 2018.

\bibitem{HoogeboomArgmax21}
E.~Hoogeboom, D.~Nielsen, P.~Jaini, P.~Forr{\'{e}}, and M.~Welling, ``Argmax
  flows and multinomial diffusion: Learning categorical distributions,'' in
  \emph{NeurIPS}, 2021, pp. 12\,454--12\,465.

\bibitem{AustinD3PM21}
J.~Austin, D.~D. Johnson, J.~Ho, D.~Tarlow, and R.~van~den Berg, ``Structured
  denoising diffusion models in discrete state-spaces,'' in \emph{NeurIPS},
  2021, pp. 17\,981--17\,993.

\bibitem{tashiro2021csdi}
Y.~Tashiro, J.~Song, Y.~Song, and S.~Ermon, ``Csdi: Conditional score-based
  diffusion models for probabilistic time series imputation,'' \emph{NeurIPS},
  vol.~34, pp. 24\,804--24\,816, 2021.

\bibitem{liu2023pristi}
M.~Liu, H.~Huang, H.~Feng, L.~Sun, B.~Du, and Y.~Fu, ``Pristi: A conditional
  diffusion framework for spatiotemporal imputation,'' \emph{ICDE}, 2023.

\bibitem{yang2022diffusion}
L.~Yang, Z.~Zhang, Y.~Song, S.~Hong, R.~Xu, Y.~Zhao, Y.~Shao, W.~Zhang, B.~Cui,
  and M.-H. Yang, ``Diffusion models: A comprehensive survey of methods and
  applications,'' \emph{arXiv preprint arXiv:2209.00796}, 2022.

\bibitem{huangMDM22}
L.~Huang, H.~Zhang, T.~Xu, and K.-C. Wong, ``Mdm: Molecular diffusion model for
  3d molecule generation,'' \emph{arXiv preprint arXiv:2209.05710}, 2022.

\bibitem{xu2022geodiff}
M.~Xu, L.~Yu, Y.~Song, C.~Shi, S.~Ermon, and J.~Tang, ``Geodiff: A geometric
  diffusion model for molecular conformation generation,'' in \emph{ICLR},
  2022.

\bibitem{torsional22}
B.~Jing, G.~Corso, J.~Chang, R.~Barzilay, and T.~S. Jaakkola, ``Torsional
  diffusion for molecular conformer generation,'' \emph{arXiv preprint
  arXiv:2206.01729}, 2022.

\bibitem{anand2022protein}
N.~Anand and T.~Achim, ``Protein structure and sequence generation with
  equivariant denoising diffusion probabilistic models,'' \emph{arXiv preprint
  arXiv:2205.15019}, 2022.

\bibitem{wu2022protein}
K.~E. Wu, K.~K. Yang, R.~v.~d. Berg, J.~Y. Zou, A.~X. Lu, and A.~P. Amini,
  ``Protein structure generation via folding diffusion,'' \emph{arXiv preprint
  arXiv:2209.15611}, 2022.

\bibitem{ingraham2022illuminating}
J.~Ingraham, M.~Baranov, Z.~Costello, V.~Frappier, A.~Ismail, S.~Tie, W.~Wang,
  V.~Xue, F.~Obermeyer, A.~Beam \emph{et~al.}, ``Illuminating protein space
  with a programmable generative model,'' \emph{bioRxiv}, pp. 2022--12, 2022.

\bibitem{watson2022broadly}
J.~L. Watson, D.~Juergens, N.~R. Bennett, B.~L. Trippe, J.~Yim, H.~E. Eisenach,
  W.~Ahern, A.~J. Borst, R.~J. Ragotte, L.~F. Milles \emph{et~al.}, ``Broadly
  applicable and accurate protein design by integrating structure prediction
  networks and diffusion generative models,'' \emph{bioRxiv}, pp. 2022--12,
  2022.

\bibitem{luo2022antigen}
S.~Luo, Y.~Su, X.~Peng, S.~Wang, J.~Peng, and J.~Ma, ``Antigen-specific
  antibody design and optimization with diffusion-based generative models,''
  \emph{bioRxiv}, pp. 2022--07, 2022.

\bibitem{corso2022diffdock}
G.~Corso, H.~St{\"a}rk, B.~Jing, R.~Barzilay, and T.~Jaakkola, ``Diffdock:
  Diffusion steps, twists, and turns for molecular docking,'' \emph{arXiv
  preprint arXiv:2210.01776}, 2022.

\bibitem{schutt2018schnet}
K.~T. Sch{\"u}tt, H.~E. Sauceda, P.-J. Kindermans, A.~Tkatchenko, and K.-R.
  M{\"u}ller, ``Schnet--a deep learning architecture for molecules and
  materials,'' \emph{The Journal of Chemical Physics}, vol. 148, no.~24, p.
  241722, 2018.

\bibitem{gasteiger2020directional}
J.~Gasteiger, J.~Gro{\ss}, and S.~G{\"u}nnemann, ``Directional message passing
  for molecular graphs,'' \emph{arXiv preprint arXiv:2003.03123}, 2020.

\bibitem{gasteiger2021gemnet}
J.~Gasteiger, F.~Becker, and S.~G{\"u}nnemann, ``Gemnet: Universal directional
  graph neural networks for molecules,'' \emph{NeurIPS}, vol.~34, pp.
  6790--6802, 2021.

\bibitem{liu2022spherical}
Y.~Liu, L.~Wang, M.~Liu, Y.~Lin, X.~Zhang, B.~Oztekin, and S.~Ji, ``Spherical
  message passing for 3d molecular graphs,'' in \emph{ICLR}, 2022.

\bibitem{schutt2021equivariant}
K.~Sch{\"u}tt, O.~Unke, and M.~Gastegger, ``Equivariant message passing for the
  prediction of tensorial properties and molecular spectra,'' in
  \emph{International Conference on Machine Learning}.\hskip 1em plus 0.5em
  minus 0.4em\relax PMLR, 2021, pp. 9377--9388.

\bibitem{jing2020learning}
B.~Jing, S.~Eismann, P.~Suriana, R.~J. Townshend, and R.~Dror, ``Learning from
  protein structure with geometric vector perceptrons,'' \emph{arXiv preprint
  arXiv:2009.01411}, 2020.

\bibitem{shi2022benchmarking}
Y.~Shi, S.~Zheng, G.~Ke, Y.~Shen, J.~You, J.~He, S.~Luo, C.~Liu, D.~He, and
  T.-Y. Liu, ``Benchmarking graphormer on large-scale molecular modeling
  datasets,'' \emph{arXiv preprint arXiv:2203.04810}, 2022.

\bibitem{graphformer21}
C.~Ying, T.~Cai, S.~Luo, S.~Zheng, G.~Ke, D.~He, Y.~Shen, and T.~Liu, ``Do
  transformers really perform badly for graph representation?'' in
  \emph{NeurIPS}, 2021, pp. 28\,877--28\,888.

\bibitem{joshi2023expressive}
C.~K. Joshi, C.~Bodnar, S.~V. Mathis, T.~Cohen, and P.~Li{\`o}, ``On the
  expressive power of geometric graph neural networks,'' \emph{arXiv preprint
  arXiv:2301.09308}, 2023.

\bibitem{luo2022one}
S.~Luo, T.~Chen, Y.~Xu, S.~Zheng, T.-Y. Liu, L.~Wang, and D.~He, ``One
  transformer can understand both 2d \& 3d molecular data,'' \emph{arXiv
  preprint arXiv:2210.01765}, 2022.

\bibitem{VDM2021}
D.~Kingma, T.~Salimans, B.~Poole, and J.~Ho, ``Variational diffusion models,''
  in \emph{NeurIPS}, 2021.

\bibitem{kohler2020equivariant}
J.~K{\"o}hler, L.~Klein, and F.~No{\'e}, ``Equivariant flows: exact likelihood
  generative learning for symmetric densities,'' in \emph{ICML}.\hskip 1em plus
  0.5em minus 0.4em\relax PMLR, 2020, pp. 5361--5370.

\bibitem{kabsch1976solution}
W.~Kabsch, ``A solution for the best rotation to relate two sets of vectors,''
  \emph{Acta Crystallographica Section A: Crystal Physics, Diffraction,
  Theoretical and General Crystallography}, vol.~32, no.~5, pp. 922--923, 1976.

\bibitem{nichol2021improved}
A.~Q. Nichol and P.~Dhariwal, ``Improved denoising diffusion probabilistic
  models,'' in \emph{ICML}.\hskip 1em plus 0.5em minus 0.4em\relax PMLR, 2021,
  pp. 8162--8171.

\bibitem{vaswani2017attention}
A.~Vaswani, N.~Shazeer, N.~Parmar, J.~Uszkoreit, L.~Jones, A.~N. Gomez,
  {\L}.~Kaiser, and I.~Polosukhin, ``Attention is all you need,'' in
  \emph{NeurIPS}, 2017.

\bibitem{qm9}
R.~Ramakrishnan, P.~O. Dral, M.~Rupp, and O.~A. von Lilienfeld, ``Quantum
  chemistry structures and properties of 134 kilo molecules,'' \emph{Scientific
  Data}, vol.~1, 2014.

\bibitem{RDKit}
\BIBentryALTinterwordspacing
G.~Landrum, ``Rdkit: Open-source cheminformatics software,'' 2016. [Online].
  Available: \url{http://www.rdkit.org}
\BIBentrySTDinterwordspacing

\bibitem{degen2008fragment}
J.~Degen, C.~Wegscheid-Gerlach, A.~Zaliani, and M.~Rarey, ``On the art of
  compiling and using'drug-like'chemical fragment spaces,'' \emph{ChemMedChem:
  Chemistry Enabling Drug Discovery}, vol.~3, no.~10, pp. 1503--1507, 2008.

\bibitem{bemis1996scaffold}
G.~W. Bemis and M.~A. Murcko, ``The properties of known drugs. 1. molecular
  frameworks,'' \emph{Journal of medicinal chemistry}, vol.~39, no.~15, pp.
  2887--2893, 1996.

\bibitem{peng2022pocket2mol}
X.~Peng, S.~Luo, J.~Guan, Q.~Xie, J.~Peng, and J.~Ma, ``Pocket2mol: Efficient
  molecular sampling based on 3d protein pockets,'' in \emph{International
  Conference on Machine Learning}.\hskip 1em plus 0.5em minus 0.4em\relax PMLR,
  2022, pp. 17\,644--17\,655.

\bibitem{gretton12MMD}
A.~Gretton, K.~M. Borgwardt, M.~J. Rasch, B.~Sch{\"o}lkopf, and A.~Smola, ``A
  kernel two-sample test,'' \emph{The Journal of Machine Learning Research},
  vol.~13, no.~1, pp. 723--773, 2012.

\bibitem{zinc250k}
J.~J. Irwin, T.~Sterling, M.~M. Mysinger, E.~S. Bolstad, and R.~G. Coleman,
  ``{ZINC:} {A} free tool to discover chemistry for biology,'' \emph{J. Chem.
  Inf. Model.}, vol.~52, no.~7, pp. 1757--1768, 2012.

\bibitem{DPMS22}
C.~Lu, Y.~Zhou, F.~Bao, J.~Chen, C.~Li, and J.~Zhu, ``Dpm-solver: {A} fast
  {ODE} solver for diffusion probabilistic model sampling in around 10 steps,''
  \emph{arXiv preprint arXiv:2206.00927}, 2022.

\bibitem{lu2022dpm++}
------, ``Dpm-solver++: Fast solver for guided sampling of diffusion
  probabilistic models,'' \emph{arXiv preprint arXiv:2211.01095}, 2022.

\bibitem{ho2022classifier-free}
J.~Ho and T.~Salimans, ``Classifier-free diffusion guidance,'' \emph{arXiv
  preprint arXiv:2207.12598}, 2022.

\bibitem{niuEDP-GNN20}
C.~Niu, Y.~Song, J.~Song, S.~Zhao, A.~Grover, and S.~Ermon, ``Permutation
  invariant graph generation via score-based generative modeling,'' in
  \emph{AISTATS}, 2020, pp. 4474--4484.

\bibitem{mercado21graphinvent}
R.~Mercado, T.~Rastemo, E.~Lindel{\"o}f, G.~Klambauer, O.~Engkvist, H.~Chen,
  and E.~J. Bjerrum, ``Graph networks for molecular design,'' \emph{Machine
  Learning: Science and Technology}, vol.~2, no.~2, p. 025023, 2021.

\end{thebibliography}
